\documentclass[particles,article,submit,moreauthors]{Definitions/mdpi} 

\renewcommand{\linenumbers}{}

\firstpage{1} 
\pubvolume{1}
\issuenum{1}
\articlenumber{0}
\pubyear{2025}
\copyrightyear{2025}
\datereceived{ } 
\daterevised{ } 
\dateaccepted{ } 
\datepublished{ } 
\hreflink{https://doi.org/} 

\def\noeq#1{(\ref{#1})}
\def\1eq#1{Eq.~(\ref{#1})}

\def\2eqs#1#2{Eqs.~(\ref{#1}) and~(\ref{#2})}
\def\3eqs#1#2#3{Eqs.~(\ref{#1}),~(\ref{#2}) and~(\ref{#3})}

\newenvironment{detailedcalc}
  {\begin{center}\begin{minipage}{0.9\linewidth}\footnotesize}
  {\end{minipage}\end{center}}

\usepackage{amsmath,amssymb}
\usepackage{graphicx}
\usepackage{array,ragged2e,booktabs}

\newcolumntype{C}[1]{>{\Centering\arraybackslash}m{#1}}
\newcolumntype{L}[1]{>{\RaggedRight\arraybackslash}p{#1}}

\newcommand{\leftflushmath}[1]{%
  \noindent\makebox[\linewidth][l]{%
    $\displaystyle
      \begin{aligned}
        #1
      \end{aligned}
    $%
  }%
}

\def\BRSTsrc#1{#1^*}
\def\aBRSTsrc#1{#1^\#}
\def\BRSTaBRSTsrc#1{\widehat{#1}}

\def\caBRSTsrc#1#2{#1^{\star#2}}

\def\caBRSTsrc#1#2{#1^{\##2}}
\def\cBRSTaBRSTsrc#1#2{\widehat{#1}^{#2}}
\def\BRSTaBRSTsrcGhost#1{\widehat{#1}}
\def\BRSTaBRSTcovD{\widehat{\cal D}^{ab}_\mu}

\def\g{\widehat\Gamma^{ \bf np}}
\def\gp{\widehat\Gamma^{ \bf p}}

\usepackage{stackengine}
\stackMath

\Title{Gauge Symmetry Beyond Perturbation Theory: BRST and anti-BRST Structure, Background Fields, and Infrared Dynamics of Yang--Mills Theory}

\TitleCitation{Gauge Symmetry Beyond Perturbation Theory}

\Author{Daniele Binosi $^{1}$\orcidA{}}

\AuthorNames{Daniele Binosi}

\isAPAStyle{%
       \AuthorCitation{Binosi, D.}
         }{%
        \isChicagoStyle{%
        \AuthorCitation{Binosi, Daniele}
        }{
        \AuthorCitation{Binosi, D.}
        }
}

\address{%
$^{1}$ \quad European Centre for Theoretical Studies in Nuclear Theory and Related Areas, Strada delle Tabarelle 286, I-38129 Trento, Italy; binosi@ectstar.eu}

\corres{Correspondence: binosi@ectstar.eu}

\abstract{We present a pedagogical and self contained account of the functional formulation of non-Abelian gauge theories, aimed at the construction of a process independent effective charge for Yang--Mills theory. Starting from the path integral quantization of gauge fields, we review gauge fixing and the emergence of Faddeev--Popov ghosts, illustrating how gauge invariance is preserved at the quantum level through Becchi--Rouet--Stora--Tyutin (BRST) symmetry. We then develop the BRST and anti-BRST formalisms and show how their simultaneous implementation leads to powerful functional identities that severely constrain the ghost and gluon sectors. Background field gauges are introduced as a natural framework in which these symmetries manifest themselves through Abelian like Ward identities, allowing for a transparent separation between quantum and background degrees of freedom. This structure makes it possible to define renormalization group invariant combinations of Green functions that generalize the QED effective charge to the non-Abelian case. The resulting effective charge is shown to be unique, gauge invariant, and process independent, providing a unified description of the theory from the ultraviolet down to the infrared. The interplay between functional identities, Dyson--Schwinger equations, and lattice results is discussed in detail, highlighting how dynamical mass generation and infrared saturation naturally emerge within this framework.
}


\begin{document}


\section{Introduction}

One of the central goals in the study of non-Abelian gauge theories is the construction of physically meaningful quantities that remain well defined beyond perturbation theory. Among these, a process independent effective charge plays a distinguished role, as it provides a unified description of the interaction strength from the ultraviolet (UV) regime, where asymptotic freedom applies, down to the infrared (IR) region, where confinement and dynamical mass generation dominate. Unlike process dependent running couplings extracted from specific observables, such an effective charge must be gauge invariant, renormalization group invariant (RGI), and uniquely defined within the theory~\cite{Deur:2023dzc}.

The construction of a process independent effective charge in Yang--Mills theory is highly nontrivial. Gauge symmetry, while essential for the consistency of the theory, obscures the direct identification of Abelian like structures familiar from quantum electrodynamics. As a result, one must rely on a combination of symmetry principles and functional identities that reorganize Green functions in a way that preserves gauge invariance while allowing for physically transparent definitions. This discussion develops a systematic framework for achieving this goal. 

In Sect.~\ref{one}, starting from the path integral formulation of $SU(N_c)$ Yang--Mills theory, we review the necessity of gauge fixing and the introduction of Faddeev--Popov ghosts. Explicit perturbative examples are used to illustrate how unphysical degrees of freedom cancel and how transversality of Green functions is maintained. 

This sets the stage for the introduction of Becchi--Rouet--Stora--Tyutin (BRST) symmetry in Sect.~\ref{two}, which is shown to replace classical gauge invariance at the quantum level and to provide the algebraic backbone for all subsequent constructions. We then explore the combined BRST and anti-BRST formulation of the theory. The simultaneous implementation of these symmetries leads to additional functional identities, including linear ones, that tightly constrain the ghost sector and play a crucial role in the IR. These identities are shown to encode information that goes well beyond perturbation theory and are essential for establishing relations among Green functions that enter the definition of the effective charge.

A pivotal role in the construction is played by background field (BF) gauges discussed in Sect.~\ref{three}, which are shown to coincide, modulo the separation between background and quantum fields, with the BRST and anti-BRST invariant theory. The linear identities are then revealed to be induced by the residual background gauge invariance, which gives rise to Ward identities (WIs) when the divergence of background legs is considered. In contrast to the Slavnov--Taylor identities (STIs) of the conventional formulation, which generally involve intricate cancellations between gluon and ghost contributions, the WIs are satisfied independently within each class of diagrams. A general proof of this property for Green functions with an arbitrary number of background and quantum gluon legs, as well as ghost and anti-ghost pairs is provided; and the way they allow for symmetry preserving truncations of the Dyson--Schwinger equations (DSEs), also introduced in this Section, discussed. Finally, the background-quantum identities (BQIs), relating background and quantum Green functions in a precise manner are also presented.

Sect.~\ref{four} contains a short presentation of (some) lattice results for the gluon propagator, the ghost dressing function and key Green functions appearing in the BQIs, evaluated in both Landau and linear covariant gauges. These lattice results are discussed in close connection with the continuum framework developed in the previous sections, showing how the observed IR behavior is consistent with the symmetry relations found. 

These simulations thereby provide direct nonperturbative evidence for the IR saturation of the gluon propagator and therefore for a mechanism that endow the gluon with a dynamically generated mass. Understanding the origin and consistency of this phenomenon is the scope of Sect.~\ref{five}. It turns out that BF gauges are particularly well suited for this purpose, as the WIs associated with background gauge invariance simplify considerably the treatment of intermediate steps, with BQIs relating the final results to conventional correlation functions. 

It is only at this point that, building on the knowledge gained, we proceed to define, in Sect.~\ref{six}, a unique, gauge invariant, and process independent effective charge, which encapsulates the dynamics of the theory from the UV to the IR.

\section{\label{one}$SU(N_c)$ Yang-Mills theory: gauge fixing and Faddeev-Popov ghosts}

The action of an $SU(N_c)$ Yang-Mills theory reads
\begin{align}
	S_{\mathrm{YM}} &= \int\!\mathrm{d}^4x\,{\cal L}_{\mathrm{YM}}=-\frac14\int\!\mathrm{d}^4x\,F^a_{\mu\nu}F^{\mu\nu}_a;&
	F^a_{\mu\nu} &=\partial_\mu A^a_\nu-\partial_\nu A^a_\mu+gf^{abc}A^b_\mu A^c_\nu,
	\label{S_YM}
\end{align}
with $A^a_\mu$ the gluon field, and $f^{abc}$ the totally antisymmetric structure constants appearing in the commutation relations satisfied by the $SU(N_c)$ generators $t^a$: $[t^a, t^b] = if^{abc}t^c$.

\noindent The action in~\1eq{S_YM} is invariant under the (infinitesimal) gauge transformations
\begin{align}
	\delta_\theta A^a_\mu&={\cal D}_\mu^{ab}\theta^b;&
	{\cal D}_\mu^{ab}&=\delta^{ab}\partial_\mu-gf^{abc}A^c_\mu,
\end{align}
with ${\cal D}_\mu^{ab}$ the covariant derivative in the adjoint representation and $\theta^b$ the infinitesimal gauge transformation parameters.

When attempting to quantize such theory, one encounters two (related) problems:
\begin{enumerate}
	\item The overcounting of physically equivalent gauge configurations, {\it i.e.}, the ones related by a local gauge transformations. Naively integrating over all field configurations $A^a_\mu$ in the path integral
	\begin{align}
		Z=\int\!{\cal D}A^a_\mu\exp\left(iS_{\mathrm{YM}}\right),
		\label{PInt}
	\end{align}
	one would end up integrating over infinitely many physically equivalent copies and the integral would diverge by acquiring an infinite multiplicative factor corresponding to the volume of the $SU(N_c)$ gauge group. 
	\item The impossibility of defining canonical equal-time commutation relations for the gluon field. The Yang-Mills lagrangian of~\1eq{S_YM} would in fact lead to canonical momenta
	\begin{align}
		\Pi^a_\mu&=\frac{\partial{\cal L}_{\mathrm{YM}}}{\partial(\partial_0A^a_\mu)}=F^a_{0\mu},
	\end{align}
	implying $\Pi^a_0=0$.   
\end{enumerate}
These difficulties appear independently of the particular quantization scheme employed: the first manifests itself in the path-integral formulation, while the second arises in canonical quantization, and both ultimately reflect the gauge redundancy of the classical Yang--Mills action.

\begin{detailedcalc}
Throughout this work we will employ Minkowski-space notation, which is convenient for discussing the structure and symmetries of Green functions and their perturbative properties. It should be understood, however, that the Yang--Mills functional integral is most rigorously defined after a Wick rotation to Euclidean space, where the weight becomes $\exp(-S_E)$. In particular, nonperturbative analyses such as those based on DSEs, bound-state equations, or lattice simulations are properly formulated in Euclidean space. Accordingly, statements concerning nonperturbative dynamics should be interpreted in this Euclidean sense, even when written in Minkowski notation for notational simplicity.
\end{detailedcalc}

It turns out that both problems are simultaneously solved by a gauge-fixing procedure. Dividing out the gauge redundancy in the path integral in~\1eq{PInt} requires selecting a single gauge configuration representative of each gauge orbit. Operationally, this is then done by introducing a gauge-fixing condition ${\cal F}^a=0$, with ${\cal F}^a={\cal F}^a[A]$ the so-called gauge-fixing function, through the    action
\begin{align}
	S_{\mathrm{GF}}&=-\frac1{2\xi}\int\!{\mathrm d}^4x\,{\cal F}^a{\cal F}^a.
	\label{S_GF}
\end{align}

\begin{detailedcalc}

To see how the gauge-fixing action of~\1eq{S_GF} emerges in the path integral quantization, one starts, at a formal level, by inserting in the path integral the unit representation
\begin{align}
	1&=\int\!{\cal D}\theta^b\,\delta({\cal F}^a[A_\theta])\mathrm{det}\left(\left.\frac{\delta{\cal F}^a[A_\theta]}{\delta\theta^b}\right\vert_{\theta=0}\right),
	\label{FP-operator}
\end{align}
where: $A_\theta$ is the gauge transformed field $A^a_{\theta\mu}=A^a_\mu+{\cal D}_\mu^{ab}\theta^b$; and the functional derivative $\frac{\delta{\cal F}^a[A_\theta]}{\delta\theta^b}$ defines the so-called Faddeev-Popov operator ${\cal M}^{ab}$:
\begin{align}
	{\cal M}^{ab}(x,y)&=\left.\frac{\delta{\cal F}^a[A_\theta(x)]}{\delta\theta^b(y)}\right\vert_{\theta=0}=\int\!\mathrm{d}^4z\,\left.\frac{\delta{\cal F}^a[A_\theta(x)]}{\delta A^c_\theta(z)}\right\vert_{\theta=0}\left.\frac{\delta A^c_\theta(z)}{\delta\theta^b(y)}\right\vert_{\theta=0}\nonumber \\
	&=\left.\frac{\delta{\cal F}^a[A_\theta(x)]}{\delta A^c_\theta(y)}\right\vert_{\theta=0}{\cal D}_\mu^{cb}[A].
	\label{FP-operator-1}
\end{align}
(Notice that in the following we will be somewhat more relaxed about the space-time dependence of functional differentiations).

The question is: what are we to do about the $\delta$ function and the determinant in the path integral? Let's treat the $\delta$ first, leaving the determinant for later. Recall that the $\delta$ function is the Fourier transform of unity, or (for each space-time point and color index)
\begin{align}
	\delta({\cal F}^a)={\cal N}\!\int\!{\cal D}b^a\exp\left(\!i\!\int b^a{\cal F}^a\right).
\end{align}

\end{detailedcalc}
\begin{detailedcalc}

Next, add and subtract a Gaussian in $b$ with a positive parameter $\xi$ to get
\begin{align}
	\delta({\cal F}^a)={\cal N}\!\int\!{\cal D}b^a\exp\left[\!i\!\int\!{\mathrm d}^4x\, \left(b^a{\cal F}^a-\frac\xi2b^ab^a\right)\right]\underbrace{\exp\left(\frac\xi2\int\!{\mathrm d}^4x\,b^ab^a\right)}_{A\mathrm{-independent\ constant}}.
\end{align}
After completing the square in the exponent
\begin{align}
	-b^a{\cal F}^a+\frac\xi2b^ab^a&=\frac\xi2\left(b^a-\frac{{\cal F}^a}\xi\right)^2-\frac1{2\xi}{\cal F}^a{\cal F}^a,
\end{align}
we can shift the $b$ integration variable $b\to b'=b-{\cal F}/\xi$ to obtain 
\begin{align}
	\delta({\cal F}^a)={\cal N}'\exp\left(\frac1{2\xi}\int\!{\mathrm d}^4x\,{\cal F}^a{\cal F}^a\right),
	\label{S_GF-derivation}
\end{align}
which gives the gauge-fixing term in~\1eq{S_GF}. Notice that as the Yang-Mills action of~\1eq{S_YM} is gauge invariant, the argument of the gauge fixing function can be changed from the $A_\theta$ in~\1eq{FP-operator} to simply $A$.

Observe, finally, that we could have decided not to integrate over the $b$ fields, in which case the gauge-fixing action would read
\begin{align}
	S_{\mathrm{GF}}&=\int\!{\mathrm d}^4x\,\left(b^a{\cal F}^a-\frac\xi2b^ab^a\right).
	\label{S_GF-NL}
\end{align}
The $b$s are the so-called Nakanishi-Lautrup (NL) multipliers~\cite{Nakanishi:1966cq,Lautrup:1966cq}; they have no dynamical content and can be eliminated through their (trivial)  equations of motion to get back the gauge-fixing action of~\1eq{S_GF}.

\end{detailedcalc}

Linear covariant gauges are defined by the gauge-fixing function~\cite{Fujikawa:1972fe}
\begin{align}
	{\cal F}^a=\partial^\mu A^a_\mu,
	\label{linear-covariant-F}
\end{align}
and give rise to the following Feynman rules:\\

\noindent
\begin{tabular}{@{} C{0.225\linewidth} L{0.775\linewidth} @{}}
\includegraphics[scale=0.6]{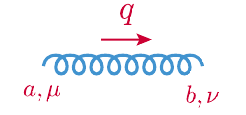}
&
\leftflushmath{
\Delta_{\mu\nu}^{ab}(q) &=
-i\delta^{ab}\frac1{q^2}\left[P_{\mu\nu}(q)-\xi \frac{q_\mu q_\nu}{q^2}\right],
}
\\
\includegraphics[scale=0.6]{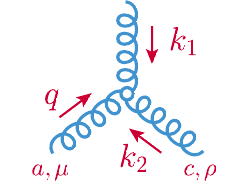}
&
\leftflushmath{
\Gamma^{(0)}_{A^a_\mu A^b_\nu A^c_\rho}(q,k_1) &=gf^{abc}\Gamma^{(0)}_{\mu\nu\rho}(q,k_1)\\&=gf^{abc}[g_{\nu\rho}(k_1-k_2)_\mu+g_{\mu\rho}(k_2-q)_\nu+g_{\mu\nu}(q-k_1)_\rho],
}
\\
\includegraphics[scale=0.6]{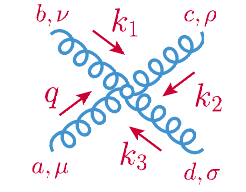}
&
\leftflushmath{
\Gamma^{(0)}_{A^a_\mu A^b_\nu A^c_\rho A^d_\sigma }(q,k_1,k_2) &=-ig^2\Gamma^{abcd(0)}_{\mu\nu\rho\sigma }(q,k_1,k_2)\\
&=-ig^2[f^{ade}f^{ecb}(g_{\mu\rho}g_{\nu\sigma}-g_{\mu\nu}g_{\rho\sigma})\\
	&\mathrel{\phantom{=}}+f^{abe}f^{edc}(g_{\mu\sigma}g_{\nu\rho}-g_{\mu\rho}g_{\nu\sigma})\\
	&\mathrel{\phantom{=}}+f^{ace}f^{edb}(g_{\mu\sigma}g_{\nu\rho}-g_{\mu\nu}g_{\rho\sigma})],
}

\end{tabular}
\medskip
where we have introduced the transverse projector
\begin{align}
	P_{\mu\nu}(q)&=g_{\mu\nu}-\frac{q_\mu q_\nu}{q^2};&
	q^\nu P_{\mu\nu}(q)&=0,
\end{align}
and, for a generic Green function $\Gamma_{\phi_1\phi_2\cdots\phi_n}$, we indicate only the momenta of the first $n-1$ fields, as the one of the $n^\mathrm{th}$ field is dictated by momentum conservation (we assume all momenta entering unless explicitly indicated, {\it e.g.}, in certain Feynman rules).

Suppose now we want to calculate the one-loop correction to the gluon propagator. This entails the calculation of the one-loop gluon self-energy, which, according to the Feynman rules stated above, comprises two terms\vspace{0.15cm}:

\noindent\makebox[\textwidth][c]{%
\includegraphics[scale=0.6]{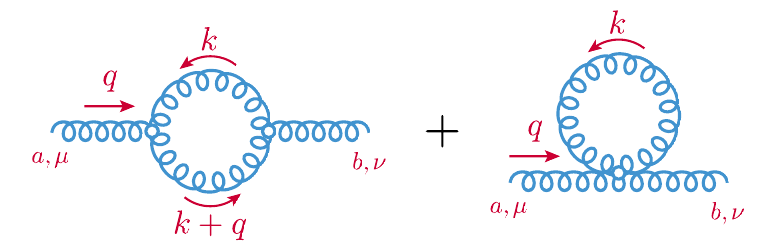}%
}
In the Feynman gauge ($\xi=1$) then we have
\begin{align}
	\Pi^{(1)}_{\mu\nu}(q;\mu)=\Pi^{(1)\mathrm{gl}}_{\mu\nu}(q;\mu)&=-\frac12g^2C_A\int_k\!\frac{\Gamma^{\rho\sigma(0)}_\mu(q,k)\Gamma^{(0)}_{\nu\rho\sigma}(q,-k-q)}{k^2(k+q)^2}-g^2C_A(d-1)\int_k\frac{g_{\mu\nu}}{k^2},
	\label{gluon-se}
\end{align}
where: the trivial color factor $\delta^{ab}$ has been omitted; $C_A$ is the Casimir eigenvalue of the adjoint representation of the $SU(N_c)$ gauge group ($C_A=N_c$); and we have introduced the dimensional regularization integral measure
\begin{align}
	\int_k=\frac{\mu^\varepsilon}{(2\pi)^d}\int\!\mathrm{d}^dk,
\end{align} 
with $d=4-\varepsilon$ the space-time dimension and $\mu$ the 't Hooft mass that ensures that the coupling constant $g$ stays dimensionless in $d$-dimensions.

At the level of the classical action $S=S_\mathrm{YM}+S_\mathrm{GF}$ the gluon is a massless gauge boson. Consequently, order by order in perturbation theory, gauge symmetry enforces the transversality of the gluon self-energy,
\begin{align}
	q^\nu\Pi^{(1)}_{\mu\nu}(q;\mu)=0\qquad\Rightarrow\qquad
	\Pi^{(1)}_{\mu\nu}(q;\mu)=P_{\mu\nu}(q)\Pi^{(1)}(q;\mu).
\end{align}
However, owing to the dimensional regularization result 
\begin{align}
	\int_k\frac1{k^2}=0,
	\label{dimreg-zero}
\end{align}
the seagull term in~\1eq{gluon-se} ({\it i.e.}, the second term) vanishes, whereas a little algebra shows that
\begin{align}
	\Gamma^{(0)}_{\mu\rho\sigma}(q,k)\Gamma^{\rho\sigma(0)}_\nu(-q,k+q)&=4q^2g_{\mu\nu}+(d-6)q_\mu q_\nu+(4d-6)k_\mu k_\nu+(2d-3)k_\mu q_\nu\nonumber \\
	&+(2d-3)q_\mu k_\nu,
\end{align}
and therefore
\begin{align}
	q^\nu\Gamma^{(0)}_{\mu\rho\sigma}(q,k)\Gamma^{\rho\sigma(0)}_\nu(-q,k+q)&=(q^2+2k{\cdot}q)[(2d-3)k_\mu+(d-2)q_\mu]+q_\mu k{\cdot}q.
	\label{qGG}
\end{align}
Writing $q^2+2k{\cdot}q=(k+q)^2-k^2$ and using the result of~\1eq{dimreg-zero}, we see that the first term in~\1eq{qGG} vanishes, so that we are left with the puzzling result
\begin{align}
	q^\nu\Pi^{(1)\mathrm{gl}}_{\mu\nu}(q;\mu)=\frac12g^2C_A\int_k\!\frac{q_\mu k{\cdot}q}{k^2(k+q)^2}\neq0.
\end{align} 

\begin{detailedcalc}
	Using the dimensional regularization results:
	\begin{subequations}
	\begin{align}
		\int_k\!\frac{1}{k^2(k+q)^2}&=
		f_\epsilon(q^2;\mu^2),\\
		\int_k\!\frac{k_\alpha}{k^2(k+q)^2}&=-\frac{q_\alpha}2f_\epsilon(q^2;\mu^2),\\
		\int_k\!\frac{k_\alpha k_\beta}{k^2(k+q)^2}&=-\frac1{12}\left(g_{\alpha\beta} q^2+4q_\alpha q_\beta\right)f_\epsilon(q^2;\mu^2),
	\end{align}
	\label{dim-reg-int}
	\end{subequations}
	where
	\begin{align}f_\epsilon(q^2;\mu^2)&=i(4\pi)^{\frac\epsilon2-2}\Gamma\left(\frac{\epsilon}2\right)\left(\frac{q^2}{\mu^2}\right)^{-\frac\epsilon2};&
	f_0(q^2;\mu^2)&=\frac{i}{4\pi^2}\left(\frac2\epsilon-\gamma-\ln4\pi-\ln\frac{q^2}{\mu^2}\right),
	\label{f-exp}
	\end{align}
it is immediate to show that indeed the 1-loop gluon contribution to the self-energy,~\1eq{gluon-se}, is not transverse, amounting to
	\begin{align}
		\Pi^{(1)\mathrm{gl}}_{\mu\nu}(q;\mu)&=\frac{g^2C_A}{12}\left(19g_{\mu\nu}-22\frac{q_\mu q_\nu}{q^2}\right)q^2f_\epsilon(q^2;\mu^2).
		\label{gl-contr}
	\end{align}
	
\end{detailedcalc}

This represents an anomaly: namely, the failure to elevate to the quantum level a classical symmetry (gauge symmetry in this particular case). 

\begin{detailedcalc}

Obviously, this result has nothing to do with our choice of performing the calculation of the gluon-self-energy in the Feynman gauge $\xi=1$. Performing it for a generic value of the gauge fixing parameter, would reveal the presence of additional terms proportional to $(1-\xi)$ and $(1-\xi)^2$, which, however, form a transverse subset. In fact, at any order in perturbation theory, the terms proportional to $(1-\xi)^n$ are individually transverse as a result of gauge fixing parameter independence.
\end{detailedcalc}

To rescue the situation we could assume that in evaluating the gluon self-energy the contribution of certain fictitious particles, that will be called ghosts, was left out~\cite{Feynman:1963}. Ghosts must contribute loops with opposite sign to ordinary bosonic fields, so that they can cancel the unphysical gluon polarizations in Feynman diagrams. This implies that ghosts must be anticommuting scalars ({\it i.e.}, Grassmann valued fields), interacting through the Feynman rules:\\
\smallskip

\noindent
 \begin{tabular}{@{} C{0.225\linewidth} L{0.775\linewidth} @{}}
\includegraphics[scale=0.6]{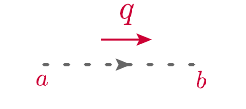}
&
\leftflushmath{
D^{ab}(q) &=
i\delta^{ab}\frac1{q^2},
}
\\
\includegraphics[scale=0.6]{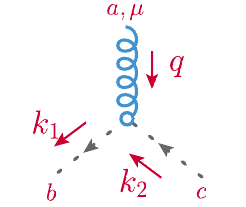}
&
\leftflushmath{
\Gamma^{(0)}_{c^cA^a_\mu \overline c^b}(k_2,q) &=gf^{abc}\Gamma^{(0)}_\mu(q,k_2)\\
    &=gf^{abc}k_{1\mu}.
}
\end{tabular}
\medskip

Then the ghost contribution to the gluon self-energy would be\vspace{0.15cm}

\noindent\makebox[\textwidth][c]{%
\includegraphics[scale=0.6]{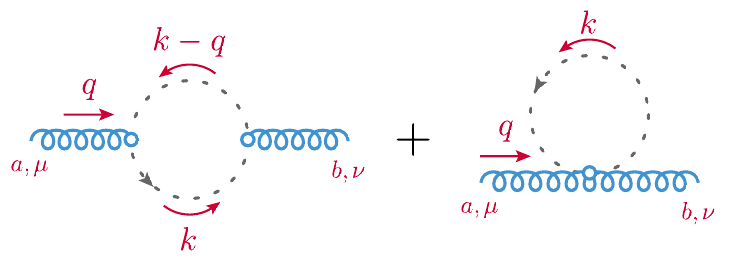}%
}
or, using the corresponding Feynman rules (and sending $k\to-k$ in the first diagram above)
\begin{align}
	\Pi^{(1)\mathrm{gh}}_{\mu\nu}(q;\mu)&=-g^2C_A\int_k\!\frac{k_\mu(k+q)_\nu}{k^2(k+q)^2}.
\end{align} 
Indeed, then the sum of the gluon and ghost contributions to the gluon self-energy is transverse:
\begin{align}
	q^\nu\Pi^{(1)}_{\mu\nu}(q;\mu)=q^\nu[\Pi^{(1)\mathrm{gl}}_{\mu\nu}(q;\mu)+\Pi^{(1)\mathrm{gh}}_{\mu\nu}(q;\mu)]=0,
	\label{cross-talk}
\end{align}
even though individually this is not true, since 
\begin{align}
	q^\nu\Pi^{(1)\mathrm{gl}}_{\mu\nu}(q;\mu)&\neq0;&
	q^\nu\Pi^{(1)\mathrm{gh}}_{\mu\nu}(q;\mu)&\neq0.
\end{align}

\begin{detailedcalc}
To check the transversality property of~\1eq{cross-talk} let's start from the (obvious) identity
\begin{align}
	\int_k\!\frac{k_\mu(k+q)_\nu}{k^2(k+q)^2}=\int_k\!\frac{k_\nu(k+q)_\mu}{k^2(k+q)^2}.
	\label{gh-id-1}
\end{align}	
Contracting with $q^\nu$ we then obtain the relation
\begin{align}
	q^2\int_k\!\frac{k_\mu}{k^2(k+q)^2}=\int_k\!\frac{q_\mu k{\cdot}q}{k^2(k+q)^2}.
	\label{gh-id-2}
\end{align}
Next, consider the right-hand side of \1eq{gh-id-1} alone to get
\begin{align}
	q^\nu\int_k\!\frac{k_\nu(k+q)_\mu}{k^2(k+q)^2}
	&=\int_k\!\frac{q_\mu k{\cdot}q}{k^2(k+q)^2}+\frac12\int_k\!\frac{k_\mu[(k+q)^2-k^2-q^2]}{k^2(k+q)^2}\nonumber \\
	&=\int_k\!\frac{q_\mu k{\cdot}q}{k^2(k+q)^2}-\frac12q^2\int_k\!\frac{k_\mu}{k^2(k+q)^2}=\frac12\int_k\!\frac{q_\mu k{\cdot}q}{k^2(k+q)^2},
\end{align}
where in the last step we have used the result of~\1eq{gh-id-2}. Thus we have
\begin{align}
	q^\nu\Pi^{(1)\mathrm{gh}}_{\mu\nu}(q;\mu)=-\frac12g^2C_A\int_k\!\frac{q_\mu k{\cdot}q}{k^2(k+q)^2}=-q^\nu\Pi^{(1)\mathrm{gl}}_{\mu\nu}(q;\mu),
\end{align}
and the transversality of the gluon self-energy is recovered.

Indeed, using the results of~\1eq{dim-reg-int}, one finds 
\begin{align}
	\Pi^{(1)\mathrm{gh}}_{\mu\nu}(q;\mu)&=\frac{g^2C_A}{12}\left(g_{\mu\nu}+2\frac{q_\mu q_\nu}{q^2}\right)q^2f_\epsilon(q^2,\mu^2),
\end{align}
which, when added to the result in~\1eq{gl-contr}, yields the transverse result
\begin{align}
	\Pi^{(1)}_{\mu\nu}(q;\mu)&=
	P_{\mu\nu}(q)\Pi^{(1)}(q^2;\mu^2);&
	\Pi^{(1)}(q^2;\mu^2)&=\frac53g^2C_Aq^2f_\epsilon(q^2;\mu^2).
	\label{conv-pi-1l}
\end{align}

\end{detailedcalc}

Close inspection of the needed Feynman rules of the ghost fields reveals that they can be obtained from the action
\begin{align}
	S_\mathrm{FP}=\int\!\mathrm{d}^4x\,\overline c^a(-\partial^\mu{\cal D}^{ab}_\mu) c^b=\int\!\mathrm{d}^4x\int\!\mathrm{d}^4y\,\overline c^a{\cal M}^{ab} c^b,
	\label{S_FP}
\end{align}
where ${\cal M}$ is the Fadeev-Popov operator introduced in~\1eq{FP-operator-1} evaluated for the covariant gauge fixing function of~\1eq{linear-covariant-F}
\begin{align}
	{\cal M}^{ab}(x,y)=-\partial^\mu{\cal D}^{ab}_\mu\delta(x-y).
\end{align} 

\begin{detailedcalc}
	Recall that when introducing the gauge-fixing condition through~\2eqs{FP-operator}{FP-operator-1}, we still had the determinant of the Faddeev-Popov operator to take care of (while we proved that the $\delta$ function gives rise to the gauge fixing action through~\1eq{S_GF-derivation}). To this end it is useful to recall that for a (complex) Grassmann algebra generated by the basis elements $\vartheta_1,\dots,\vartheta_n$, with $\{\vartheta_i,\vartheta_j\}=0$, one has the result
	\begin{align}
		\int\!\mathrm{d}^n\overline\vartheta\,\mathrm{d}^n\vartheta\exp\left(\overline\vartheta{\cal M}\vartheta\right)=\mathrm{det}({\cal M}).
	\end{align}
	This means that the determinant in~\1eq{FP-operator} can be written as
	\begin{align}
		\mathrm{det}\left({\cal M}^{ab}\right)=\int\!{\cal D}c{\cal D}\overline c\exp\left(\int\!\mathrm{d}^4x\!\int\!\mathrm{d}^4y\, \overline c^a{\cal M}^{ab}c^b\right),
	\end{align}
	which, in the case of the gauge fixing function~\1eq{linear-covariant-F}, is exactly the same Faddeev-Popov ghost action introduced in~\1eq{S_FP}.
\end{detailedcalc}

Notice finally that there might be gauges for which the Faddeev-Popov operator of~\1eq{FP-operator-1} does not depend on the gluon field $A$, giving rise to a trivial ghost sector. As an example of such a {\it ghost-free} gauge, consider the light-cone gauge~\cite{Leibbrandt:1984}, defined by the gauge fixing function
\begin{align}
	{\cal F}^a&=n^\mu A^a_\mu;& n^2&=0.
\end{align}  
Then~\1eq{FP-operator-1} implies
\begin{align}
	{\cal M}^{ab}=n^\mu{\cal D}^{ab}_\mu=\delta^{ab}n^\mu\partial_\mu,
\end{align}  
where in the last step we made use of the gauge-fixing condition $n^\mu A^a_\mu=0$. So ghosts decouple from the theory in this gauge; but the price we have to pay is that the gluon propagator has the form 

\begin{tabular}{@{} C{0.25\linewidth} L{0.75\linewidth} @{}}
\includegraphics[scale=0.6]{gluon-propagator.pdf}
&
\leftflushmath{
\Delta_{\mu\nu}^{ab}(q) &=
-i\delta^{ab}\frac1{q^2}\left[g_{\mu\nu}-\frac{q_\mu n_\nu+n_\mu q_\nu}{n{\cdot}q}+ \frac{q^2n_\mu n_\nu}{(q{\cdot}n)^2}\right],
}
\end{tabular}
which is transverse both in $n$ and in $q$. However, notice the presence of the spurious pole in $1/(q{\cdot}n)$ (and its square); the correct treatment of such poles need a precise prescription~\cite{BassettoNardelliSoldati:1991}, a fact that complicates enormously the calculation already at the perturbative level beyond one-loop, let alone at the non-perturbative one (which is what we are really interested in). 

In what follows we will thus restrict our attention to the linear covariant gauges of~\1eq{linear-covariant-F}.

\begin{detailedcalc}

In so doing  we will, for the moment, ignore the issue of Gribov copies~\cite{Gribov:1977wm,Singer:1978dk}. Strictly speaking, the complication described below arises within naive analyses of the gauge-fixing procedure, where the standard Faddeev--Popov construction does not guarantee a unique representative of each gauge orbit. As we will discuss at the end of our presentation, within the dynamical mass generation framework that will emerge later, the difficulties associated with Gribov copies are expected to be naturally avoided~\cite{Gao:2017uox}.

	To illustrate how this problem arises in the standard formulation, consider the geometric interpretation of the gauge-fixing condition. In principle, the hyper-surface defined by ${\cal F}^a=\partial^\mu A^a_\mu=0$ should intersect each gauge orbit only once. Let us therefore consider a gauge transformed field
\begin{align}
A^a_{\theta\mu}=A^a_\mu+{\cal D}^{ab}_\mu\theta^b,
\end{align}
and ask under which conditions it satisfies the same gauge condition. One finds that $\partial^\mu A^a_{\theta\mu}=0$ provided that
\begin{align}
\partial^\mu{\cal D}^{ab}_\mu\theta^b=0\qquad\Rightarrow\qquad {\cal M}^{ab}\theta^b=0,
\end{align}
where ${\cal M}$ is the Faddeev--Popov operator. Thus, if ${\cal M}$ possesses zero-modes, there exist non-trivial $\vartheta^b$ such that $A^a_\mu$ and $A^a_{\vartheta\mu}$ are distinct fields satisfying the same gauge condition. In this case the gauge orbit intersects the gauge-fixing hyper-plane more than once, and the path integral still integrates multiple times over the same orbit. Notice that in Abelian gauge theories (or in non-covariant gauges) the Faddeev-Popov operator is field independent and thus zero-modes can be removed through suitable boundary conditions. 
	
	If $\mathcal{M}$ is positive definite, then the gauge fixing is locally unique in the neighborhood of that configuration. This observation leads to the definition of the first Gribov region $\Omega$: it is the set of all transverse gauge fields for which the Faddeev--Popov operator is strictly positive,
\begin{align}
	\Omega = \left\{ A_\mu \; \big| \; \partial_\mu A_\mu = 0 \ \text{and} \ \mathcal{M} > 0 \right\} .
\end{align}
The boundary of this region, known as the Gribov horizon $\partial \Omega$, is reached when the smallest eigenvalue of $\mathcal{M}$ vanishes. At this point, the gauge fixing hypersurface becomes tangent to the gauge orbit, and, as we have seen above, an infinitesimal gauge transformation exists that preserves the gauge condition. This signals the onset of Gribov copies: beyond the horizon, the Faddeev--Popov determinant changes sign and multiple gauge equivalent configurations proliferate. Although configurations inside $\Omega$ are still not completely free of copies, the region is convex, bounded in all directions, and contains the perturbative vacuum, making it a natural and well defined domain to which the functional integral can be restricted. From a dynamical point of view, the Gribov horizon plays a crucial role in the IR. As the functional integral samples larger field amplitudes, it becomes increasingly sensitive to the accumulation of near zero modes of the Faddeev--Popov operator, causing typical configurations to lie close to $\partial \Omega $. 

	As anticipated, however, within the dynamical mass generation framework that we will develop, the difficulties associated with Gribov copies are avoided.
\end{detailedcalc}

\section{\label{two}Becchi--Rouet--Stora--Tyutin (BRST) symmetry, anti-BRST symmetry, and the ghost sector of $SU(N_c)$ Yang-Mills theories}
\subsection{Becchi--Rouet--Stora--Tyutin symmetry}

As we have seen, the quantization of a Yang-Mills theory requires fixing a gauge; and this operation modifies the action as 
\begin{align}
	S=S_\mathrm{YM}+S_\mathrm{GF}+S_\mathrm{FP},
	\label{S_tot}
\end{align}
with each term given in~\3eqs{S_YM}{S_GF}{S_FP} respectively. As already noticed, however, the gauge-fixing term can be written as in~\1eq{S_GF-NL}, since the NL multipliers satisfy the trivial equations of motion:
\begin{align}
	\frac{\partial{\cal L}}{\partial b^a}=0\quad\Rightarrow\qquad
	b^a=\frac1\xi{\cal F}^a,
\end{align} 
which gives back~\1eq{S_GF} once inserted in~\1eq{S_GF-NL}.

Now let us introduce the Becchi--Rouet--Stora--Tyutin (BRST) operator $s$~\cite{Becchi:1974md,Becchi:1975nq,Tyutin:1975qk} such that 
\begin{align}
	sA^a_\mu={\cal D}^{ab}_\mu c^b.
\end{align}  
Then, as the action of $s$ on the gauge field coincides with its gauge transformation where the gauge parameter has been replace by the ghost field $c$, the gauge invariance of the Yang-Mills action,~\1eq{S_YM}, implies also BRST invariance: 
\begin{align}
	sS_\mathrm{YM}=0.
\end{align}
On the other hand, setting
\begin{align}
	s\overline c^a&=b^a;&
	sb^a=0,
\end{align}
implies that the sum of gauge fixing and Faddeev-Popov ghost actions can be rewritten as a total BRST variation
\begin{align}
	S_\mathrm{GF}+S_\mathrm{FP}&=s\Psi;&
	\Psi&=\int\!\mathrm{d}^4x\,\overline c^a\left({\cal F}^a-\frac\xi2b^a
	\right),
	\label{GFFerm}
\end{align}
where $\Psi$ is the so-called gauge-fixing fermion, and 
\begin{align}
	S_\mathrm{GF}&=\int\!\mathrm{d}^4x\left(b^a{\cal F}^a-\frac\xi2b^ab^a
	\right);&
	S_\mathrm{FP}&=-\int\!\mathrm{d}^4x\,\overline c^as{\cal F}^a.
	\label{S_GF_and_S_FP}
\end{align}

What is finally missing is to establish the BRST variation of the ghost field. This can be obtained considering the closure of the gauge (Lie) algebra, which states that the commutator of two gauge transformations must itself be a gauge transformation, with
\begin{align}
	[\delta_{\theta_1},\delta_{\theta_2}]&=\delta_{[\theta_1,\theta_2]}=\delta_{\theta_3};& 
	\theta^a_3&=gf^{abc}\theta_1^b\theta_2^c.
\end{align}
As $s$ is Grassmann-odd, this implies the nilpotency of the BRST operator:
\begin{align}
	\{s,s\}&=0\qquad\Rightarrow\qquad s^2=0,
\end{align}
so that requiring $s^2A^a_\mu=0$ yields the ghost transformation
\begin{align}
	sc^a=-\frac12gf^{abc}c^bc^c.
\end{align}

Thus, the gauge-fixing procedure necessary to quantize the Yang-Mills action,~\1eq{S_YM}, has generated the new action $S$,~\1eq{S_tot}, that is invariant under BRST transformations: $sS=0$, with $s^2=0$ and 
\begin{align}
	sA^a_\mu&={\cal D}^{ab}_\mu c^b;&
	sc^a&=-\frac12gf^{abc}c^bc^c;&
	s\overline c^a&=b^a;&
	s b^a=0.
	\label{BRST}
\end{align}
In other words the gauge symmetry of the original action,~\1eq{S_YM}, has been traded for the BRST symmetry of the gauge-fixed action,~\1eq{S_tot}.
\begin{detailedcalc}
	The path integral measure is also invariant under the BRST transformations of~\1eq{BRST}. To show that this is indeed the case, let's consider an infinitesimal BRST change of variable
	\begin{align}
		\Phi\to\Phi'&=\Phi+\epsilon s\Phi;& \Phi&=\{A^a_\mu,c^a,\overline c^a,b^a\},
	\end{align}
	where $\epsilon$ must be Grassmann-odd: $\epsilon^2=0$.
	Then the measure transforms according to ${\cal D}\Phi'=J{\cal D}\Phi$ where the Jacobian $J$ is given by
	\begin{align}
		 J=\mathrm{det}\left(\frac{\delta\Phi'}{\delta\Phi}\right)=\mathrm{det}\left(1+\epsilon\frac{\delta s\Phi}{\delta\Phi}\right)=1+\epsilon\mathrm{Tr}\left(\frac{\delta s\Phi}{\delta\Phi}\right).
	\end{align}
	As we want to show that the $J=1$ we need to prove that the condition	
	\begin{align}
		\mathrm{Tr}\left(\frac{\delta s\Phi}{\delta\Phi}\right)=0,
	\end{align}
	holds true for all $\Phi$.  Indeed:
	\begin{subequations}
	\begin{align}
    	A^a_\mu: &\qquad 
    	\mathrm{Tr}\!\left( \frac{\delta\, (s A^a_\mu)}{\delta A^b_\nu} \right)
    	= \left. g f^{acd} \frac{\delta A^c_\mu}{\delta A^b_\nu} c^d \,\right|_{\substack{a=b \\ \mu=\nu}}=gf^{aad}g^\mu_\mu c^d=0, \\
    	c^a: &\qquad \mathrm{Tr}\!\left( \frac{\delta\, (s c^a)}{\delta c^b} \right)
    	= -\frac12gf^{acd}\left.\frac{\delta (c^dc^d)}{\delta c^b}\right\vert_{a=b}=-\frac12g(f^{aad}c^d+f^{aca}c^c)=0, \\
    	\overline c^a: &\qquad \mathrm{Tr}\!\left( \frac{\delta\, (s \overline c^a)}{\delta \overline c^b} \right)
    	= \left.\frac{\delta\, b^a}{\delta \overline c^b}\right\vert_{a=b}=0, \\
    	b^a: &\qquad \mathrm{Tr}\!\left( \frac{\delta\, (s b^a)}{\delta  b^b} \right)=0.
	\end{align}
	\end{subequations}
	Notice that if it turns out that $J\neq1$, we get a non invariant measure ${\cal D}\Phi'=(1+\epsilon\Delta){\cal D}\Phi$ with $\Delta$ a local functional; inserting it back in the path integral produces a breaking term
	\begin{align}
		sS\sim\int\!\mathrm{d}^4x\Delta,
	\end{align}
	which represents a gauge anomaly. The most famous example is the Adler-Bell-Jackiw (chiral) anomaly~\cite{Adler:1969gk,Bell:1969ts}, which is due to the non-unity value Jacobian from a chiral rotation, giving a non-vanishing $\Delta\sim\mathrm{Tr}(\gamma_5)$~\cite{Dittrich:1985tr}.
	
	In our case, however, the theory is non-anomalous and the invariance of the path integral measure has the important consequence that a total BRST variation integrates to zero. In fact, consider the path integral of a generic functional $F[\Phi]$; then under a BRST transformation
	\begin{align}
		\int\!{\cal D}\Phi\,F[\Phi]=\int\!{\cal D}\Phi'\,F[\Phi']=\int\!{\cal D}\Phi\,F[\Phi+\epsilon s\Phi]=\int\!{\cal D}\Phi\,F[\Phi]+\epsilon\int\!{\cal D}\Phi\, sF[\Phi],
	\end{align}
	and therefore
	\begin{align}
		\int\!{\cal D}\Phi\, sF[\Phi]=0.
		\label{total_BRST_PI}
	\end{align} 			
	As we will see, this result will have important consequences in what follows.	
\end{detailedcalc}
The BRST symmetry of the gauge-fixed Yang-Mills action is most easily exposed through the so-called Batalin-Vilkoviski (BV) formalism~\cite{Batalin:1977pb,Batalin:1983ggl}. To this end, we introduce for every field $\Phi=\{A^a_\mu,c^a,\overline c^a,b^a\}$ an antifield $\Phi^*=\{A^{*a}_\mu,c^{*a},\overline c^{*a},b^{*a}\}$ with: opposite Grassmann parity of the corresponding field; ghost charge $\mathrm{gh}(\Phi^*)=1+\mathrm{gh}(\Phi)$, where $\mathrm{gh}(\Phi)=\mathrm{gh}(A^a_\mu,c^a,\overline c^a,b^a)=(0,1,-1,0)$; and, finally, $s\Phi^*=0$. 

Next we add to the action in~\1eq{S_tot} a term coupling the antifields with the (BRST variation of the) corresponding field, obtaining the complete action
\begin{align}
	S_\mathrm{C}&=S+S_\mathrm{BV};& S_\mathrm{BV}&=
		\int\!\mathrm{d}^4x\,\sum \Phi^*s\Phi=\int\!\mathrm{d}^4x\,\left(A^{*a}_\mu sA^{a\mu}+c^{*a}sc^a+\overline c^{*a}b^a\right).
	\label{S_C}
\end{align}
Notice that as $sb^a=0$ there is no need for an antifield associated to the NL multipliers; in addition we can also write
\begin{align}
	S_\mathrm{C}&=S_\mathrm{YM}+sX;& X&= \int\! \mathrm{d}^4 x\, \left [ \sum(-1)^{\mathrm{gh}(\Phi^*)}\Phi^*\Phi+\overline c^a{\cal F}^a-\frac\xi2\overline c^ab^a \right ].
	\label{Xx}
\end{align}

\begin{detailedcalc}
	Both the BV action $S_\mathrm{BV}$ and the sum of the gauge-fixing and Faddeev-Popov actions $S_\mathrm{GF+FP}=S_\mathrm{GF}+S_\mathrm{FP}$ are $s$-exact (or pure gauge), {\it i.e.}, there exists a functional $X_i$ such that $S_i=s X_i$ (obviously $X=\sum X_i$), see, in particular, \1eq{Xx} above and \1eq{GFFerm}. While this condition automatically ensures that $S_i$ is also $s$-closed (or gauge invariant), {\it i.e.}, $sS_i=0$, it is a stronger condition than the latter. Being gauge invariant does not mean being pure gauge: think about  $S_\mathrm{YM}$, which is clearly $s$-closed but not $s$-exact. 
	
	$s$-exact terms do not change the theory's partition function. In fact, with $\lambda \in \mathbb{R}$ consider 
	\begin{align}
		Z_\lambda &= \int \mathcal D\Phi\,\exp\left[i(S + \lambda sX)\right];&
		\frac{dZ_\lambda}{d\lambda}
			&= i\! \int\! \mathcal D\Phi\,(sX)\, 
   			\exp \left[i(S + \lambda sX)\right].
	\end{align}
	Since $sS=0$ and $s^2=0$, we have $s\left\{X\exp\left[i(S + \lambda sX)\right]\right\}
		= (sX)\exp\left[i(S + \lambda sX)\right]$
	and therefore
	\begin{align}
		\frac{dZ_\lambda}{d\lambda}
		&= i\! \int\! \mathcal D\Phi\,s\left\{X\exp\left[i(S + \lambda sX)\right]\right\}.
	\end{align}
	
	On the other hand, we know from~\1eq{total_BRST_PI} that the path integral of a total BRST variation is zero; thus
	\begin{align}
		\frac{dZ_\lambda}{d\lambda}=0\qquad\Rightarrow\qquad
		Z_\lambda=Z_0=\int \mathcal D\Phi\,\exp(iS).
	\end{align}
	This is sometimes expressed by saying that physical theories are determined solely by the cohomology class of the action (in ghost number zero); see, {\it e.g.},~\cite{Henneaux:1992ig,Piguet:1995er}. Any two actions differing by an $s$-exact term are in the same cohomology class:
	\begin{align}
		S'=S+sX \qquad\Rightarrow\qquad [S']=[S]\in H^0(s),
	\end{align}
	and so we are free to shift the Yang-Mills action by $s$-exact terms, as done in~\2eqs{S_tot}{S_C}.  
	
	The same applies to the observables of the theory.  Consider, in fact, the expectation value of an $s$-closed functional ${\cal O}[\Phi]$, and let's show that it does not change when the action is shifted by an $s$-exact term. Defining
	\begin{align}
		\langle{\cal O}\rangle_\lambda=\frac1{Z_\lambda}\int \mathcal D\Phi\,{\cal O}\exp\left[i(S + \lambda sX)\right],
	\end{align} 
	the same reasoning above applied in the case of the partition function $Z_\lambda$, implies that whenever $s{\cal O}=0$
	\begin{align}
		\frac{d\langle{\cal O}\rangle_\lambda}{d\lambda}=0\qquad\Rightarrow\qquad
		\langle{\cal O}\rangle_\lambda=\langle{\cal O}\rangle_0=\langle{\cal O}\rangle.
	\end{align}
	Thus, within a BRST formulation of gauge theories: physical observables ${\cal O}$ are $s$-closed, $s{\cal O}=0$; and since two observables ${\cal O}$ and ${\cal O}'$ differing by an $s$-exact term ${\cal O}'={\cal O}+s X$ have the same expectation value, they are physically equivalent. Thus also observables live in the (zero ghost charge) cohomology of the BRST operator: $[{\cal O}]\in H^0(s)$.   
\end{detailedcalc}   

\noindent The complete action of~\1eq{S_C} satisfies the master equation
\begin{align}
	\int\!\mathrm{d}^4x\,\sum\frac{\delta S_\mathrm{C}}{\delta\Phi^*}\frac{\delta S_\mathrm{C}}{\delta\Phi}=0.
	\label{BV_ME}
\end{align}

\begin{detailedcalc}
	To verify the result above, observe that, the BRST differential can be written as
	\begin{align}
		{\cal S}&=\sum \int\! \mathrm{d}^4 x\,s\Phi(x)\frac{\delta}{\delta\Phi(x)}.
		\label{BRST-diff}
	\end{align}
	Then we have
	\begin{align}
		{\cal S}(S_\mathrm{C})&=\int\mathrm{d}^4x\,\sum(s\Phi)\frac{\delta S_\mathrm{C}}{\delta\Phi}=\int\!\mathrm{d}^4x\,\sum\frac{\delta S_\mathrm{C}}{\delta\Phi^*}\frac{\delta S_\mathrm{C}}{\delta\Phi}=\int\!\mathrm{d}^4x\,\sum\frac{\delta S_\mathrm{C}}{\delta\Phi^*}\frac{\delta S_\mathrm{YM}}{\delta\Phi}+\int\!\mathrm{d}^4x\,\sum\frac{\delta S_\mathrm{C}}{\delta\Phi^*}\frac{\delta S_\mathrm{BV}}{\delta\Phi}.
	\end{align}

	Now, on the one hand the first term on the right-hand side of the equation above is zero due to the BRST invariance of the Yang-Mills action
	\begin{align}
		\int\mathrm{d}^4x\,\sum(s\Phi)\frac{\delta S_\mathrm{C}}{\delta\Phi}=\int\mathrm{d}^4x\,(s S_\mathrm{C})=0;
	\end{align} 
		\end{detailedcalc}
\begin{detailedcalc}

	on the other hand, the second term is zero due to the nilpotency of the BRST:
	\begin{align}
		\int\mathrm{d}^4x\,\sum\Phi^*\!
		\sum(s\Phi')\frac{\delta (s\Phi)}{\delta\Phi'}=
			\int\mathrm{d}^4x\,\sum\Phi^*s^2\Phi=0.
	\end{align}
	Overall, we have then have ${\cal S}(S_\mathrm{C})=0$, which is just~\1eq{BV_ME}.  
	
\end{detailedcalc}

To elevate the BRST symmetry to the quantum level, one then invokes the quantum action principle (QAP) requiring~\cite{PiguetSorella:1995}
\begin{align}
	{\cal S}(\Gamma)=\int\!\mathrm{d}^4x\left(\frac{\delta\Gamma}{\delta A^{*a}_\mu}\frac{\delta\Gamma}{\delta A^{a\mu}}+\frac{\delta\Gamma}{\delta c^{*a}}\frac{\delta\Gamma}{\delta c^a}+b^a\frac{\delta\Gamma}{\delta \overline c^a}
	\right)=0,
	\label{STF}
\end{align} 
where ${\cal S}$ is the so-called Slavnov--Taylor identity (STI) functional and $\Gamma$ is the full quantum  effective action.

\begin{detailedcalc}
	The QAP represents the statement that classical local variations ({\it i.e.}, gauge transformations) of the action correspond to local operator insertions ({\it i.e.}, BRST variations) in the theory's quantum Green functions. It is what ensures that the symmetries and equations of motion survive renormalization; or tells exactly how they break when anomalies appear. 
\end{detailedcalc}

The structure of the STI functional in~\1eq{STF} can be simplified by noticing that the anti-ghost $\overline c$ and NL multiplier $b$ have a linear BRST transformation; thus, the complete action of~\1eq{S_C} can be split into the sum of a minimal sector, which depends on the minimal variables $\{A^a_\mu,c^a,A^{*a}_\mu,c^{*a}\}$, and a trivial sector, which depends on $\{\overline c^a, \overline c^{*a}, b^a\}$:
\begin{align}
	S_\mathrm{C}=S_\mathrm{min}+\overline c^{*a}b^a,
	\label{Smin}
\end{align}
and similarly for the effective action $\Gamma$. Variables like $\overline c$ and $b$, which are such that $s\overline c=b$ and $sb=0$, are called BRST doublets or trivial pairs. The last term in~\1eq{Smin} has no effect on the master equation of~\1eq{BV_ME}, which is satisfied by $S_\mathrm{min}$ alone; and in what follows we will only consider the minimal sector of the action, with the STI functional given by
\begin{align}
	{\cal S}(\Gamma)&=\int\!\mathrm{d}^4x\left(\frac{\delta\Gamma_\mathrm{min}}{\delta A^{*a}_\mu}\frac{\delta\Gamma_\mathrm{min}}{\delta A^{a\mu}}+\frac{\delta\Gamma_\mathrm{min}}{\delta c^{*a}}\frac{\delta\Gamma_\mathrm{min}}{\delta c^a}\right)=0.
		\label{STFmin}
\end{align}
  
\begin{detailedcalc}
	Thus, the BRST differential can be decomposed as
	\begin{align}
		{\cal S}(\Gamma)&=\sum_\mathrm{min}\int\!\mathrm{d}^4x\,\frac{\delta\Gamma}{\delta\Phi^*}\frac{\delta\Gamma}{\delta \Phi}+\sum_\mathrm{doublets}\hspace{-0.05cm}\int\!\mathrm{d}^4x\,(s\phi)\frac{\delta\Gamma}{\delta \phi}.
	\end{align}	
\end{detailedcalc} 

\1eq{STFmin} is one of a set of functional equations that we will need to understand the dynamics of the infra-red (IR) sector of Yang-Mills theories. Its functional differentiations with respect to combinations of different fields eventually evaluated at zero fields and antifields, give rise to the STIs, which encode existing symmetry relations among the theory's Green functions.

\begin{detailedcalc}
		In order to reach meaningful (nonzero) STIs one needs to keep in mind that: ${\cal S}(\Gamma)$ is Grassmann-odd; and Green functions with nonzero ghost charge must vanish. This implies that nonzero expressions are obtained when taking differentiations with respect to field combinations containing: one ghost field; two ghost fields and one antifield; or a ghost antifield and three ghost fields.  	
\end{detailedcalc}
In particular, as the linear part of the BRST transformation of a gauge boson fields is proportional to the divergence of a ghost field, $sA^a_\mu\sim\partial_\mu c^a$, STIs involving two or more gauge bosons are obtained by differentiating ${\cal S}(\Gamma)$ with respect to a combination of fields in which one of the gluons is replaced by a ghost field.
\begin{detailedcalc}
	As a first example of an STI, let's study the gauge boson 2-point function $\Gamma_{AA}$ (the self-energy). According to what is stated above we then have to consider the following functional differentiation:
	\begin{align}
		\left.\frac{\delta^2{\cal S}(\Gamma)}{\delta A^a_\mu\delta c^b}\right\vert_{A,c=0}=0.
	\end{align}
	Then, using the minimal sector only,
	\begin{align}
		\frac{\delta{\cal S}(\Gamma)}{\delta c^b}=\int\!\mathrm{d}^4x\,\frac{\delta^2\Gamma_\mathrm{min}}{\delta c^b \delta A^{*c}_\rho}\frac{\delta\Gamma_\mathrm{min}}{\delta A^{c\rho}}+\cdots,
	\end{align}
	where the dots indicate terms that would give vanishing contributions to the final result. Next
		\begin{align}
		\frac{\delta^2{\cal S}(\Gamma)}{\delta A^a_\mu\delta c^b}=\int\!\mathrm{d}^4x\,\frac{\delta^2\Gamma_\mathrm{min}}{\delta c^b \delta A^{*c}_\rho}\frac{\delta^2\Gamma_\mathrm{min}}{\delta A^a_\mu \delta A^{c\rho}}+\cdots.
		\label{2-p-derivation}
	\end{align}	
	Then, setting to zero all fields and antifields, we obtain
	\begin{align}
		\Gamma^\mathrm{min}_{c^bA^{*c}_\rho}(q)\Gamma^\mathrm{min}_{A^{c\rho} A^a_\mu}(q)=0.
		\label{2-point-STI}
	\end{align}
	We now use Lorenz invariance to write $\Gamma_{c^bA^{*c}_\rho}(q)=q_\rho\delta^{ac}\Gamma_{cA^*}(q^2)$; and since $\Gamma_{cA^*}(q^2)$ is non vanishing beyond tree-level, we obtain the transversality of gauge-boson 2-point function
	\begin{align}
		q^\nu\Gamma^\mathrm{min}_{A^a_\mu A^b_\nu}(q)=0\qquad\Rightarrow\qquad\Gamma^\mathrm{min}_{A^a_\mu A^b_\nu}(q)=i\delta^{ab}P_{\mu\nu}(q)\Pi(q^2).
	\end{align}  
\end{detailedcalc}

Additional functional equations can be obtained by studying the dependence of the complete action on the NL multiplier $b$ and the anti-ghost field $\overline c$, both depending crucially on the form of the gauge-fixing function ${\cal F}$. In the first case, we immediately obtain the equation of motion of the $b$-field
\begin{align}
	\frac{\delta\Gamma}{\delta b^a}+\xi b^a-{\cal F}^a=0.
	\label{b-eq}
\end{align}
In the second case, using the BRST differential of~\1eq{BRST-diff} we find that for any functional $F=F[\varphi]$ such that $\mathrm{gh}(sF)=0$, the nilpotency of the BRST operator implies
\begin{align}
	\left\{\frac{\delta}{\delta\overline c^a},{\cal S}\right\}F&=\sum\int\!\mathrm{d}^4x\left[\frac{\delta}{\delta\overline c^a}s\varphi(x)\right]\frac{\delta F}{\delta\varphi(x)}.
\end{align}
As the only dependence on the anti-ghost field is in the $s$-exact term of the action, one has
\begin{align}
	\frac{\delta\Gamma}{\delta \overline c^a}=\frac{\delta}{\delta \overline c^a}(s X)=\sum
\int\!\mathrm{d}^4x \, \left[\frac{\delta}{\delta \overline c^a}\,s\varphi(x)\right]\frac{\delta X}{\delta\varphi(x)}-s\,\frac{\delta X}{\delta \overline c^a}. 
	\label{fpe-intermediate}
\end{align}
Using~\1eq{Xx}, we then see that the first term on the right-hand side of the equation above is zero whereas for the second term we obtain
\begin{align}
	s\,\frac{\delta X}{\delta \overline c^a}&=s{\cal F}^a=\partial^\mu (sA^a_\mu)=\partial^\mu \frac{\delta \Gamma}{\delta A^{*a}_\mu},
\end{align} 
yielding the final ghost equation
\begin{align}
	{\cal G}^a\Gamma&=0;&
	{\cal G}^a&=\frac{\delta}{\delta \overline c^a}+\partial^\mu \frac{\delta }{\delta A^{*a}_\mu}.
	\label{FP-eq}	
\end{align}

\begin{detailedcalc}
	A more expeditious way to obtain the ghost equation~\1eq{FP-eq} is, in this case, to observe that~\1eq{S_GF_and_S_FP} implies that the entire action's dependence on the anti-ghost field is contained in the term
\begin{align}
	\overline c^as{\cal F}^a=\overline c^a\partial^\mu (sA^a_\mu)=\overline c^a\partial^\mu\frac{\delta S_\mathrm{C}}{\delta A^{*a}_\mu}.
\end{align}
Passing then to the full quantum effective action $\Gamma$ one obtains the same ghost equation of~\1eq{FP-eq}, namely
\begin{align}
	\frac{\delta\Gamma}{\delta \overline c^a}+\partial^\mu \frac{\delta \Gamma}{\delta A^{*a}_\mu}=0.	
\end{align}
\end{detailedcalc}
	
The ghost equation of~\1eq{FP-eq} can be used to understand the difference between considering the minimal or complete action functional when deriving STIs. In practice, STIs generated from the minimal action coincide with the ones obtained from the complete action after the implementation of~\1eq{FP-eq}.

\begin{detailedcalc}
	To see this, let's consider again the transversality of the gauge-boson two-point function,~\1eq{2-point-STI}. The trivial-pair sector in the complete action gives rise to an additional contribution to the right-hand side of~\1eq{2-p-derivation}:
	\begin{align}
		\frac{\delta^2{\cal S}(\Gamma)}{\delta A^a_\mu\delta c^b}=\int\!\mathrm{d}^4x\,\left(\frac{\delta^2\Gamma_\mathrm{C}}{\delta c^b \delta A^{*c}_\rho}\frac{\delta^2\Gamma_\mathrm{C}}{\delta A^{c\rho}\delta A^a_\mu} + \frac{\delta b^c}{\delta A^a_\mu}\frac{\delta^2\Gamma_\mathrm{C}}{\delta c^b\delta\overline c^c}\right)+\cdots,
	\end{align}
	yielding, after setting all the fields to zero,
	\begin{align}
		\Gamma^\mathrm{C}_{c^bA^{*c}_\rho}(q)\Gamma^\mathrm{C}_{A^{c\rho} A^a_\mu}(q)+\frac{q_\mu}{\xi}\Gamma^\mathrm{C}_{c^b\overline c^a}(q)=0.
		\label{interm}
	\end{align}
	Differentiating the ghost equation with respect to $c^b$ we obtain
	\begin{align}
		\Gamma^\mathrm{C}_{c^b\overline c^a}(q)+iq^\mu\Gamma^\mathrm{C}_{c^bA^{*a}_\mu}(q)=0,
	\end{align}
	which, inserted back into~\1eq{interm} gives
	\begin{align}
		\Gamma^\mathrm{C}_{c^bA^{*c}_\rho}(q)\Gamma^\mathrm{C}_{A^{c\rho} A^a_\mu}(q)-i\frac{q_\mu q^\rho}{\xi}\Gamma^\mathrm{C}_{c^bA^{*a}_\rho}(q)=0\qquad\Rightarrow\qquad
		q^\nu\Gamma^\mathrm{C}_{A^a_\mu A^b_\nu}(q)-i\delta^{ab}\frac{q_\mu q^2}{\xi}=0,
	\end{align}
	which extends the STI of~\1eq{2-point-STI} at tree-level, since 
	\begin{align}
		\Gamma^\mathrm{C}_{A^a_\mu A^b_\nu}(q)&=\delta^{ab}\Delta^{-1}_{\mu\nu}(q);&
		\Delta^{-1}_{\mu\nu}(q)&=i\left\{P_{\mu\nu}(q)\Delta^{-1}(q^2)+\frac1{\xi}q_\mu q
		_\nu\right\};&
		\Delta(q^2)&=\frac1{q^2+i\Pi(q^2)}.
		\label{GammaDelta}
	\end{align}
	Thus, we see that Green functions generated by the minimal action coincides with the ones obtained from the complete action up to tree-level terms proportional to the gauge fixing parameter. In what follows we will often ignore this tree-level difference between the complete and minimal Green functions dropping all the ``min'' and ''C'' superscripts distinguishing them.
	
	Notice finally that, when defining the scalar factor $\Delta$, we have explicitly factored out an $i$ in front of $\Pi$; therefore the latter quantity is simply given by the corresponding Feynman diagrams in Minkowski space.
	 
\end{detailedcalc}

\subsection{Anti-BRST}

In the BRST formulation of gauge theories, the ghost field~$c$ plays a particularly distinguished role. It effectively takes the place of the infinitesimal gauge transformation parameter that appears in the classical gauge symmetry. As we have glimpsed, the transformation properties of~$c$ can be understood as reflecting the cohomological structure of the Lie algebra associated with the gauge group (see, {\it e.g.},~\cite{Nakanishi:1977ae,Barnich:2000zw}). 

By contrast, the anti-ghost~$\overline c$ and its BRST partner, the Nakanishi--Lautrup field~$b$, enter the theory for a rather different reason. They act as auxiliary fields that implement the gauge-fixing condition ${\cal F}=0$ and its BRST transform $s{\cal F}=0$ at the level of the functional integral. In this sense, they serve as Lagrange multipliers ensuring the consistency of the chosen gauge. It is also important to note that, unlike the ghost field, the antighost~$\overline c$ is not the Hermitian conjugate of the ghost~$c$ and satisfies an equation of motion that differs from the one satisfied by $c$; the two are independent Grassmann variables. 

Given these differences, one might expect that~$c$ and~$\overline c$ could never be interchanged in a meaningful symmetry. Surprisingly, however, it was shown long ago~\cite{Curci:1976bt,Ojima:1980da,Baulieu:1981sb}
that the BRST algebra admits an additional nilpotent symmetry, known as the anti-BRST symmetry, in which the roles of ghost and anti-ghost fields are precisely exchanged. In fact, the anti-BRST transformations can be obtained directly from the BRST ones of Eq.~\eqref{BRST} by simply swapping~$c$ and~$\overline c$;
 	that is one has
	\begin{align}
		\overline s A^a_\mu&={\cal D}^{ab}_\mu \overline c^b;&
		\overline s\, \overline c^a&=-\frac12gf^{abc}\overline c^b\overline c^c;&
		\overline s c^a&=\overline b^a;&
		\overline s \overline b^a&=0.
		\label{antiBRST}
	\end{align}
Additionally, to close the $s$ and $\overline s$ algebra two more transformations are needed: 
	\begin{align}
		s\overline b^a&=gf^{abc}\overline b^b c^c;&
		\overline s b^a&=gf^{abc} b^b\overline c^c.
		\label{additional}
	\end{align}

Clearly both $s$ and $\overline s$ are nilpotent: $s^2$=$\overline s^2=0$; then, the additional (natural) requirement that their sum is also nilpotent (or that $\{s,\overline s\}=0$), results in the constraint~\cite{Curci:1976bt}
	\begin{align}
		\overline b^a=-b^a-gf^{abc}c^b\overline c^c.
		\label{constr}
	\end{align}
Use of the Jacobi identity, $f^{ade} f^{bce}+ f^{bde} f^{cae}+ f^{cde} f^{abe}= 0$, shows that this constraint is consistent with the algebra closure of~\1eq{additional}. Finally, the nontrivial BRST-antiBRST transformations of the fields read
	\begin{align}
		s\overline s A^a_\mu&={\cal D}^{ab}_\mu b^b+gf^{abc}\left({\cal D}_\mu^{bd}c^d\right)\overline c^c;&
		s\overline s c^a&=s\overline b^a;&
		s\overline s\, \overline c^a&=-\overline s b^a.
		\label{98}
	\end{align}

The generalization of the BV action of~\1eq{S_C} requires, in the non-minimal sector, 9 sources~\cite{Binosi:2013cea}: the three usual antifields $A^*$, $c^*$ and $\overline c^*$; the four anti-BRST sources $\aBRSTsrc{A}$, $\aBRSTsrc{c}$, $\aBRSTsrc{\overline c\hspace{0.05cm}}$, $\aBRSTsrc{b}$; and, finally, the mixed BRST/anti-BRST sources $\BRSTaBRSTsrc{A}$ and $\BRSTaBRSTsrcGhost{c}$. There's no need to add  any source associated to the BRST variation of $\overline b$: due to the constraint of~\1eq{constr}, $s\overline b$ can be completely recovered from the corresponding transformations of $b$, $c$, and $\overline c$. The  ghost charge assignments are
	\begin{align}
	\mathrm{gh}(\Phi^*)&=-\mathrm{gh}(\Phi)-1;&
	\mathrm{gh}(\aBRSTsrc{\Phi})&=-\mathrm{gh}(\Phi)+1;&
	\mathrm{gh}(\BRSTaBRSTsrc{\Phi})&=-\mathrm{gh}(\Phi),&
	\end{align}
	with the corresponding BRST, anti-BRST and mixed BRST/anti-BRST transformations being given by:
	\begin{align}
		s\,\Phi^*&=\overline s\,\Phi^*=0;& s\,\aBRSTsrc{b}_a&=\aBRSTsrc{\overline c}_a;& 
		s\,\BRSTaBRSTsrc{\Phi}&=\aBRSTsrc{\Phi};\nonumber \\
		s\,\aBRSTsrc{\Phi}&=\overline s\,\aBRSTsrc{\Phi}=0;&
		\overline s\,\aBRSTsrc{b}_a&=0;&
		\overline s\,\BRSTaBRSTsrc{\Phi}&=-\Phi^*.
		\label{brst.abrst.srcs}
	\end{align}
The $s$- and $\overline s$-exact generalization of the complete action of~\1eq{S_C} is then given by
\begin{align}
	S'_\mathrm{C}&=S_\mathrm{YM}+S_\mathrm{GF}+S_\mathrm{FP}+S'_\mathrm{BV};&
	S'_\mathrm{BV}&=\int\!\mathrm{d}^4x\,\sum\left(\Phi^*s\,\Phi+\aBRSTsrc{\Phi}\overline s\,\Phi+\BRSTaBRSTsrc{\Phi}\,s\overline s\,\Phi\right).
	\label{S'_BV}
\end{align}
Indeed, one can write
\begin{align}
		S'_\mathrm{C}&=S_\mathrm{YM}+sX=S_\mathrm{YM}+\overline s Y,
\end{align}
with
\begin{subequations}
	\begin{align}
		X&= \int\! \mathrm{d}^4 x\, \left [ \sum\left((-1)^{\mathrm{gh}(\Phi^*)}\Phi^*\Phi+\BRSTaBRSTsrc{\Phi}\overline s \Phi\right)+\overline c^a{\cal F}^a-\frac\xi2\overline c^ab^a \right ],
			\label{Xgen}\\
		Y&= \int\! \mathrm{d}^4x \, \left [ \sum\left((-1)^{\mathrm{gh}(\aBRSTsrc{\Phi})}\aBRSTsrc{\Phi}\Phi-\BRSTaBRSTsrc{\Phi} s \Phi\right)-c^a{\cal F}^a+\frac\xi2 c^ab^a \right ],
			\label{Y}
	\end{align}
\end{subequations}	
which allows us to prove that $S'_\mathrm{BV}$ is both $s$- and $\overline s$-exact.

The simultaneous closure of the action under both BRST and anti-BRST symmetries, gives rise to a plethora of identities that we are going to discuss in the following. To begin with, one has the generalization of the STI functional of~\1eq{STF}, which reads
\begin{align}
	{\cal S}(\Gamma)&=\int\!\mathrm{d}^4x\Bigg[\underbrace{\frac{\delta\Gamma}{\delta A^{*a}_\mu}\frac{\delta\Gamma}{\delta A^{a\mu}}+\frac{\delta\Gamma}{\delta c^{*a}}\frac{\delta\Gamma}{\delta c^a}}_\mathrm{minimal}\underbrace{+
		b^a\frac{\delta\Gamma}{\delta\overline c^a}+\caBRSTsrc{A}{a}_\mu\frac{\delta\Gamma}{\delta\cBRSTaBRSTsrc{A}{a\mu}}+\caBRSTsrc{c}{a}\frac{\delta\Gamma}{\delta\BRSTaBRSTsrcGhost{c}^a}+\aBRSTsrc{\overline c}_a\frac{\delta\Gamma}{\delta\caBRSTsrc{b}{a}}}_\mathrm{doublets}
		\Bigg]=0.
	\label{STF-gen}
\end{align}
The anti-BRST differential
\begin{align}
	\overline{\cal S}&=\sum \int\! \mathrm{d}^4 x\,\overline{s}\Phi(x)\frac{\delta}{\delta\Phi(x)},
		\label{aBRST-diff}
\end{align} 
gives rise instead to the anti-ST identity functional:
\begin{align}
	\overline{\cal S}(\Gamma)&=\int\!\mathrm{d}^4x\Bigg[\underbrace{\frac{\delta\Gamma}{\delta \caBRSTsrc{A}{a}_\mu}\frac{\delta\Gamma}{\delta A^{a\mu}}+\frac{\delta\Gamma}{\delta \caBRSTsrc{c}{a}}\frac{\delta\Gamma}{\delta c^a}+\frac{\delta\Gamma}{\delta \caBRSTsrc{\overline c\hspace{0.05cm}}{a}}\frac{\delta\Gamma}{\delta \overline c^a}+\frac{\delta\Gamma}{\delta \caBRSTsrc{b}{a}}\frac{\delta\Gamma}{\delta b^a}}_\mathrm{minimal}\underbrace{-A^{*a}_\mu\frac{\delta\Gamma}{\delta\cBRSTaBRSTsrc{A}{a\mu}}-c^{*}_a\frac{\delta\Gamma}{\delta\BRSTaBRSTsrcGhost{c}^a}}_\mathrm{doublets}
		\Bigg]=0.\hspace{0.5cm}
	\label{antiSTF-gen}
\end{align}
Next, the $s$-exact term of~\1eq{Xgen} gives rise to a generalization of the ghost equation of~\1eq{FP-eq}, which, following the same steps that lead to the determination of the equation in the BRST invariant case, now reads
 \begin{align}
	{\cal G}^a \Gamma&={\cal D}^{ab}_\mu \caBRSTsrc{A}{b\mu}
	+gf^{abc} \caBRSTsrc{c}{b} c^c+gf^{abc}\caBRSTsrc{\overline c\hspace{0.05cm}}{b}\overline c^c-gf^{abc}\caBRSTsrc{b}{b}b^c;\nonumber \\
	{\cal G}^a&=\frac{\delta}{\delta \overline c^a} 
	+\BRSTaBRSTcovD\frac{\delta}{\delta A^{*b}_\mu}+gf^{abc}\BRSTaBRSTsrcGhost{c}^b\frac{\delta}{\delta c^*_c}.
	\label{GE-xi}
\end{align}
The $\overline s$-exact term in~\1eq{Y} provides instead a new equation that was not present in the purely BRST invariant case: the anti-ghost equation. This can be obtained by observing that, in full analogy with~\1eq{fpe-intermediate}, we have
\begin{align}
	\frac{\delta\Gamma}{\delta c^a}=
	\frac{\delta}{\delta c^a}(\overline s\, Y)=\sum \int \mathrm{d}^4 x \, \left[\frac{\delta}{\delta c^a}\overline s\,\varphi( x)\right]\frac{\delta Y}{\delta\varphi(x)}-
	\overline s\,\frac{\delta Y}{\delta c^a}.
\end{align}
Then a lengthy (but relatively straightforward) calculation shows that
\begin{align}
	\overline {\cal G}^a \Gamma &={\cal D}^{ab}_\mu A^{*b\mu} +gf^{abc}c^{*b}c^c;\nonumber \\
	 \overline {\cal G}^a&=\frac{\delta}{\delta c^a} 
+ gf^{abc}\overline c^c\frac{\delta}{\delta b^b}+\xi\frac{\delta}{\delta \caBRSTsrc{b}{a}}-\BRSTaBRSTcovD\frac{\delta}{\delta \caBRSTsrc{A}{b\mu}}
-gf^{abc}\BRSTaBRSTsrcGhost{c}^b\frac{\delta}{\delta \caBRSTsrc{c}{c}}-gf^{abc}\caBRSTsrc{b}{b}\frac{\delta}{\delta \caBRSTsrc{\overline c\hspace{0.05cm}}{c}},
	\label{local-anti-ghost}
\end{align}
where $\BRSTaBRSTcovD=\partial_\mu\delta^{ab}+f^{acb}\cBRSTaBRSTsrc{A}{c}_\mu$ is the covariant derivative with respect to the BRST/anti-BRST source $\BRSTaBRSTsrc{A}$. Finally, the $b$-equation now assumes the form
\begin{align}
	\frac{\delta\Gamma}{\delta b^a}=\BRSTaBRSTcovD(A^{b\mu}-\cBRSTaBRSTsrc{A}{b\mu})-\xi b^a-gf^{abc}\caBRSTsrc{b}{b}\overline c^c-\caBRSTsrc{c}{a}-gf^{abc}\BRSTaBRSTsrcGhost{c}^bc^c.
	\label{bEgen}
\end{align}

The presence of the local anti-ghost equation implies, when coupled to the STI functional of~\1eq{STF-gen}, the existence of a Ward identity (WI) functional, since
\begin{align}
	{\cal W}^a\Gamma&= {\cal S}_\Gamma ( \overline {\cal G}^a\Gamma  - {\cal D}^{ab}_\mu A^{*b\mu} -gf^{abc}c^{*b}c^c ) + {\cal G}^a {\cal S}(\Gamma)=0,
	\label{WId}
\end{align}
where ${\cal S}_\Gamma$ is the linearized STI operator
\begin{align}
{\cal S}_\Gamma =  \int\!\mathrm{d}^4x & \left [
 \frac{\delta\Gamma}{\delta A^{*a}_\mu}\frac{\delta}{\delta A^{a\mu}}
+\frac{\delta\Gamma}{\delta A^{a\mu}} \frac{\delta}{\delta A^{*a}_\mu}
+\frac{\delta\Gamma}{\delta c^{*a}}\frac{\delta}{\delta c^a}
+\frac{\delta\Gamma}{\delta c^a}\frac{\delta}{\delta c^{*a}}+
b^a\frac{\delta}{\delta\overline c^a} \right.&
\nonumber\\
& \left.+\caBRSTsrc{A}{a}_\mu\frac{\delta}{\delta\cBRSTaBRSTsrc{A}{a\mu}}+\caBRSTsrc{c}{a}\frac{\delta}{\delta\BRSTaBRSTsrcGhost{c}^{a}}+\caBRSTsrc{\overline c}{a}\frac{\delta}{\delta\caBRSTsrc{b}{a}} 
\right ], &
\end{align}
while the WI operator ${\cal W}$ reads
\begin{align}
{\cal W}^a = & -{\cal D}^{ab}_\mu \frac{\delta}{\delta A^b_{\mu}}
-\BRSTaBRSTcovD \frac{\delta}{\delta \BRSTaBRSTsrc{A}^b_{\mu}}+gf^{abc}\sum\phi^c\frac{\delta}{\delta\phi^b};&
\phi&=\{c,\overline c, b,\BRSTsrc{A}, \BRSTsrc{c},\aBRSTsrc{A},\aBRSTsrc{c},\aBRSTsrc{\overline c},\aBRSTsrc{b},\BRSTaBRSTsrc{c}\}.
\label{WI} 
\end{align}

\begin{detailedcalc}
	There is no shortcut for establishing~\1eq{WId}; only knowing that there is an anti-ghost equation and a lengthy brute-force calculation. 
\end{detailedcalc}

\subsection{Yang-Mills ghost sector and confinement}

Rendering the Yang-Mills action both $s$- and $\bar s$-closed allows one to fully constrain the ghost sector of Yang-Mills theories for any value of the gauge-fixing parameter $\xi$. In the ghost sector, four Green functions are superficially divergent: $\Gamma_{c^a\bar c^b}$, $\Gamma_{\bar c^a \caBRSTsrc{A}{b}_\mu}$, $\Gamma_{c^aA^{*b}_\mu}$ and, finally, $\Gamma_{\caBRSTsrc{A}{a}_\mu A^{*b}_\nu}$. Then, factoring out the (trivial) color factor, we have:
\begin{subequations}
\begin{align}
	\left.\frac{\delta}{\delta c}{\cal G} \Gamma\right\vert_{\Phi,\Phi^*,\Phi^\#=0}=0&;&\Gamma_{c\overline{c}}(q)+iq^\mu\Gamma_{cA^*_\mu}(q)=0;
	\label{g1}\\
	\left.\frac{\delta}{\delta\aBRSTsrc{A}_\alpha}{\cal G} \Gamma\right\vert_{\Phi,\Phi^*,\Phi^\#=0}=0&;&i\Gamma_{\aBRSTsrc{A}_\alpha\overline{c}}(q)-q^\nu\Gamma_{\aBRSTsrc{A}_\mu A^*_\nu}(q)=iq_\alpha;
	\label{g2}\\
	\left.\frac{\delta}{\delta \overline{c}}\overline{\cal G} \Gamma\right\vert_{\Phi,\Phi^*,\Phi^\#=0}=0&;&i\Gamma_{\overline{c}c}(q)+q^\mu\Gamma_{\overline{c}\aBRSTsrc{A}_\mu}(q)+i\xi\Gamma_{\overline{c}\aBRSTsrc{b}}(q)=0;
	\label{ag1}\\
	\left.\frac{\delta}{\delta A^*_\mu}\overline{\cal G} \Gamma\right\vert_{\Phi,\Phi^*,\Phi^\#=0}=0&;&i\Gamma_{A^*_\mu c}(q)+q^\nu\Gamma_{A^*_\mu \aBRSTsrc{A}_\nu}(q)+i\xi\Gamma_{A^*_\mu\aBRSTsrc{b}}(q)=iq_\mu.
	\label{ag2}
\end{align}
\end{subequations}
In addition
\begin{align}
	\left.\frac{\delta}{\delta\aBRSTsrc{b}}{\cal G} \Gamma\right\vert_{\Phi,\Phi^*,\Phi^\#=0}=0&;&\Gamma_{\aBRSTsrc{b}\overline{c}}(q)+iq^\mu\Gamma_{\aBRSTsrc{b} A^*_\mu}(q)=0.
	\label{g3}
\end{align}
If we now contract~\1eq{ag2} with $q^\alpha$ and then use~\2eqs{g1}{g3} we obtain
\begin{align}
	i\Gamma_{c\bar c}(q)=q^2-iq^\mu q^\nu\Gamma_{\aBRSTsrc{A}_\nu A^*_\mu}(q)-i\xi\Gamma_{\aBRSTsrc{b}\overline{c}}(q).
	\label{rel1}
\end{align}

\begin{detailedcalc}
	Notice that for functions which are linear in the momentum $q$ one has, for example,
	\begin{align}
		\Gamma_{A^*_\mu c}(q)=\Gamma_{A^*_\mu c}(q,-q)=-\Gamma_{c A^*_\mu}(-q,q)=\Gamma_{c A^*_\mu}(q,-q)=\Gamma_{c A^*_\mu}(q),
	\end{align}
	where the first minus is due to reversing the order of the functional differentiation of $\Gamma$ for Grassmann-odd variables, while the second one comes from reversing the sign in the momentum flowing $q$. The same results evidently holds for $\Gamma_{\aBRSTsrc{A}_\alpha \overline c}$, whereas one has
	\begin{align}
		\Gamma_{A^*_\mu \aBRSTsrc{A}_\nu}(q)=\Gamma_{A^*_\mu \aBRSTsrc{A}_\nu}(q,-q)=-\Gamma_{\aBRSTsrc{A}_\nu A^*_\mu}(-q,q)=-\Gamma_{\aBRSTsrc{A}_\nu A^*_\mu}(q,-q)=-\Gamma_{\aBRSTsrc{A}_\nu A^*_\mu}(q)=-\Gamma_{\aBRSTsrc{A}_\mu A^*_\nu}(q),
	\end{align} 
	since Lorenz invariance demands $\Gamma_{A^*_\mu \aBRSTsrc{A}_\nu}(q)\sim  g_{\mu\nu}A(q^2)+q_\mu q_\nu B(q^2)$.
\end{detailedcalc}

Let us now introduce the ghost dressing function $F$ through
\begin{align}
	\Gamma_{c\bar c}(q)&=-iq^2F^{-1}(q^2);& D_{c\bar c}(q)&=i\frac{F(q^2)}{q^2},
	\label{ghdressF}
\end{align}
and, in addition, use Lorenz invariance to introduce the scalar functions
\begin{subequations}
\begin{align}
	\Gamma_{\aBRSTsrc{A}_\mu A^*_\nu}(q)&=ig_{\mu\nu}G(q^2)+i\frac{q_\mu q_\nu}{q^2}L(q^2);&
	\Gamma_{\aBRSTsrc{b}\overline{c}}(q)&=iq^2K(q^2);\\
	\Gamma_{cA^*_\mu}(q)&=q_\mu C(q^2);&
	\Gamma_{\overline{c}\aBRSTsrc{A}_\mu}(q)&=q_\mu E(q^2).
\end{align}
\label{defs}
\end{subequations}
Then,~\1eq{rel1} gives~\cite{Binosi:2013cea} 
\begin{align}
	F^{-1}(q^2)&=1+G(q^2)+L(q^2)+\xi K(q^2).
	\label{ghost-sector-constraint}
\end{align}
whereas we have, from~\2eqs{g1}{ag1}
\begin{subequations}
\begin{align}
	i\Gamma_{c\bar c}(q)&=q^\mu\Gamma_{cA^*_\mu}(q)& &\qquad\Rightarrow\qquad& F^{-1}(q^2)&\equiv C(q^2);\\
	i\Gamma_{c\bar c}(q)&=q^\mu\Gamma_{\overline{c}\aBRSTsrc{A}_\mu}(q)+i\xi\Gamma_{\overline{c}\aBRSTsrc{b}}(q)&
	&\qquad\Rightarrow\qquad& F^{-1}(q^2)&\equiv E(q^2)+\xi K(q^2).
\end{align}
\end{subequations}
When $\xi=0$ one has the simpler relation
\begin{align}
	F^{-1}(q^2)=1+G(q^2)+L(q^2);
	\label{F1GL}
\end{align}
and since it is relatively easy to prove that $L(q^2\to0)\to0$~\cite{Aguilar:2009nf}, one finds that the ghost dressing and $G$ functions are related in the IR:
\begin{align}
	F^{-1}(0)=1+G(0).
	\label{IRG}
\end{align}

Particularly interesting is the fact that the function $G$ can be obtained by considering the correlation function corresponding to the time ordered product of two covariant derivatives, one acting on a ghost field, and one on an anti-ghost field~\cite{Kugo:1979gm,Kugo:1995km}:
\begin{align}
	G^{ab}_{\mu\nu}(y-x)&=\langle T[({\cal D}^{bc}_\nu c^c)_x({\cal D}^{ad}_\mu\overline{c}^d)_y]\rangle.
\end{align} 
As the covariant derivative of a ghost (anti-ghost) field  is coupled to the BRST source $A^*$ (the anti-BRST source $\aBRSTsrc{A}$) one has
\begin{align}
	G^{ab}_{\mu\nu}(y-x)&=\frac{\delta^2W}{\delta\caBRSTsrc{A}{a}_\mu(y)\delta A^{*b}_\nu(x)},
	\label{KO-function} 
\end{align} 
where $W$ is the generating functional of the theory's connected graphs.

\begin{detailedcalc}
	Thus, $W$ represents the Legendre transform of the 1-particle irreducible generator $\Gamma$ with respect to $\Phi$:
	\begin{align}
	W&=\Gamma+\int\!\mathrm{d}^4x\,J_\Phi\Phi;&
	J_\Phi&=-(-1)^{\epsilon(\Phi)}\frac{\delta\Gamma}{\delta\Phi};& 
	\Phi&=\frac{\delta W}{\delta J_\Phi};&
	\frac{\delta W}{\delta\zeta}&=\frac{\delta\Gamma}{\delta\zeta}, \quad \zeta=\{\Phi^*,\aBRSTsrc{\Phi},\BRSTaBRSTsrc{\Phi}\},
	\end{align}
	where $J_\Phi$ denotes the source of the quantum field $\Phi$ and $\epsilon(\Phi)$ is its statistics, with $\epsilon(\Phi)=0$ (1) if $\Phi$ is a boson (fermion). 
\end{detailedcalc}

Now, there are only two possible connected contributions to~\1eq{KO-function}; and, factoring out the color structure, one has (in momentum space)
\begin{align}
	iG_{\mu\nu}(q)=\Gamma_{\aBRSTsrc{A}_\mu A^*_\nu}(q)+\Gamma_{\aBRSTsrc{A}_\mu\overline{c}}(q)D_{\overline{c}c}(q)\Gamma_{cA^*_\nu}(q).
\end{align} 
Using then~\2eqs{defs}{ghost-sector-constraint} we get
\begin{align}
	G_{\mu\nu}(q)&=G(q^2)g_{\mu\nu}+\frac{q_\mu q_\nu}{q^2}L(q^2)-\frac{q_\mu q_\nu}{q^2}E(q^2)F(q^2)C(q^2)\nonumber \\
	&=P_{\mu\nu}(q)G(q^2)-\frac{q_\mu q_\nu}{q^2},
\end{align} 
which represents a generalization to any value of the gauge-fixing parameter $\xi$ of a result known to be valid in the Landau gauge ($\xi=0$), where $G$ represents the so-called Kugo-Ojima function (usually indicated with $u$). The correlator in~\1eq{KO-function} is in fact accessible on the lattice, where it has delivered the IR behavior of the function $G$ in this gauge.

\begin{detailedcalc}
	To understand why the IR behavior of the $G$ function is so important we need to go back to the cohomology of the BRST operator. Recall that the physical observables of a Yang-Mills theory live in the zero ghost charge cohomology of the BRST operator $H^0(s)$; this is also true for the asymptotic particles (the spectrum) of the theory.   
	
	Owing to Noether's theorem, we know that associated to the BRST symmetry there is a conserved charge $Q_B$ so that $Q_B^2=0$ and for any physical state $\vert\mathrm{phys}\rangle$ one has
	\begin{align}
		Q_B\vert\mathrm{phys}\rangle=0.
	\end{align}
 	In this way, two physical states that differ by a $Q_B$ exact term are in the same cohomology class:
 	\begin{align}
 		Q_B\vert\mathrm{phys}\rangle'=Q_B\left(\vert\mathrm{phys}\rangle+Q_B\vert\chi\rangle\right)=Q_B\vert\mathrm{phys}\rangle=0.
 	\end{align}
 	\end{detailedcalc}
 	\begin{detailedcalc}
 	A way to realize color confinement is then through the construction of a color global charge $Q^a$ such that for every physical state
 	\begin{align}
 		Q^a\vert\mathrm{phys}\rangle=0,
 	\end{align}
 	which would then imply that all states are color singlets. In particular, if we could write
 	\begin{align}
 		Q^a=\{ Q_B,{\cal N}^a\},
 		\label{colcharge}
 	\end{align}
 	for some ${\cal N}^a$ then
 	\begin{align}
 		Q^a\vert\mathrm{phys}\rangle=Q_B{\cal N}^a\vert\mathrm{phys}\rangle+{\cal N}^aQ_B\vert\mathrm{phys}\rangle=Q_B\left({\cal N}^a\vert\mathrm{phys}\rangle\right),
 	\end{align}
 	which implies that $Q^a$ is BRST exact and therefore trivial in cohomology space.
 	\end{detailedcalc}
 	\begin{detailedcalc}
 	Now it turns out that the equation of motion of the gauge field can be written as 
 	\begin{align}
 		J^a_\mu=\partial^\nu F^a_{\mu\nu}+\{Q_B,{\cal D}^{ab}_\mu\bar c^b\},
 	\end{align}
 	where $J^a_\mu$ is the Noether current associated with color symmetry. Additionally, the Noether current is not uniquely defined, since one can always add a term  of the form $\partial_\nu f^{\mu\nu}$, where $f^{\mu\nu}$ is an arbitrary local antisymmetric tensor. In other words, the modified current $J^{\prime a}_{\mu} \equiv J^{a}_{\mu} + \partial_\nu f^{\mu\nu}$ is still conserved; and, moreover, the corresponding conserved charge generates the correct infinitesimal color transformation on any field operator.
 	Thus we can define the color charge $Q^a$ as
 	\begin{align}
 		Q^a&=\int\!{\mathrm d}^3x\,\left(J^a_0-\partial^\nu F^a_{0\nu}\right)=\int\!{\mathrm d}^3x\,\{Q_B,{\cal D}^{ab}_0\bar c^b\},
 		\label{cchr}
 	\end{align}
 	which is exactly of the needed form of~\1eq{colcharge} once we proceed to the identification ${\cal N}^a=\{Q_B,{\cal D}^{ab}_0\bar c^b\}$.
 	\end{detailedcalc}
 	\begin{detailedcalc}
 	
 	The problem here is that the definition in~\1eq{cchr} does not lead to a well-defined color charge operator owing to the presence of massless one-particle poles, which render the 3-volume integration divergent. To see this, observe that the $b$-equation,~\1eq{bEgen}, implies the relation
 	\begin{align}
 		\Gamma_{A^a_\mu b^b}=-iq_\mu\delta^{ab}\qquad\Rightarrow\qquad
 		\langle T(A^a_\mu b^b)\rangle=i\frac{q_\mu}{q^2}\delta^{ab}&.
 	\end{align} 
 	In addition, ghost charge conservation implies that $\langle T(A^a_\mu \bar c^b)\rangle=0$ so that
 	\begin{align}
 		s\langle T(A^a_\mu \bar c^b)\rangle=0\qquad\Rightarrow\qquad
		\langle T[{\cal D}^{ac}_\mu c^c)\bar c^b]\rangle=-\langle T(A^a_\mu b^b)\rangle=-i\frac{q_\mu}{q^2}\delta^{ab}.
 	\end{align}
 	As the relations above are {\it exact}, they imply the existence of massless asympotic fields, the so-called quartet mechanism, see~\cite{Kugo:1979gm}. Cancelling these poles requires a certain relation between the weights with which such fields contribute to the charge in~\1eq{cchr}. Indeed, one finds~\cite{Kugo:1979gm,Kugo:1995km}
 	\begin{align}
 		Q^a&=\int\!{\mathrm d}^3x\,\left(J^a_0+\frac vw\partial^\nu F^a_{0\nu}\right),
 		\label{welldcc}
 	\end{align}  
 	subjected to the constraint 
 	\begin{align}
 		v=-w+(1+u).
 	\end{align}
 	Then, if $1+u=1+G(0)=0$, $v/w=-1$ and one is left with the well defined color charge in~\1eq{welldcc}, that coincides with~\1eq{cchr}.   
 	
	\end{detailedcalc}

\section{\label{three}Background field gauges}

We have shown that rendering the $SU(N_c)$ Yang-Mills action simultaneously invariant under both BRST and anti-BRST symmetry entails the appearance of new functional identities, the most important being the local anti-ghost equation,~\1eq{local-anti-ghost}, and the WI functional,~\1eq{WI}. It turns out that there are gauges in which both a WI functional and a local anti-ghost equation (at least in the Landau gauge) also appear: background field (BF) gauges~\cite{Abbott:1980hw,Abbott:1981ke}.  

In these gauges, one starts by splitting the gauge field into a background part, $\BRSTaBRSTsrc{A}$, and a quantum part, $Q$, according to
\begin{align}
	A^a_\mu=\cBRSTaBRSTsrc{A}{a}_\mu+Q_\mu^a.
\end{align}
Each term will then encode different gauge transformations:
\begin{itemize}
	\item The quantum field $Q$ represents the variable in the path integral, so that we can define a gauge transformation that acts on $Q$ (and leaves the path integral measure invariant) keeping $\BRSTaBRSTsrc{A}$ fixed
	\begin{align}
		\delta_\theta\cBRSTaBRSTsrc{A}{a}_\mu&=0;&
		\delta_\theta Q^a_\mu&={\cal D}_\mu^{ab}\theta^b.
		\label{qgi}	
	\end{align}
	\item The background field $\BRSTaBRSTsrc{A}$ represents instead a classical source, so that gauge invariance with respect to the field $A$ can be thought of as a simultaneoeus transformation of both $\BRSTaBRSTsrc{A}$ and $Q$, according to
	\begin{align}
		\delta_\omega\cBRSTaBRSTsrc{A}{a}_\mu&=\BRSTaBRSTcovD\omega^b;&
		\delta_\omega Q^a_\mu&=gf^{abc}Q^b_\mu \omega^c,
		\label{bgi}
	\end{align}
	which implies the usual gauge transformation on the total field $A$: $\delta_\omega A^a_\mu={\cal D}_\mu^{ab}\omega^b$. Evidently, under a background gauge transformation the quantum part of the gauge field transforms as would a matter field in the adjoint representation.    
\end{itemize}  
Classically, ignoring gauge fixing, the path integral would be invariant under both kinds of transformations. In practice, however, we know that we need to fix the gauge; but we can do that only by breaking the quantum gauge invariance,~\1eq{qgi}, leaving the background gauge invariance of the action in~\1eq{bgi} intact. 

This can be achieved by choosing the gauge fixing function
\begin{align}
	\widehat{\cal F}^a=\BRSTaBRSTcovD Q^\mu_b,
	\label{gff}
\end{align}
so that, under the background gauge transformation of~\1eq{bgi}, one finds the adjoint transformation
\begin{align}
	\delta_\omega\widehat{\cal F}^a=gf^{abc}\widehat{\cal F}^b\omega^c.
	\label{Fgt}
\end{align}

\begin{detailedcalc}
	To explicitly check~\1eq{Fgt}, observe that
	\begin{align}
		\delta_\omega\BRSTaBRSTcovD &=gf^{abc}\widehat{\cal D}^{cd}_\mu\omega^d;&
		\delta_\omega Q^b_\mu&=gf^{bde}Q^d_\mu\omega^e,
	\end{align}
	which implies
	\begin{align}
		\delta_\omega\widehat{F}^a&=gf^{abc}(\partial^\mu Q_\mu^b)\omega^c+g^2(f^{abe}f^{bdc}+f^{bec}f^{adb})\cBRSTaBRSTsrc{A}{d}_\mu\omega^cQ^e_\mu\nonumber \\
		&=gf^{abc}(\partial^\mu Q_\mu^b+gf^{bde}\cBRSTaBRSTsrc{A}{d}_\mu Q^e_\mu)\omega^c\nonumber\\
		&=gf^{abc}\widehat{\cal F}^b\omega^c.
	\end{align}
\end{detailedcalc}
Then, the gauge fixing action corresponding to~\1eq{gff} is invariant under background gauge transformations, since
\begin{align}
	\delta_\omega(\widehat{{\cal F}}^a\widehat{{\cal F}}^a)\sim gf^{abc}\widehat{{\cal F}}^a\widehat{{\cal F}}^b\omega^c=0.
\end{align}

The ghost sector associated with the breaking of the quantum gauge invariance can be obtained by introducing the gauge-fixing fermion
\begin{align}
	\widehat\Psi&=\int\!\mathrm{d}^4x\,\overline{c}^a\left(\widehat{\cal F}^a-\frac\xi2b^a\right);&
	\widehat{S}_\mathrm{GF}+\widehat{S}_\mathrm{FP}&=s\widehat{\Psi}.	
\end{align}
with the additional BRST transformations
\begin{align}
	s\cBRSTaBRSTsrc{A}{a}_\mu&=\caBRSTsrc{A}{a}_\mu;& s\caBRSTsrc{A}{a}_\mu&=0;&
	sQ^a_\mu&={\cal D}^{ab}_\mu c^b-\caBRSTsrc{A}{a}_\mu,
	\label{add-back-BRST}
\end{align}	
where $\BRSTaBRSTsrc{A}$ is an anticommuting source which implements the equation of motion of the background part of the gauge field at the quantum level.  As $\BRSTaBRSTsrc{A}$ and $\aBRSTsrc{A}$ are introduced as a BRST-doublet/trivial pair, the physical observables of the theory are not affected. Background gauge invariance is then preserved if
\begin{align}
	\delta_\omega\varphi&=gf^{abc}\varphi^b\omega^c;&
	\varphi&=\{c,\overline{c},b,\aBRSTsrc{A}\}.
\end{align}

\begin{detailedcalc}
	Indeed, one has (omitting the space-time integral)
	\begin{align}
		\delta_\omega\widehat{\Psi}&=
		\delta_\omega\overline{c}^a\left({\cal F}^a-\frac\xi2b^a\right)+\overline{c}^a\delta_\omega{\cal F}^a-\frac\xi2\overline{c}^a\delta_\omega b^a\nonumber\\
		&=gf^{abc}\overline{c}^b\omega^c{\cal F}^a-g\frac\xi2gf^{abc}\overline{c}^b\omega^cb^a+\overline{c}^agf^{abc}\overline{c}^a{\cal F}^b\omega^c-\frac\xi2\overline{c}^agf^{abc}b^b\omega^c=0,
	\end{align}
	which implies $\delta_\omega s\widehat{\Psi}=s\delta_\omega\widehat{\Psi}=0$.
\end{detailedcalc}
Thus we have that the action
\begin{align}
	\widehat{S}=S_\mathrm{YM}+\widehat{S}_\mathrm{GF}+\widehat{S}_\mathrm{FP},
	\label{back-action}
\end{align}
is both BRST and background gauge invariant. 
 
In addition to the usual Feynman rules involving quantum fields (which are obtained from the previously introduced ones through the replacement $A\to Q$),
$\widehat{S}$ gives rise to the following Feynman rules involving background fields:  
\medskip

\noindent
\begin{tabular}{@{} C{0.225\linewidth} L{0.775\linewidth} @{}}
\includegraphics[scale=0.6]{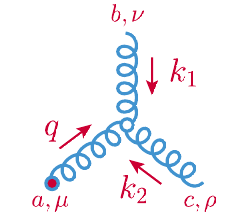}
&
\leftflushmath{
\Gamma^{(0)}_{\cBRSTaBRSTsrc{A}{a}_\mu A^b_\nu A^c_\rho}(q,k_1) &=gf^{abc}\widehat\Gamma^{(0)}_{\mu\nu\rho}(q,k_1)\\
&=gf^{abc}[g_{\nu\rho}(k_1-k_2)_\mu+g_{\mu\rho}(k_2-q+\frac{k_1}{\xi})_\nu\\
&\mathrel{\phantom{=}}+g_{\mu\nu}(q-k_1-\frac{k_2}{\xi} )_\rho],
}\\

\includegraphics[scale=0.6]{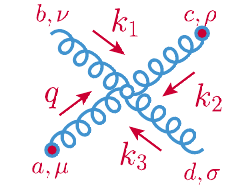}
&
\leftflushmath{
\Gamma^{(0)}_{\cBRSTaBRSTsrc{A}{a}_\mu A^b_\nu \cBRSTaBRSTsrc{A}{c}_\rho A^d_\sigma }(q,k_1,k_2)
    &=-ig^2[
	f^{ade}f^{ecb}(g_{\mu\rho}g_{\nu\sigma}-g_{\mu\nu}g_{\rho\sigma}+\frac1\xi g_{\mu\sigma}g_{\nu\rho})\\
	&\mathrel{\phantom{=}}+f^{abe}f^{edc}(g_{\mu\sigma}g_{\nu\rho}-g_{\mu\rho}g_{\nu\sigma}-\frac1\xi g_{\mu\nu}g_{\rho\sigma})\\
	&\mathrel{\phantom{=}}+f^{ace}f^{edb}(g_{\mu\sigma}g_{\nu\rho}-g_{\mu\nu}g_{\rho\sigma})],
}\\
\includegraphics[scale=0.6]{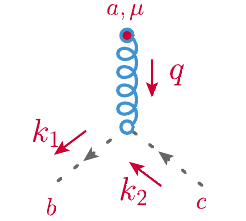}
&
\leftflushmath{
\Gamma^{(0)}_{c^c\cBRSTaBRSTsrc{A}{a}_\mu \overline c^b}(k_2,q)&=gf^{abc}\widehat\Gamma_\mu(q,k_2)\\ 
	&=gf^{abc}(k_1+k_2)_{\mu},
}
\end{tabular}

\noindent
\begin{tabular}{@{} C{0.225\linewidth} L{0.775\linewidth} @{}}
\includegraphics[scale=0.6]{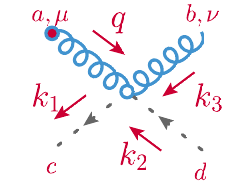}
&
\leftflushmath{
\Gamma^{(0)}_{c^d\cBRSTaBRSTsrc{A}{a}_\mu A^b_\nu \overline c^c}(k_2,q,k_3) &=ig^2g_{\mu\nu}f^{ace}f^{ebd},
}
\\
\includegraphics[scale=0.6]{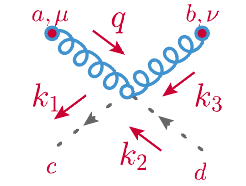}
&
\leftflushmath{
\Gamma^{(0)}_{c^d\cBRSTaBRSTsrc{A}{a}_\mu \cBRSTaBRSTsrc{A}{b}_\nu \overline c^c}(k_2,q,k_3) &=ig^2g_{\alpha\beta}(f^{ace}f^{ebd}+f^{bce}f^{ead}).
} 

\end{tabular}
\medskip
Notice that there is a final vertex
\begin{align}
	\Gamma_{\widehat{A}^a_\mu Q^b_\nu Q^c_\rho Q^d_\sigma}(q,k_1,k_2)&=-ig^2\widehat{\Gamma}^{abcd}_{\mu\nu\rho\sigma}(q,k_1,k_2); 
\end{align}
at tree level, however, one has the identity
$\Gamma^{(0)}_{\widehat{A}^a_\mu Q^b_\nu Q^c_\rho Q^d_\sigma}\equiv\Gamma^{(0)}_{Q^a_\mu Q^b_\nu Q^c_\rho Q^d_\sigma}$.

Equipped with these rules, we can calculate the Green functions involving background fields; and particularly relevant is the self-energy of two background fields, which, according to the Feynman rules above, has the following two one-loop gluon contributions

\medskip
\noindent\makebox[\textwidth][c]{%
\includegraphics[scale=0.6]{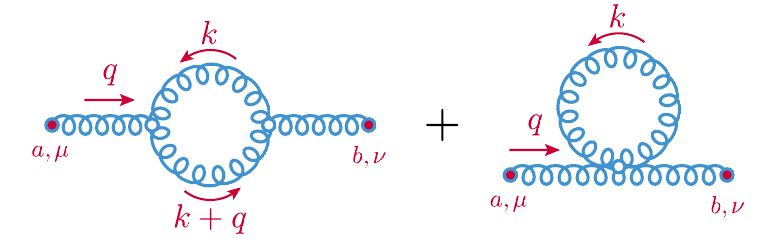}
}
The seagull term vanishes as in the calculation of the conventional self-energy owing to the dimensional regularization result of~\1eq{dimreg-zero}; the first diagram yields instead (in the Feynman gauge)
\begin{align}
	\widehat{\widehat{\Pi}}\mbox{}^{(1)\mathrm{gl}}_{\mu\nu}(q;\mu)&=\frac12g^2C_A\int_k\!\frac{\widehat\Gamma^{(0)}_{\mu\rho\sigma}(q,k)\widehat\Gamma^{\rho\sigma(0)}_\nu(-q,k+q)}{k^2(k+q)^2},
	\label{BB-gluon-se}
\end{align}
with
\begin{align}
	\widehat\Gamma^{(0)}_{\mu\rho\sigma}(q,k)\widehat\Gamma^{(0)\rho\sigma}_\nu(-q,k+q)&=d(2k+q)_\mu (2k+q)_\nu+8q^2P_{\mu\nu}(q).
	\label{r1}
\end{align}
Notice that in the equations above the number of `hats' on the corresponding Green function indicates the number of background fields we are considering.

The ghost contributions are instead given by the diagrams
\medskip

\noindent\makebox[\textwidth][c]{%
\includegraphics[scale=0.6]{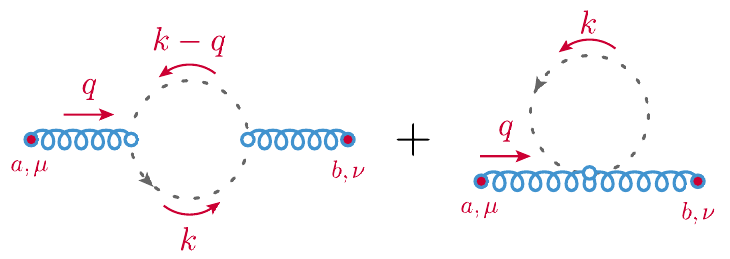}
}
Again the seagull term vanishes in dimensional regularization, whereas the first term reads
\begin{align}
	\widehat{\widehat{\Pi}}\mbox{}^{(1)\mathrm{gh}}_{\mu\nu}(q;\mu)&=-g^2C_A\int_k\!\frac{\widehat\Gamma^{(0)}_{\mu}(q,k-q)\widehat\Gamma^{(0)}_\nu(-q,k)}{k^2(k-q)^2}\nonumber \\
	&=-g^2C_A\int_k\!\frac{(2k+q)_\mu (2k+q)_\nu}{k(k+q)^2},
	\label{BB-gluon-se}
\end{align}
where in the last step we have made the replacement $k\to-k$.

It is then immediate to realize that the transversality of the background gluon 2-point function is realized in a ``block-wise'' fashion~\cite{Aguilar:2006gr,Binosi:2009qm}, {\it i.e.}, independently for each gluon and ghost contributions,
\begin{align}
	q^\nu\widehat{\widehat{\Pi}}\mbox{}^{(1)\mathrm{gl}}_{\mu\nu}(q;\mu)&=0;&
	q^\nu\widehat{\widehat{\Pi}}\mbox{}^{(1)\mathrm{gh}}_{\mu\nu}(q;\mu)&=0,
	\label{BB-blockwise}
\end{align}
as opposed to the ``cross-talk'' realization happening in the conventional case (where the cancellation happens because the gluon and ghost terms cancel each other). We will return to this important concept (and its generalization) later on.

\begin{detailedcalc}
	To check \1eq{BB-blockwise} observe that it is sufficient to prove that the ghost term is transverse, as~\1eq{r1} implies
	\begin{align}
		\widehat{\widehat{\Pi}}\mbox{}^{(1)\mathrm{gl}}_{\mu\nu}(q;\mu)&\sim
		\widehat{\widehat{\Pi}}\mbox{}^{(1)\mathrm{gh}}_{\mu\nu}(q;\mu)+\mathrm{transverse\ terms}.
	\end{align}
	Then one has
	\begin{align}
		q^\nu\widehat{\widehat{\Pi}}\mbox{}^{(1)\mathrm{gh}}_{\mu\nu}(q;\mu)\propto\int_k\!\frac{(2k+q)_\mu(q^2+2k{\cdot}q)}{k^2(k+q)^2}=\int_k\!\frac{(2k+q)_\mu[(k+q)^2-k^2)}{k^2(k+q)^2}=2q_\mu\int\!\frac1{k^2}=0.
	\end{align}
	Since $\widehat{\widehat{\Pi}}^{(1)\mathrm{gh}}$ is transverse we then have
	\begin{align}
		\widehat{\widehat{\Pi}}\mbox{}^{(1)\mathrm{gh}}_{\mu\nu}(q;\mu)&=P_{\mu\nu}(q)\widehat{\widehat{\Pi}}\mbox{}^{(1)\mathrm{gh}}(q^2;\mu^2),
	\end{align}
	with 
	\begin{align}
		\widehat{\widehat{\Pi}}\mbox{}^{(1)\mathrm{gh}}(q^2;\mu^2)=-g^2C_A\frac1{d-1}\int_k\!\frac{(2k+q)^2}{k^2(k+q)^2}=g^2C_A\frac{q^2}{d-1}\int_k\!\frac{1}{k^2(k+q)^2}.
	\end{align}
	Thus, also using~\1eq{r1}, we get
	\begin{subequations}
	\begin{align}
		\widehat{\widehat{\Pi}}\mbox{}^{(1)\mathrm{gl}}_{\mu\nu}(q;\mu)&=\frac{g^2C_A}2\left(\frac{7d-8}{d-1}\right)q^2P_{\mu\nu}(q)\int_k\!\frac{1}{k^2(k+q)^2};\\
		\widehat{\widehat{\Pi}}\mbox{}^{(1)\mathrm{gh}}_{\mu\nu}(q;\mu)&=\frac{g^2C_A}{d-1}q^2P_{\mu\nu}(q)\int_k\!\frac{1}{k^2(k+q)^2},
	\end{align}
	\end{subequations}
	which yield the final results 
	\begin{align}
		\widehat{\widehat{\Pi}}\mbox{}^{(1)\mathrm{gl}}_{\mu\nu}(q;\mu)&=\frac{10}3g^2C_Aq^2P_{\mu\nu}(q)f_\epsilon(q^2,\mu^2);&
		\widehat{\widehat{\Pi}}\mbox{}^{(1)\mathrm{gh}}_{\mu\nu}(q;\mu)&=\frac{1}3g^2C_Aq^2P_{\mu\nu}(q)f_\epsilon(q^2,\mu^2).
		\label{1lBB}	
	\end{align}
	 
\end{detailedcalc}  
Notice also that the conventional and background gluon one-loop self-energies are related through the equation
\begin{align}
	\Pi_{\mu\nu}^{(1)}(q;\mu)&=\widehat{\widehat{\Pi}}\mbox{}^{(1)}_{\mu\nu}(q;\mu)-2g^2C_Aq^2P_{\mu\nu}(q)\int_k\frac1{k^2(k+q)^2},
	\label{BQI-in-disguise}
\end{align} 
which represents the one-loop expansion of a set of identities that will be presented later on and go under the name of background-quantum (BQ) identities~\cite{Binosi:2002ez,Grassi:1999tp}. Using then the results~\2eqs{f-exp}{1lBB}, we obtain
\begin{align}
	\widehat{\widehat{\Pi}}\mbox{}^{(1)}_{\mu\nu}(q;\mu)&=P_{\mu\nu}(q)\widehat{\widehat{\Pi}}\mbox{}^{(1)}(q^2;\mu^2);& \widehat{\widehat{\Pi}}\mbox{}^{(1)}(q^2;\mu^2)&=i\alpha(\mu^2)\frac{\beta_0}{4\pi}q^2\left(\frac2\epsilon-\gamma-\ln4\pi-\ln\frac{q^2}{\mu^2}\right),
	\label{1-loop-BB-se}
\end{align}
where $\beta_0=(4\pi)^2b=\frac{11}{3}C_A$ is the one-loop coefficient of the QCD $\beta$ function in the absence of quarks loops. Notice that, while we have obtained the result reported in~\1eq{1-loop-BB-se} in the Feynman gauge, the latter is completely general. Then, dropping the divergent part (that will renormalize the tree-level propagator), we obtain at one-loop order in perturbation theory
\begin{align}
q^2\hspace{0.13cm} \widehat{\hspace{-0.13cm}\widehat{d}}\mbox{\ }^{(1)}(q^2)=\alpha(\mu^2)q^2\widehat{\widehat{\Delta}}\mbox{}^{(1)}(q^2;\mu^2)&=\frac{\alpha(\mu^2)}{1+\frac i{q^2}\widehat{\widehat{\Pi}}\mbox{}^{(1)}(q^2;\mu^2)}=\frac{\alpha(\mu^2)}{1+\alpha(\mu^2)\frac{\beta_0}{4\pi}\ln\frac{q^2}{\mu^2}} .
\end{align}
The denominator vanishes at the scale defined by
$\alpha(\mu^2)\frac{\beta_0}{4\pi}\ln\frac{\Lambda_\mathrm{QCD}^2}{\mu^2}=-1$,
which introduces the RGI scale $\Lambda_\mathrm{QCD}$. Expressing the result in terms of this scale, one obtains
\begin{align}
	q^2\hspace{0.13cm}\widehat{\hspace{-0.13cm}\widehat{d}}\mbox{\ }^{(1)}(q^2)&=\frac{1}{\frac{\beta_0}{4\pi}\ln\frac{q^2}{\Lambda_\mathrm{QCD}^2}}.
	\label{thed}
\end{align}

Thus, at least at one-loop, the $q^2\hspace{0.13cm}\widehat{\hspace{-0.13cm}\widehat{d}}$ combination involving the background propagator, is RGI and coincides with the theory's effective charge, exactly as it happens in the case of quantum electro-dynamics, where the photon propagator at one-loop provides $\beta_0=-4/3$. As we will soon see, this is a completely general result solely owing to the background gauge invariance of the action and the corresponding WI this residual invariance generates.    

\begin{detailedcalc}
	To be sure this does not happen for the conventional (quantum) propagator, for which the expansion of~\1eq{conv-pi-1l} yields
	\begin{align}
		\Pi^{(1)}(q^2;\mu^2)&=i\alpha\frac{5C_A}{12\pi}q^2\left(\frac2\epsilon-\gamma-\ln4\pi-\ln\frac{q^2}{\mu^2}\right).
	\end{align}
	In this case, however, RGI quantities are dictated by the vertex structure of the action and its BRST invariance, which leads to the relations
	\begin{align}
		Z_{Ac^2}&=Z_gZ_A^{\frac12}Z_c;&
		Z_{A^3}&=Z_gZ_A^{\frac32};&
		Z_{A^4}&=Z^2_gZ_A^2;
	\end{align}	
	where $Z_{Ac^2}$, $Z_{A^3}$ and $Z_{A^4}$ are the renormalization constants of the ghost-gluon, three-gluon and four-gluon vertices respectively and 
	\begin{align}
		A_0&=Z_A^\frac12A;&
		c_0&=Z_cc;&
		g_0&=Z_gg,
	\end{align}
	with the `0' subscript indicating bare quantities. Standard calculations give (again in the Feynman gauge, for consistency)
	\begin{align}
		F^{(1)}(q^2;\mu^2)&=i\alpha\frac{C_A}{8\pi}q^2\left(\frac2\epsilon-\gamma-\ln4\pi-\ln\frac{q^2}{\mu^2}\right)=\Gamma^{(1)}_{Ac^2}(q^2;\mu^2),
	\end{align}
	where $\Gamma_{Ac^2}$ is the ghost-gluon vertex dressing function. Then, looking this time at the (momentum-independent) renormalization constants rather than to the (momentum-dependent) finite pieces of the dressing functions, we get the one-loop result
	\begin{align}
		Z_\alpha&=Z^2_g=Z^2_{Ac^2}Z_A^{-1}Z_c^{-2}=1-\alpha\frac{\beta_0}{4\pi}\frac2{\epsilon},
	\end{align}
	which, through the (one-loop) beta function $\beta=-bg^3$ gives rise to the effective coupling 
	\begin{align}
		\alpha(q^2)=\frac1{\frac{\beta_0}{4\pi}\ln\frac{q^2}{\Lambda^2_\mathrm{QCD}}},
	\end{align}
	as previously found in~\1eq{thed}.

\end{detailedcalc}

Similarly to what was done in the covariant gauges case, from the action of~\1eq{back-action} we can construct the complete action according to
\begin{align}
	\widehat{S}_\mathrm{C}=\widehat{S}+S_\mathrm{BV},
\end{align}
with the antifields transforming in the adjoint representation with respect to the background gauge transformations, so that $\widehat{S}_\mathrm{C}$ is  background gauge invariant too:
\begin{align}
	\delta_\omega\Phi^{*a}=gf^{abc}\Phi^{*b}\omega^c\qquad\Rightarrow\qquad \delta_\omega\widehat{S}_\mathrm{C}=0.
\end{align}
Given the additional BRST transformations~\1eq{add-back-BRST}, the effective background action $\widehat{\Gamma}$ satisfies the STI
	\begin{align}
		\widehat{\cal S}(\widehat\Gamma)&=0;&
	\widehat{\cal S}(\widehat\Gamma)&=\int\!\mathrm{d}^4x\left[\frac{\delta\widehat\Gamma}{\delta A^{*a}_\mu}\frac{\delta\widehat\Gamma}{\delta Q^{a\mu}}+\frac{\delta\widehat\Gamma}{\delta c^{*a}}\frac{\delta\widehat\Gamma}{\delta c^a}+
		b^a\frac{\delta\widehat\Gamma}{\delta\overline c^a}+\caBRSTsrc{A}{a}_\mu\left(\frac{\delta\widehat\Gamma}{\delta\cBRSTaBRSTsrc{A}{a\mu}}-\frac{\delta\widehat\Gamma}{\delta Q^{a\mu}}\right)\right],
		\label{backgroundSTI}
	\end{align}
whereas the residual background gauge invariance gives rise to the WI
\begin{align}
	\widehat{\cal W}^a\widehat\Gamma&=0;&
	\widehat{\cal W}^a&=-{\cal D}^{ab}_\mu \frac{\delta}{\delta Q^b_{\mu}}
-\BRSTaBRSTcovD \frac{\delta}{\delta \BRSTaBRSTsrc{A}^b_{\mu}}+gf^{abc}\sum\phi^c\frac{\delta}{\delta\phi^b},
\label{backgroundWI} 
\end{align}
with $\phi=\{c,\overline c, b,\BRSTsrc{A}, \BRSTsrc{c},\aBRSTsrc{A}\}$.

\subsection{BF gauges and anti-BRST}

The reader might have noticed that the background field, $\BRSTaBRSTsrc{A}$, and the associated source, $\aBRSTsrc{A}$, have been denoted with exactly the same letters as, respectively, the gluon field BRST/anti-BRST and anti-BRST sources (see the discussion below~\1eq{98}). To understand why we did that, observe that the BF gauge fixing function of~\1eq{gff} can be rewritten as
\begin{align}
	\widehat{\cal F}^a&={\cal F}^a-{\cal D}^{ab}_\mu\cBRSTaBRSTsrc{A}{b\mu}.
\end{align} 
Then we find~\cite{Binosi:2013cea}
\begin{align}
	s\widehat{\Psi}&=s\left(\overline{c}^a{\cal F}^a-\frac{\xi}2\overline{c}^ab^a\right)-s\left(\overline{c}^a{\cal D}^{ab}_\mu\cBRSTaBRSTsrc{A}{b\mu}\right)=S_\mathrm{GF}+S_\mathrm{FP}+\cBRSTaBRSTsrc{A}{a}_\mu s\overline{s}A^a_\mu+\caBRSTsrc{A}{a}_\mu\overline{s}A^{a\mu}.
	\label{BRSTaBRST}
\end{align}

\begin{detailedcalc}
	To prove~\1eq{BRSTaBRST} we start by observing that, neglecting surface terms, one has
	\begin{align}
		\Phi^a{\cal D}^{ab}_\mu\Phi^{\prime b}&=-(\partial_\mu\Phi^a)\Phi^{\prime b}+gf^{abc}A^c_\mu\Phi^b\Phi^{\prime a}=({\cal D}^{ab}_\mu\Phi^a)\Phi^{\prime b},
	\end{align}
	which immediately provides the results
	\begin{align}
		b^a{\cal D}^{ab}_\mu\cBRSTaBRSTsrc{A}{b\mu}&=-\cBRSTaBRSTsrc{A}{a\mu}{\cal D}^{ab}_\mu b^b;&
		\overline{c}^a{\cal D}^{ab}_\mu\caBRSTsrc{A}{b\mu}&=\caBRSTsrc{A}{a\mu}{\cal D}^{ab}_\mu \overline{c}^b,
	\end{align}
	where we have duly taken into account the statistics of the fields. Then, using the identities above, we find
	\begin{align}
		-s(\overline{c}^a{\cal D}^{ab}_\mu\cBRSTaBRSTsrc{A}{b\mu})&=-b^a{\cal D}^{ab}_\mu\cBRSTaBRSTsrc{A}{b\mu}+\overline{c}^a[s{\cal D}^{ab}_\mu]\cBRSTaBRSTsrc{A}{b\mu}+\overline{c}^a{\cal D}^{ab}_\mu\caBRSTsrc{A}{b\mu}\nonumber \\
		&=\cBRSTaBRSTsrc{A}{a\mu}{\cal D}^{ab}_\mu b^b+gf^{abc}\overline{c}^b{\cal D}^{cd}c^d+\caBRSTsrc{A}{a\mu}{\cal D}^{ab}_\mu \overline{c}^b=\cBRSTaBRSTsrc{A}{a}_\mu s\overline{s}A^a_\mu+\caBRSTsrc{A}{a}_\mu\overline{s}A^{a\mu}.
	\end{align}
\end{detailedcalc}
This is a somewhat unexpected result as we see BF gauges emerge from the requirements of the simultaneous BRST and anti-BRST invariance of the Yang-Mills action:
\begin{align}
	\underbrace{S_\mathrm{YM}+s\Psi+S'_\mathrm{BV}}_\text{BRST/anti-BRST\ invariant}&=\underbrace{S_\mathrm{YM}+s\widehat{\Psi}+S_\mathrm{BV}}_\text{BF\ gauge-fixed}+\int\!\mathrm{d}^4x\bigg(\underbrace{\caBRSTsrc{c}{a}\overline{s} c^a+\caBRSTsrc{b}{a}\overline{s} b^a+\caBRSTsrc{\overline{c}}{a}\overline{s} \overline{c}^a+\cBRSTaBRSTsrc{c}{a}s\overline{s}c^a}_\text{leftovers}\bigg).
\end{align}
The background field $\BRSTaBRSTsrc{A}$ is therefore identified with the source of the BRST/anti-BRST variation of the gluon field, whereas its doublet partner $\aBRSTsrc{A}$ is identified with the source associated with the anti-BRST variation of the gluon field.  This explains why the STI in~\1eq{backgroundSTI} is identical to the one in~\1eq{STF-gen} when the  background-quantum  splitting is ``undone'', and the WI~\1eq{backgroundWI} perfectly matches the one found in~\1eq{WI}.  

Since in the following we will not be concerned with the sectors of the theory where $\aBRSTsrc{c}$, $\aBRSTsrc{b}$, $\aBRSTsrc{\overline{c}}$ or $\BRSTaBRSTsrc{c}$ play some role, these terms can be put to zero and one is left with the identity~\cite{Binosi:2013cea}
\begin{align}
	\text{BRST/anti-BRST\ invariance}\equiv\text{BF\ gauges}.
	\label{mbeq}
\end{align}    
Not only does this equation give a meaning to the somewhat mysterious invariance of the action under the anti-BRST symmetry~\cite{Weinberg:1996kr}; it will also show that rendering the Yang-Mills action simultaneously invariant under both BRST and anti-BRST symmetry provides new and better ways to understand its IR dynamics, as we have already seen in the case of the ghost sector.    

\subsection{Background-quantum and Ward identities}

Green functions involving background fields are related to the corresponding ones where one or more of those fields are replaced by quantum fields through a set of identities that go under the name of background-quantum identities  (BQIs)~\cite{Binosi:2002ez,Grassi:1999tp}. The most important of these relations are those satisfied by the gluon two-point function. To get them, we take the functional differentiation of the STI~\1eq{backgroundSTI}, with respect to one background source and one background field (which, incidentally, is also a {\it source}) or quantum field. Then, omitting the hat on both the STI and the effective action, owing to~\1eq{mbeq}, we find the BQIs
\begin{subequations}
\begin{align}
	\left.\frac{\delta^2}{\delta\aBRSTsrc{A}\delta\BRSTaBRSTsrc{A}}{\cal S}(\Gamma)\right\vert_{\Phi,\Phi^\#,\Phi^*=0}&=0;&
	i\Gamma_{\cBRSTaBRSTsrc{A}{a}_{\mu}\cBRSTaBRSTsrc{A}{b}_{\nu}}(q)&=[ig^\rho_\mu\delta^{ac}+\Gamma_{\caBRSTsrc{A}{a}_{\mu} A^{*c}_\rho}(q)]\Gamma_{Q^{c\rho}\cBRSTaBRSTsrc{A}{b}_{\nu}}(q).\\
	\left.\frac{\delta^2}{\delta\aBRSTsrc{A}\delta Q}{\cal S}(\Gamma)\right\vert_{\Phi,\Phi^\#,\Phi^*=0}&=0;&
	i\Gamma_{\cBRSTaBRSTsrc{A}{a}_{\mu}Q^{b}_{\nu}}(q)&=[ig^\rho_\mu\delta^{ac}+\Gamma_{\caBRSTsrc{A}{a}_{\mu} A^{*c}_\rho}(q)]\Gamma_{Q^{c\rho}Q^{b}_{\nu}}(q).
\end{align}
\label{2-pointBQI}
\end{subequations}
Using the transversality of the gluon two-point functions, we see that the $L$ term in the decomposition of~\1eq{defs} of the function $\Gamma_{\aBRSTsrc{A}A^*}$ drops out and the BQIs above simplify to
\begin{align}
	\Gamma_{\BRSTaBRSTsrc{A}_{\mu}\BRSTaBRSTsrc{A}_{\nu}}(q)&=[1+G(q^2)]\Gamma_{Q_{\mu}\BRSTaBRSTsrc{A}_{\nu}}(q);&
	\Gamma_{\BRSTaBRSTsrc{A}_{\mu}Q_{\nu}}(q)&=[1+G(q^2)]\Gamma_{Q_{\mu}Q_{\nu}}(q).
	\label{intBQI}
\end{align}
The two identities above can be then combined in a unique relation:
\begin{align}
	\Gamma_{\BRSTaBRSTsrc{A}_{\mu}\BRSTaBRSTsrc{A}_{\nu}}(q)=[1+G(q^2)]^2\Gamma_{Q_{\mu}Q_{\nu}}(q),
	\label{BQIwithG}
\end{align}
which shows that the same function governing the dynamics of the ghost sector determines also how the gluon 2-point sector behaves. Notice that using~\1eq{GammaDelta} we can finally relate the background propagator with the conventional one:
\begin{align}
	\Delta^{-1}(q^2)=\frac{\widehat{\Delta}\mbox{}^{-1}(q^2)}{1+G(q^2)}=\frac{\widehat{\widehat{\Delta}}\mbox{}^{-1}(q^2)}{[1+G(q^2)]^2}.
	\label{BQIprop}
\end{align} 
The importance of this equation will be soon clear.

\begin{detailedcalc}

	Expanding~\1eq{BQIwithG} at one-loop and observing that $G^{(0)}=0$, we find
	\begin{align}
		\Gamma^{(1)}_{\BRSTaBRSTsrc{A}_{\mu}\BRSTaBRSTsrc{A}_{\nu}}(q;\mu)=\Gamma^{(1)}_{Q_{\mu}Q_{\nu}}(q;\mu)+2G^{(1)}(q^2;\mu^2)\Gamma^{(0)}_{Q_{\mu}Q_{\nu}}(q).
	\end{align}
 	Using next~\1eq{GammaDelta}, which extends, {\it mutatis mutandis}, to the background two-point functions, we obtain
	\begin{align}
		\widehat{\widehat{\Pi}}\mbox{}^{(1)}_{\mu\nu}(q;\mu)&=\Pi_{\mu\nu}^{(1)}(q;\mu)+2G^{(1)}(q^2;\mu^2)q^2P_{\mu\nu}(q).
		\label{one-random-elation}
	\end{align}
	
	Now, the function $\Gamma_{\aBRSTsrc{A}A^*}$ is, in fact, calculable, both in perturbation theory and beyond that. To this end we need the Feynman rules:
\medskip	
	
\begin{tabular}{@{} C{0.225\linewidth} L{0.775\linewidth} @{}}
\includegraphics[scale=0.6]{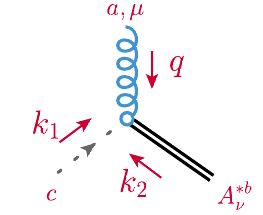}
&
\leftflushmath{
\Gamma^{(0)}_{c^c A^a_\mu A^{*b}_\nu}(q,k_1) &=-igf^{abc}g_{\mu\nu}
,}
\\
\includegraphics[scale=0.6]{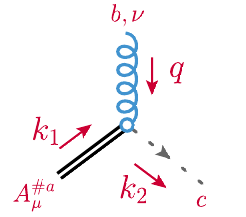}
&
\leftflushmath{
\Gamma^{(0)}_{A^{\#a}_\mu A^b_\nu \overline c^c  }(k_1,q) &=-igf^{abc}g_{\mu\nu}.}

\end{tabular}
\end{detailedcalc}

\begin{detailedcalc}
Thus, at the one-loop level there is only a single Feynman diagram contributing to the amplitude $\Gamma_{\aBRSTsrc{A}A^*}$, and namely

\noindent\makebox[\textwidth][c]{%
\includegraphics[scale=0.6]{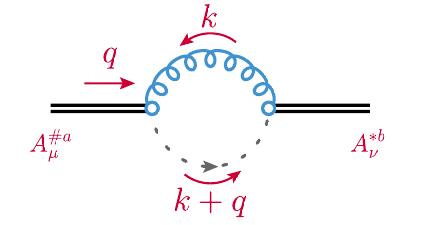}
}
which yields
\begin{align}
	\Gamma^{(1)}_{A^{\#a}_\mu A^{*b}_\nu}(q;\mu)&=iC_A\delta^{ab}g_{\mu\nu}\int_k\frac1{k^2(k+q)^2};& G^{(1)}(q^2;\mu^2)&=C_A\int_k\frac1{k^2(k+q)^2}.
\end{align}
As anticipated, combining this result with the one in~\1eq{one-random-elation} gives back~\1eq{BQI-in-disguise} previously found when confronting the one-loop gluon self-energy evaluated in the covariant and BF (Feynman) gauges. 

\end{detailedcalc}

Functional differentiations of the WI~\1eq{backgroundWI} provide identities expressing the residual invariance of the action under background gauge transformations. In particular, differentiations with respect to two quantum fields, a ghost and an anti-ghost field, and three quantum fields, gives the identities
\begin{subequations}
	\begin{align}
	q^\mu\Gamma_{\cBRSTaBRSTsrc{A}{a}_\mu Q^b_\nu Q^c_\rho}(q,k_1)&=igf^{aec}\Gamma_{Q^b_\nu Q^e_\rho}(k_1)+igf^{aeb}\Gamma_{Q^c_\rho Q^e_\nu}(k_2);& q+k_1+k_2&=0;\label{WI3g}\\ 
	q^\mu\Gamma_{c^c\cBRSTaBRSTsrc{A}{a}_\mu \overline c^b}(k_2,q)&=igf^{aec}\Gamma_{c^e \overline c^b}(k_1)+igf^{aeb}\Gamma_{c^c \overline c^e}(k_2);& q-k_1+k_2&=0;\\
	q^\mu\Gamma_{\cBRSTaBRSTsrc{A}{a}_\mu Q^b_\nu Q^c_\rho Q^d_\sigma}(q,k_1,k_2)&=
	igf^{ade}\Gamma_{Q^b_\nu Q^c_\rho Q^e_\sigma}(k_1,k_2)\nonumber \\
	&+igf^{abe}\Gamma_{ Q^c_\rho Q^d_\sigma Q^e_\nu}(k_2,k_3)\nonumber \\
	&+igf^{ace}\Gamma_{ Q^d_\sigma Q^b_\nu Q^e_\rho}(k_3,k_1);& q+k_1+k_2+k_3&=0.
	\end{align}
	\label{hatWIs} 
\end{subequations}

\begin{detailedcalc}
	These results can easily be generalized to provide the  divergence with respect to the background field of a generic Green function containing $n$ quantum fields $Q^{b_i}_{\nu_i}(q_i)$ and $m$ ghost/anti-ghost pairs $c^{c_j}(k_{1j})\overline c^{d_j}(k_{2j})$; one has
	\begin{align}
		(-1)^n q^\mu\Gamma_{\widehat A^a_\mu(q)\Pi_i Q^{b_i}_{\nu_i}(q_i)\Pi_jc^{c_j}(k_{1j})\overline c^{d_j}(k_{2j})}&=ig\sum_if^{ab_ie}\Gamma_{\Pi_{k\neq i } Q^{b_k}_{\nu_k}(q_k)Q^e_{\nu_i}(q+q_i)\Pi_jc^{c_j}(k_{1j})\overline c^{d_j}(k_{2j})}\nonumber\\
		&+ig\sum_jf^{ac_je}\Gamma_{\Pi_i  Q^{b_i}_{\nu_i}(q_i)c^e(q+k_{1j})\overline{c}^{d_j}(k_{2j})\Pi_{k\neq j}c^{c_k}(k_{1k})\overline c^{d_k}(k_{2k})}\nonumber\\
		&+ig\sum_jf^{ad_je}\Gamma_{\Pi_i  Q^{b_i}_{\nu_i}(q_i)c^{c_j}(k_{1j})\overline{c}^{e}(q+k_{2j})\Pi_{k\neq j}c^{c_k}(k_{1k})\overline c^{d_k}(k_{2k})},
		\label{fullygenWI}
	\end{align}
	where we have temporarily adopted a notation in which all field momenta are explicitly shown in the subscript of the Green function, rather than written as its arguments.  
\end{detailedcalc}

The first identity in~\1eq{hatWIs} represents a transposition of what we are familiar with from QED, where the divergence of the photon-lepton vertex equates the difference of two inverse lepton's propagators; 
	\label{WI3gII}
and, as in the QED case, it will imply a relation between the coupling constant and the background-gluon renormalization factors. Indeed if we indicate with $Z_{\widehat{A}Q^2}$ the renormalization constant of the $\widehat{A}QQ$ vertex, the from the action we get the relations
\begin{align}
	Z_{\widehat{A}Q^2}&=Z_gZ_{\widehat{A}}^{\frac12}Z_Q;&
	\widehat{A}_0&=Z_{\widehat{A}}^{\frac12}\widehat{A};& Q_0&=Z^{\frac12}_QQ,&
\end{align} 
whereas the WI~\1eq{WI3gII} gives
\begin{align}
	Z_{\widehat{A}Q^2}=Z_Q \qquad\Rightarrow\qquad Z_gZ_{\widehat{A}}^{\frac12}=1.
\end{align} 
This then implies that, exactly as it happens in QED, one can form the RGI combination~\cite{Binosi:2002vk,Aguilar:2009nf} 
\begin{align}
	\widehat{\hspace{-0.13cm}\widehat{d}}\mbox{\ }(q^2)&=\alpha(\mu^2)\widehat{\widehat{\Delta}}(q^2;\mu^2)=\alpha(\mu^2)\frac{\Delta(q^2;\mu^2)}{[1+G(q^2;\mu^2)]^2},
	\label{dhathat}
\end{align}
where in the last step we made use of the BQI of~\1eq{BQIprop}. This combination, which carries the dimension of an inverse mass squared, will be the key to constructing a unique and process independent IR extension of the theory's effective charge, the ultimate goal of this discussion.

\subsection{Dyson--Schwinger equations and block-wise realization of Ward identities}

The functional approach developed so far has already allowed us to derive relations and establish results which are generally valid, even beyond the usual perturbative expansion. To proceed further along this fruitful path we need to introduce the DSEs~\cite{Dyson:1949ha,Dyson:1949hb,Schwinger:1951,Schwinger:1951II}, which, when a theory is quantized within the path integral formalism that we have adopted, provide the quantum equations of motion of the theory's fields. A detailed treatment on how to derive these equations from the generating functional is outside the scope of this discussion; we will therefore content ourselves with developing a somewhat sketchy approach~\cite{Eichmann:HadronPhysicsNotes} that allows to derive them (and can be made, in fact, rigorous).  

Coming from the generating functional, it should be no surprise that the derivation of such equations involves the use of (the by now familiar) functional differentiations; however, rather than providing their formal action, we give the following graphical rules for a generic field $\Phi$:
  
\medskip
\begin{equation}
\raisebox{-0.48\height}{%
  \includegraphics[scale=0.65]{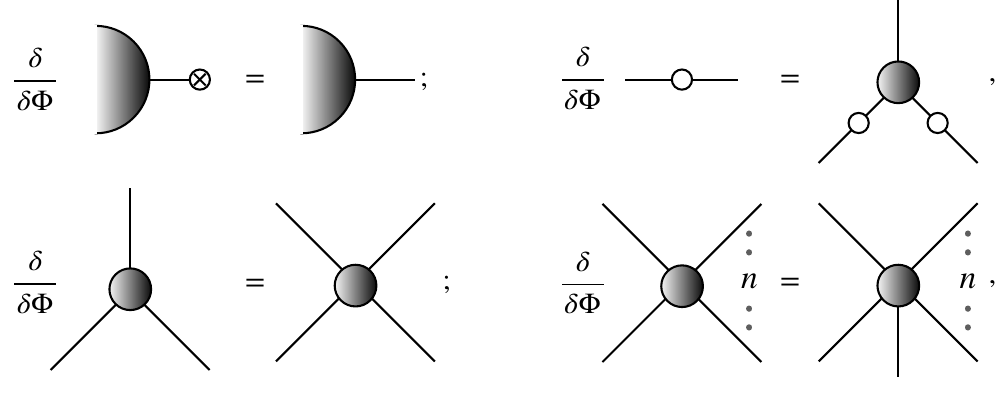}%
}
\label{DSE-rules}
\end{equation}

\noindent where white (gray) blobs represent fully dressed propagators (one-particle irreducible vertices). Notice that in providing these rules we have suppressed all signs, $i$ factors and multiplicities that derive from having to differentiate the different legs.

Then (topologies contributing to the) Yang-Mills DSEs can be obtained by taking successive functional differentiations of the field's quantum equation of motion  

\begin{equation}
\raisebox{-0.48\height}{%
  \includegraphics[scale=0.65]{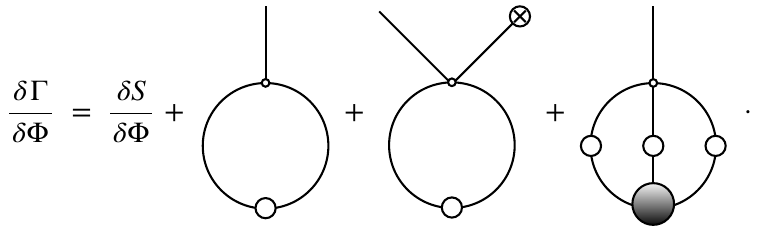}%
}
\label{PhiDSE}
\end{equation}

\noindent One sees immediately that the topologies characterizing~\1eq{PhiDSE} can be divided in two groups according to the number of loops present: besides the tree level term (the first entry on the right-hand side) one has one-loop dressed diagrams (second and third term) and two-loop dressed diagrams (the fourth term). As the differentiation rules~\1eq{DSE-rules} provided are local, not only does this characterization persist across all DSEs, but also one-loop (respectively, two-loop) dressed diagrams in the DSE can be generated only from functional differentiations  of one-loop (respectively, two-loop) dressed diagrams.

Next, the topologies contributing to the DSE satisfied by the two-point function are obtained from~\1eq{PhiDSE} by taking another functional differentiation and then setting all the fields to zero, thus obtaining 

\begin{equation}
\raisebox{-0.48\height}{%
  \includegraphics[scale=0.65]{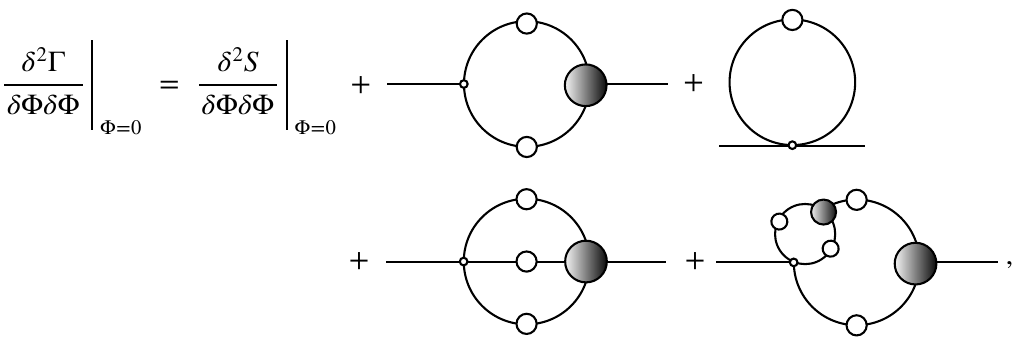}%
}
\label{PhiPhiDSE}
\end{equation}
where the first (second) line contains the one-loop (two-loop) dressed topologies. Notice that these topologies might or might not appear in a particular realization of the DSE: for example, for the gluon two-point function all topologies are active, whereas for the ghost two-point functions only the first one-loop dressed topology appears.

Another differentiation gives rise to the topologies for the three-point functions

\begin{equation}
\raisebox{-0.48\height}{%
  \includegraphics[scale=0.65]{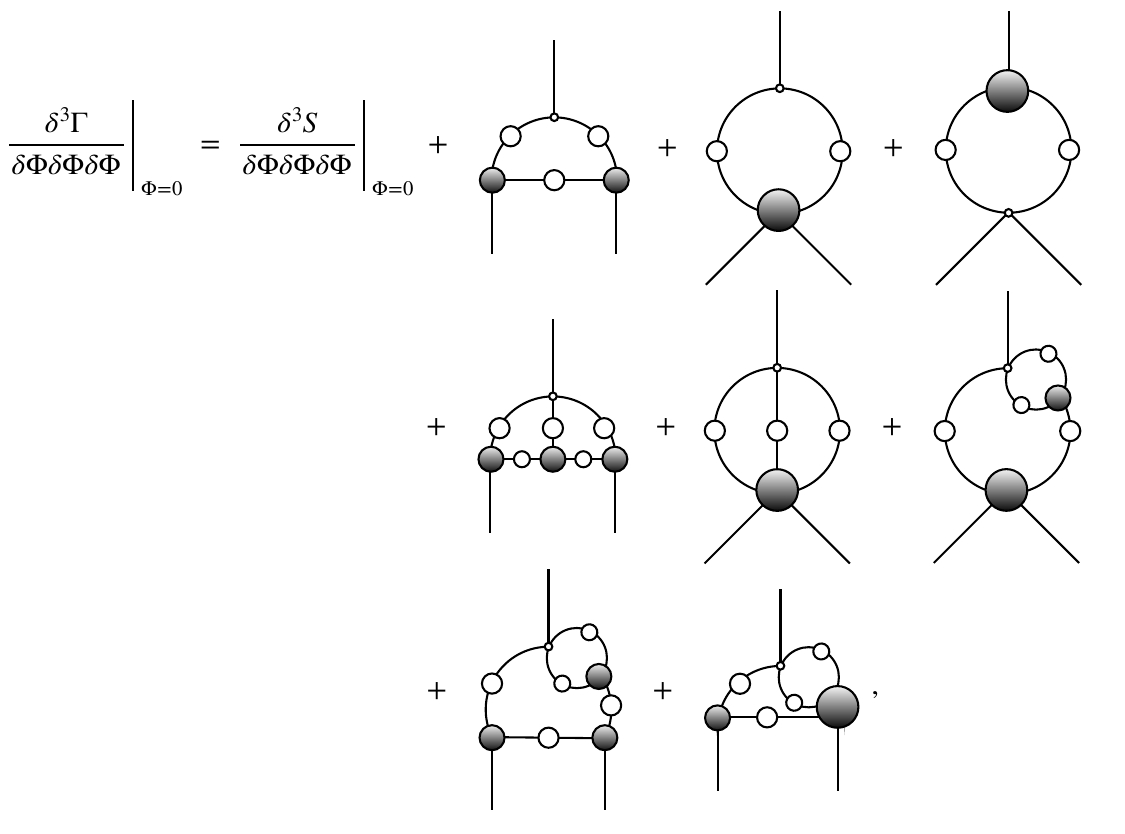}%
}
\label{PhiPhiPhiDSE}
\end{equation}

\noindent with the first line containing the one-loop dressed topologies.
Notice that this equation can be written in a more compact notation as
\begin{equation}
\raisebox{-0.48\height}{%
  \includegraphics[scale=0.65]{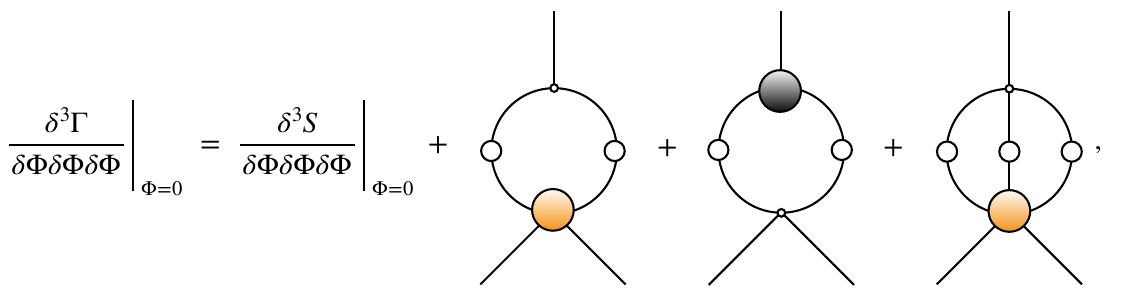}%
}
\label{PhiPhiPhiDSE-compact}
\end{equation}
where the orange blobs are DSE kernels whose skeleton expansion in terms of one-particle irreducible (1-PI) vertices generates all the diagrams in~\1eq{PhiPhiPhiDSE} above.

When specializing DSEs to gluons or ghosts, a second classification naturally arises, based on the types of legs circulating inside the loops (without decomposing full propagators or vertices). If all internal legs are gluons, we refer to \emph{gluon diagrams}; otherwise, we have \emph{ghost diagrams} (note that the number of ghost legs must be even---one $c$ and one $\bar c$---in order to conserve ghost charge). Thus, for any given DSE, there are four possible classes of contributions: one-loop dressed gluon/ghost, and two-loop dressed gluon/ghost contributions.

Accordingly, one may introduce the notion of a \emph{block-wise} realization of a WI or STI: such an identity is said to be realized \emph{block-wise} if it is satisfied separately within each class of diagrams, without any cross-talk between different sectors. The reader may recall our earlier computation of the one-loop gluon self-energy, where we found that $\widehat{\widehat{\Pi}}\mbox{}^{(1)}$ is block-wise transverse, see~\1eq{BB-blockwise}; the conventional self-energy $\Pi^{(1)}$ is not. 

This property is actually valid to all orders~\cite{Aguilar:2006gr}: the proof in this case is particularly simple, and proceeds by contracting the diagrams in the four blocks by the background gluon momentum $q$ from the side of the fully dressed vertices, thus triggering the corresponding WIs~\1eq{hatWIs}. Much more complicated is the proof that the WI satisfied by the vertex $\Gamma_{\widehat{A}QQ}$ is realized according to exactly the same pattern~\cite{Aguilar:2022exk}. 

Luckily, however, we do not need that: as we will see, it turns out that one can relatively easily prove that {\it all} the background WI are realized in a block-wise fashion. To this end, let us consider a generic Green function involving one background field, $n$ quantum fields, and $m$ ghost/anti-ghost pairs, and let us write it schematically as $\Gamma_{\widehat{A}_\mu Q^n(c\overline{c})^m}$. Then, such a function has four possible contributions
\begin{align}
	\Gamma_{\widehat{A}_\mu Q^n(c\overline{c})^m}=\Gamma^{1\ell;\rm{gl}}_{\widehat{A}_\mu Q^n(c\overline{c})^m}+\Gamma^{1\ell;\rm{gh}}_{\widehat{A}_\mu Q^n(c\overline{c})^m}+\Gamma^{2\ell;\rm{gl}}_{\widehat{A}_\mu Q^n(c\overline{c})^m}+\Gamma^{2\ell;\rm{gh}}_{\widehat{A}_\mu Q^n(c\overline{c})^m}.
	\label{deco2}
\end{align} 
Now, we also know that $\Gamma_{\widehat{A}_\mu Q^n(c\overline{c})^m}$ satisfies the general WI of~\1eq{fullygenWI}, which can be written, without loss of generality, in the schematic form
\begin{align}
	q^\mu\Gamma_{\widehat{A}_\mu Q^n(c\overline{c})^m}=ig\sum f^{\cdots}\Gamma_{Q^n(c\overline{c})^m}.
\end{align}
Now the right-hand side of the identity above can also be unambiguously decomposed into 1- and 2-loop dressed gluon and ghost contributions:
\begin{align}
	ig\sum f^{\cdots}\Gamma_{Q^n(c\overline{c})^m}&=ig\sum f^{\cdots}\Gamma^{1\ell;\rm{gl}}_{Q^n(c\overline{c})^m}+ig\sum f^{\cdots}\Gamma^{1\ell;\rm{gh}}_{Q^n(c\overline{c})^m}+ig\sum f^{\cdots}\Gamma^{2\ell;\rm{gl}}_{Q^n(c\overline{c})^m}\nonumber \\
	&+ig\sum f^{\cdots}\Gamma^{2\ell;\rm{gh}}_{Q^n(c\overline{c})^m}.
	\label{deco1}
\end{align}
For the left-hand side, we can instead contract both sides of~\1eq{deco2} with the momentum of the background leg, to obtain
\begin{align}
	q^\mu\Gamma_{\widehat{A}_\mu Q^n(c\overline{c})^m}=q^\mu\Gamma^{1\ell;\rm{gl}}_{\widehat{A}_\mu Q^n(c\overline{c})^m}+q^\mu\Gamma^{1\ell;\rm{gh}}_{\widehat{A}_\mu Q^n(c\overline{c})^m}+q^\mu\Gamma^{2\ell;\rm{gl}}_{\widehat{A}_\mu Q^n(c\overline{c})^m}+q^\mu\Gamma^{2\ell;\rm{gh}}_{\widehat{A}_\mu Q^n(c\overline{c})^m}.
	\label{deco}
\end{align}
 
Now, for each group on the right-hand side of~\1eq{deco} one proceeds by contracting its composing diagrams with the background gluon momentum $q$ from the side of the fully dressed vertices, triggering additional WIs of the type~\1eq{fullygenWI}. Observe however that the latter identities are: linear, so there cannot be cross-talk between 1- and 2-loop dressed group of diagrams; ghost free, so that the particle content inside loops is preserved and no additional ghost line can be possibly generated besides the ones already present within each group. 

Therefore, even after triggering all the possible WIs with the background momentum $q$, each sector in~\1eq{deco} will generate a result that is still within that sector: No cancellations have occurred between the different sectors and no diagram originating in one group has been reassigned to any other group. Thus:
\begin{align}
	q^\mu\Gamma^{1\ell;\rm{gl}}_{\widehat{A}_\mu Q^n(c\overline{c})^m}&={\cal R}^{1\ell;\rm{gl}}_{Q^n(c\overline{c})^m};&
	q^\mu\Gamma^{1\ell;\rm{gh}}_{\widehat{A}_\mu Q^n(c\overline{c})^m}&={\cal R}^{1\ell;\rm{gh}}_{Q^n(c\overline{c})^m},\nonumber \\
	q^\mu\Gamma^{2\ell;\rm{gl}}_{\widehat{A}_\mu Q^n(c\overline{c})^m}&={\cal R}^{2\ell;\rm{gl}}_{Q^n(c\overline{c})^m};&
	q^\mu\Gamma^{2\ell;\rm{gh}}_{\widehat{A}_\mu Q^n(c\overline{c})^m}&={\cal R}^{2\ell;\rm{gh}}_{Q^n(c\overline{c})^m}.&
\end{align}

But these results must also match the decomposition of~\1eq{deco1}, meaning in turn that 
\begin{align}
	{\cal R}^{1\ell;\rm{gl}}_{Q^n(c\overline{c})^m}&=ig\sum f^{\cdots}\Gamma^{1\ell;\rm{gl}}_{Q^n(c\overline{c})^m};&
	{\cal R}^{1\ell;\rm{gh}}_{Q^n(c\overline{c})^m}&=ig\sum f^{\cdots}\Gamma^{1\ell;\rm{gh}}_{Q^n(c\overline{c})^m},\nonumber \\
	{\cal R}^{2\ell;\rm{gl}}_{Q^n(c\overline{c})^m}&=ig\sum f^{\cdots}\Gamma^{2\ell;\rm{gl}}_{Q^n(c\overline{c})^m};&
	{\cal R}^{2\ell;\rm{gh}}_{Q^n(c\overline{c})^m}&=ig\sum f^{\cdots}\Gamma^{2\ell;\rm{gh}}_{Q^n(c\overline{c})^m},
\end{align}
finally provideing the claimed block-wise realization of the WI.

\begin{detailedcalc}
	The non-linearity and the presence of ghosts in general STIs force crosstalk between  the different blocks. Perhaps the easiest way to realize this is through the so-called Pinch Technique (PT)~\cite{Cornwall:1981zr,Cornwall:1989gv,Binosi:2002ft,Binosi:2003rr,Binosi:2004qe,Aguilar:2006gr,Binosi:2007pi,Binosi:2008qk,Binosi:2009qm}, a field-theoretic method in non-Abelian gauge theories that systematically reorganizes perturbative amplitudes so as to extract effective Green functions that are gauge independent, process independent, and satisfy QED-like WIs. At all orders, PT Green functions are known to coincide with those computed in the Feynman gauge of the BF gauges, providing a formal foundation for the construction and explaining their Abelian-like properties despite the non-Abelian dynamics of the underlying theory.
	
	The starting point of the PT construction is the splitting of the conventional three-gluon vertex
	\begin{subequations}
	\label{GP}
	\begin{align}
		\Gamma^{(0)}_{\alpha \mu \nu}(q,k_1)&=\widehat{\Gamma}^{(0)}_{\alpha \mu \nu}(q,k_1)+\Gamma_{\alpha \mu \nu}^{{\rm P}}(q,k_1),\\
		\widehat{\Gamma}^{(0)}_{\alpha \mu \nu}(q,k_1) &=
		(k_1-k_2)_{\alpha} g_{\mu\nu} + 2q_{\nu}g_{\alpha\mu} 
		- 2q_{\mu}g_{\alpha\nu}, \label{GF}\\
		\Gamma_{\alpha \mu \nu}^{{\rm P}}(q,k_1) &= 
 		k_{2\nu} g_{\alpha\mu} - k_{1\mu}g_{\alpha\nu},
	\end{align}
	\end{subequations}
where $\widehat{\Gamma}$ coincides with the BF gauge vertex $\Gamma_{\widehat{A}QQ}$ evaluated in the Feynman gauge, whereas $\Gamma^{{\rm P}}$ contains the longitudinal momenta that trigger all sorts of STIs inside gluon diagrams (in the PT construction gluon propagators are dynamically projected to their value in the Feynman gauge). 
  	
  	Now, consider the DSE of the conventional $\Gamma_{QQQ}$ vertex. Then, the only diagram that admits the split~\1eq{GP} is the first one in~\1eq{PhiPhiPhiDSE-compact} when all the lines are gluons. On the other hand, we know that the PT procedure dynamically generates the BF gauge vertex $\Gamma_{\widehat{A}QQ}$, so that the STIs triggered by longitudinal momenta present in $\Gamma^\mathrm{P}$ must be such that 
	\begin{equation}
	\raisebox{-0.48\height}{%
  	\includegraphics[scale=0.6]{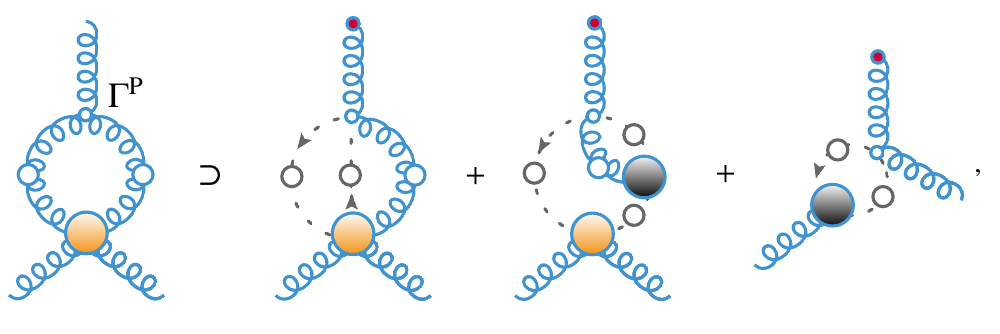}%
	}
	\label{PT-QQQ}
	\end{equation}
	which shows that STIs can generate from a 1-loop dressed gluon diagram both 1- and 2-loop dressed ghost diagrams.
	\end{detailedcalc}

The combination of the BQI~\1eq{BQIprop} and the block-wise property is extremely powerful. Because of the latter property the DSE for the background field propagator is easier to handle than the conventional one. Recall the one-loop case: block-wise transversality implies that the gluon and ghost contributions to the gluon self-energy are individually transverse. Consequently, omitting one of them still yields a transverse tensor structure, albeit with an incorrect prefactor. This should be contrasted with the conventional formulation, where neglecting the ghost contribution leads to an unphysical, non transverse result. The same is true for the full DSE; for example we can consider retaining only the diagrams belonging to the 1-loop dressed gluon class, thus obtaining the (truncated) DSE
\begin{align}
\widehat{\widehat{\Delta}}\mbox{}^{-1}(q^2)\,P_{\mu\nu}(q)
= q^2 P_{\mu\nu}(q)+i\widehat{\widehat{\Pi}}\mbox{}^{1\ell;\mathrm{gl}}_{\mu\nu}(q).
\label{eq:851}
\end{align}
and from the block-wise transversality property we know that the one-loop dressed gluon diagrams are transverse, namely
\begin{align}
	\widehat{\widehat{\Pi}}\mbox{}^{1\ell;\mathrm{gl}}_{\mu\nu}(q)=P_{\mu\nu}(q)\widehat{\widehat{\Pi}}\mbox{}^{1\ell;\mathrm{gl}}(q^2).
\end{align}
Consequently, the transverse projector  appears identically on both sides of \1eq{eq:851}, allowing one to isolate the scalar coefficients and arrive at the scalar equation
\begin{align}
\widehat{\widehat{\Delta}}\mbox{}^{-1}(q^2)
= q^2+i\widehat{\widehat{\Pi}}\mbox{}^{1\ell;\mathrm{gl}}(q^2).
\label{eq:853}
\end{align}
An analogous truncated equation may be written for any of the remaining classes previously identified, or for arbitrary combinations thereof, without spoiling the transversality of the result. The only problem with this line of reasoning, is that these DSEs do not constitute a genuine dynamical equation, since background fields act as sources and therefore cannot propagate inside loops. As a consequence, the propagator appearing on the left-hand side of~\1eq{eq:853} does not coincide with the one on its right-hand side, which is the conventional propagator associated with the quantum fields. Schematically,
\begin{align}
\widehat{\widehat{\Delta}}\mbox{}^{-1}(q^2) = q^2 + \int_k \Delta(k)\,\Delta(k+q) \cdots+\int_k\Delta(k)\cdots,
\end{align}
and one cannot solve for either $\Delta$ or $\widehat{\widehat{\Delta}}$ in isolation. This shortcoming however is evidently remedied by the BQI of~\1eq{BQIprop}, since one can then write
\begin{align}
	{\Delta}^{-1}(q^2)=\frac{\widehat{\widehat{\Delta}}\mbox{}^{-1}(q^2)}{[1+G(q^2)]^2}=\frac{q^2+i\widehat{\widehat{\Pi}}\mbox{}^{1\ell;\mathrm{gl}}(q^2)}{[1+G(q^2)]^2}.
	\label{giDSE}
\end{align}

Thus, the block-wise realization of WIs coupled to the propagator BQI, provides a systematic, gauge-invariant truncation scheme for the propagator's DSE that preserves the transversality of the result. As a consequence, the number of coupled DSEs that must be retained in order to maintain gauge (or BRST) symmetry can drastically be reduced.

\begin{figure}
	\includegraphics[scale=0.345]{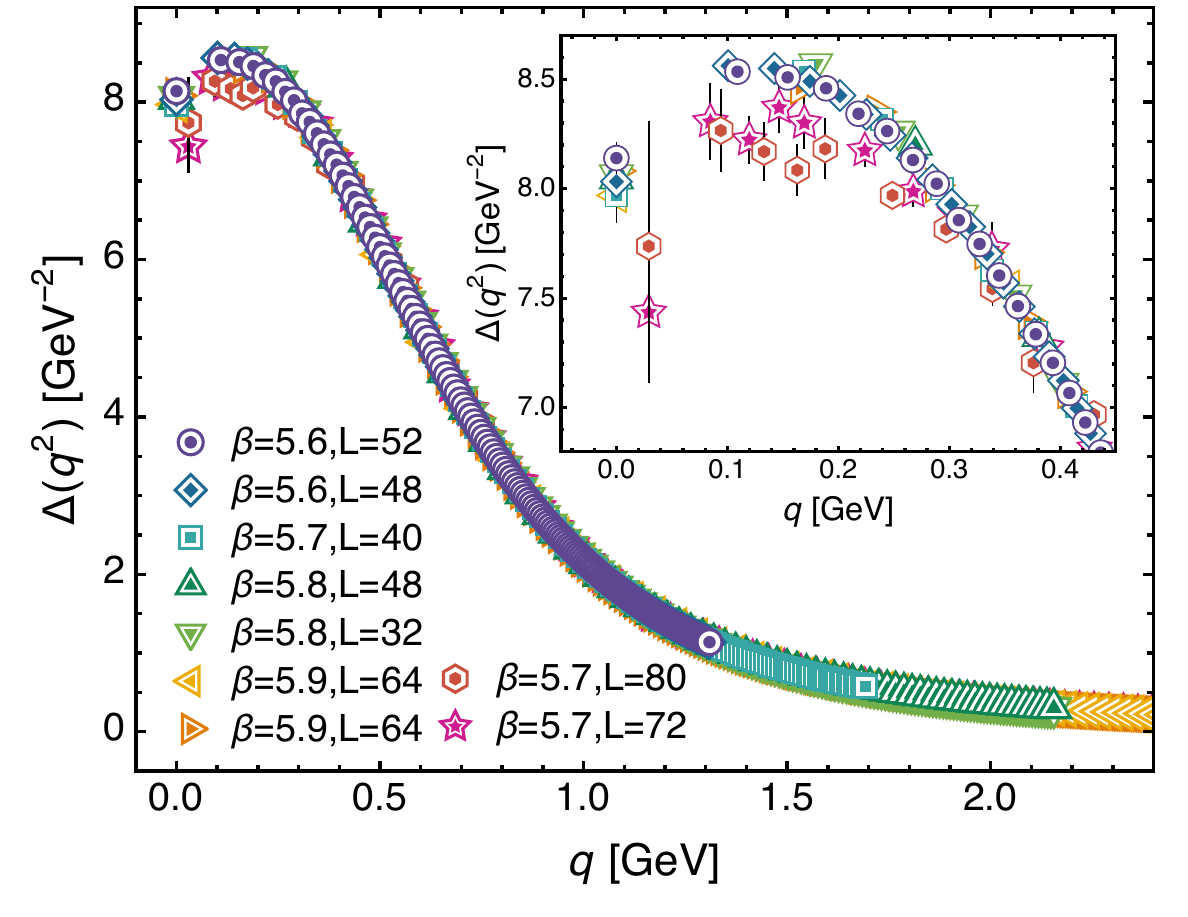}
	\includegraphics[scale=0.345]{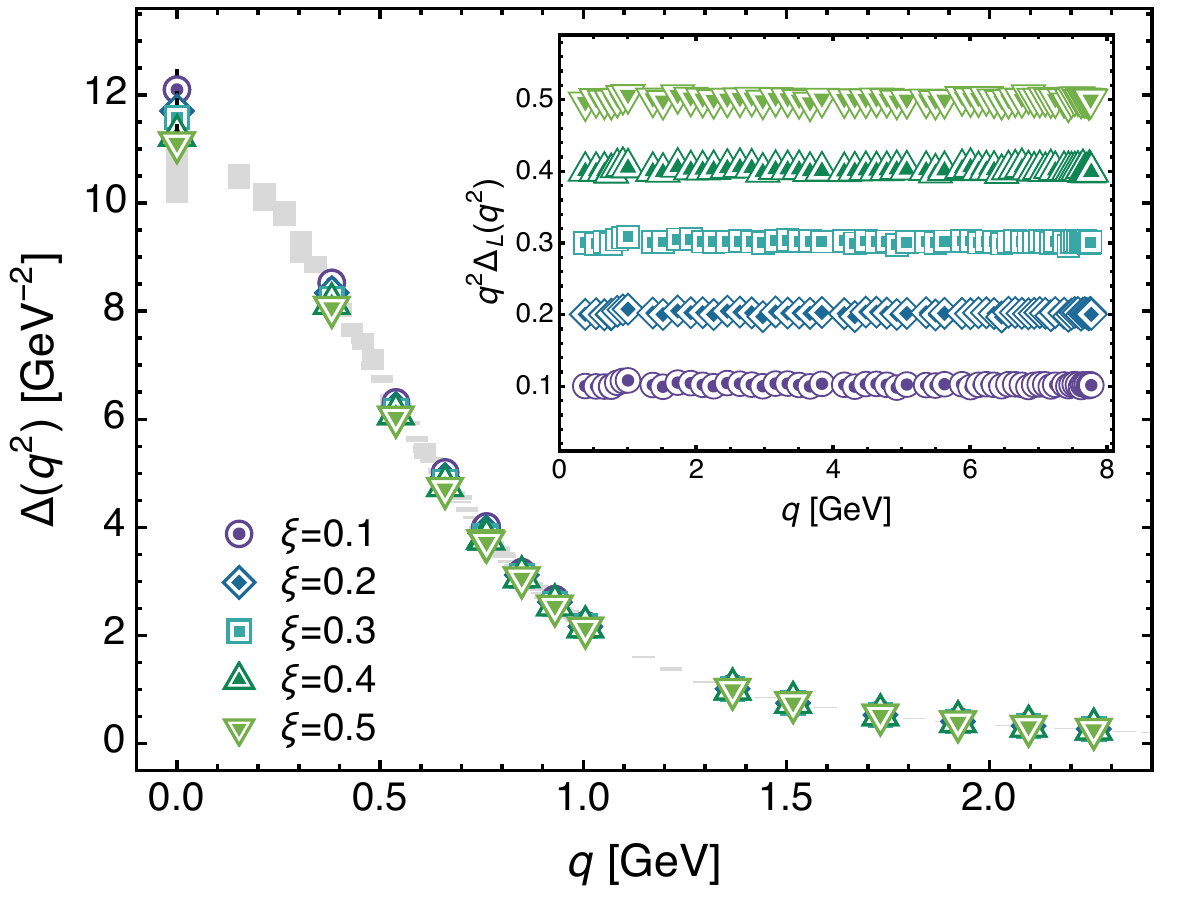}
	\caption{\label{fig:gluon} ({\it Left panel}) Compilation of SU$(3)$ quenched lattice results for the Landau gauge gluon propagator, corrected for discretization artifacts~\cite{Aguilar:2021okw}. The inset highlights the IR region, where the propagator saturates to a finite, nonzero value. The maximum observed in the deep IR arises from the dynamical interplay between gluons and ghosts. ({\it Right panel}) SU$(3)$ quenched lattice results for the gluon propagator for different values of the gauge fixing parameter $\xi$ at $\beta=6.0$ and $L=32$. The gray rectangles provide an estimate of the expected volume effects. The inset shows the behavior of the longitudinal form factor $q^2\Delta_L(q^2)$ which evidently shows no deviation with respect to the expected result $q^2\Delta_L(q^2)=\xi$, see~\1eq{GammaDelta}. All results are renormalized at $\mu = 4.3\,\mathrm{GeV}$.}
\end{figure}

\section{\label{four}Lattice results}

The IR sector of non-Abelian gauge theories has been extensively investigated using lattice simulations, providing direct nonperturbative access to the behavior of (Euclidean) Green functions at very low momenta~\cite{SumanSchilling:1996,Cucchieri:2006tf,Cucchieri:2007md,Cucchieri:2007rg,Bowman:2007du,Bogolubsky:2007ud,Cucchieri:2008qm,Boucaud:2008ky,Bogolubsky:2009dc,Oliveira:2009eh,Cucchieri:2009zt,Cucchieri:2010xr,Ayala:2012pb,Binosi:2016xxu,Bicudo:2015rma,Athenodorou:2016oyh,Boucaud:2017obn,Duarte:2016ieu,Aguilar:2021lke}.  

In the case of the gluon propagator, lattice simulations in three and four dimensions, performed on very large physical volumes and for several lattice spacings, show that the scalar form factor $\Delta(q^2)$ approaches a finite nonvanishing value in the limit $q^{2} \to 0$, see Fig.~\ref{fig:gluon} (left panel).
	The IR finiteness of the gluon propagator has been observed both in quenched simulations and in the presence of dynamical quarks, where it has been found that unquenching effects modify the propagator at intermediate momenta but do not alter its qualitative deep IR behavior. In addition, finite volume and discretization effects have been carefully analyzed and shown not to account for the observed saturation. In $R_{\xi}$ gauges lattice implementations indicate that the transverse gluon propagator remains IR finite for small values of the gauge fixing parameter $\xi$, see Fig.~\ref{fig:gluon}, right. Its zero momentum limit varies smoothly with $\xi$ and continuously connects to the Landau gauge result as $\xi \to 0$; additionally, the longitudinal component follows the tree level behavior imposed by the gauge condition and does not display nontrivial IR dynamics, see the inset on the right panel of the same figure.

\begin{figure}
\centering
	\includegraphics[scale=0.337]{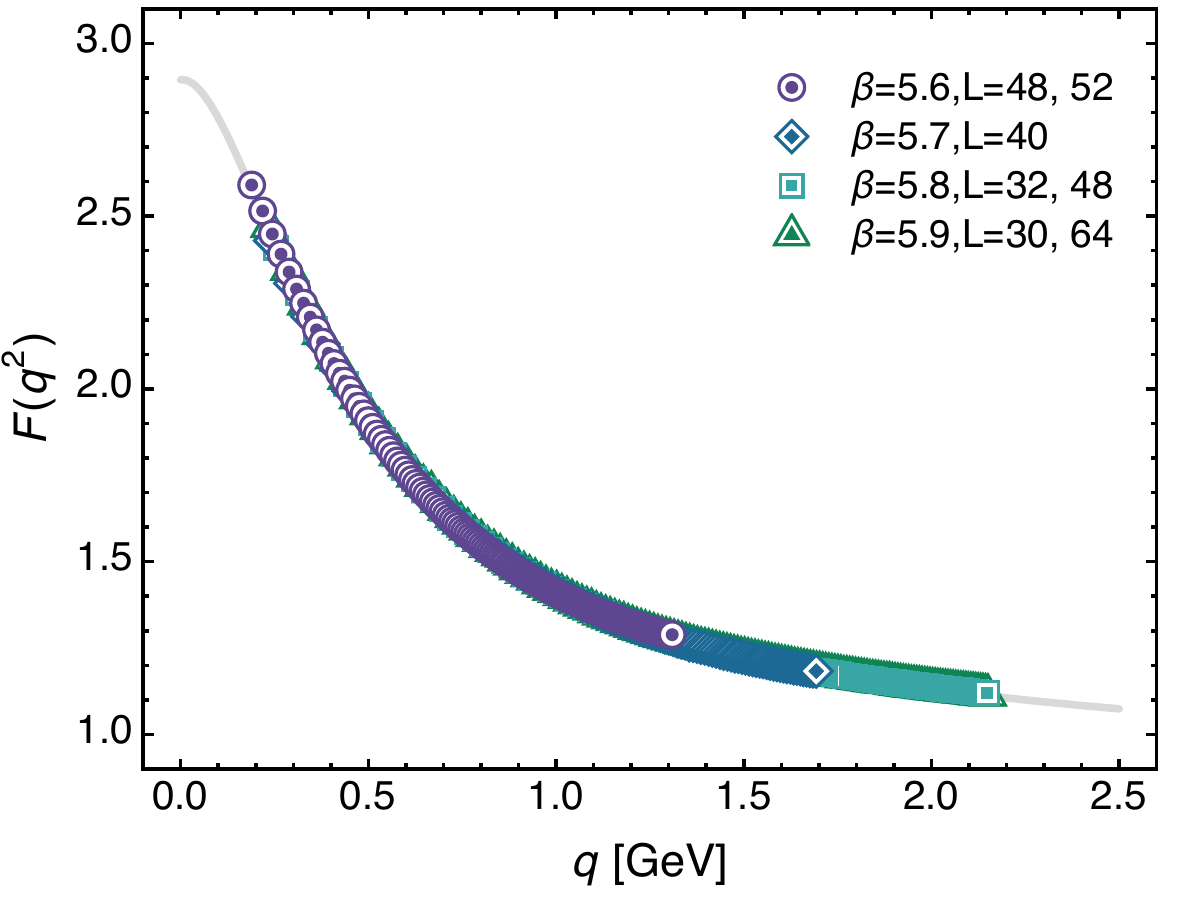}
	\includegraphics[scale=0.337]{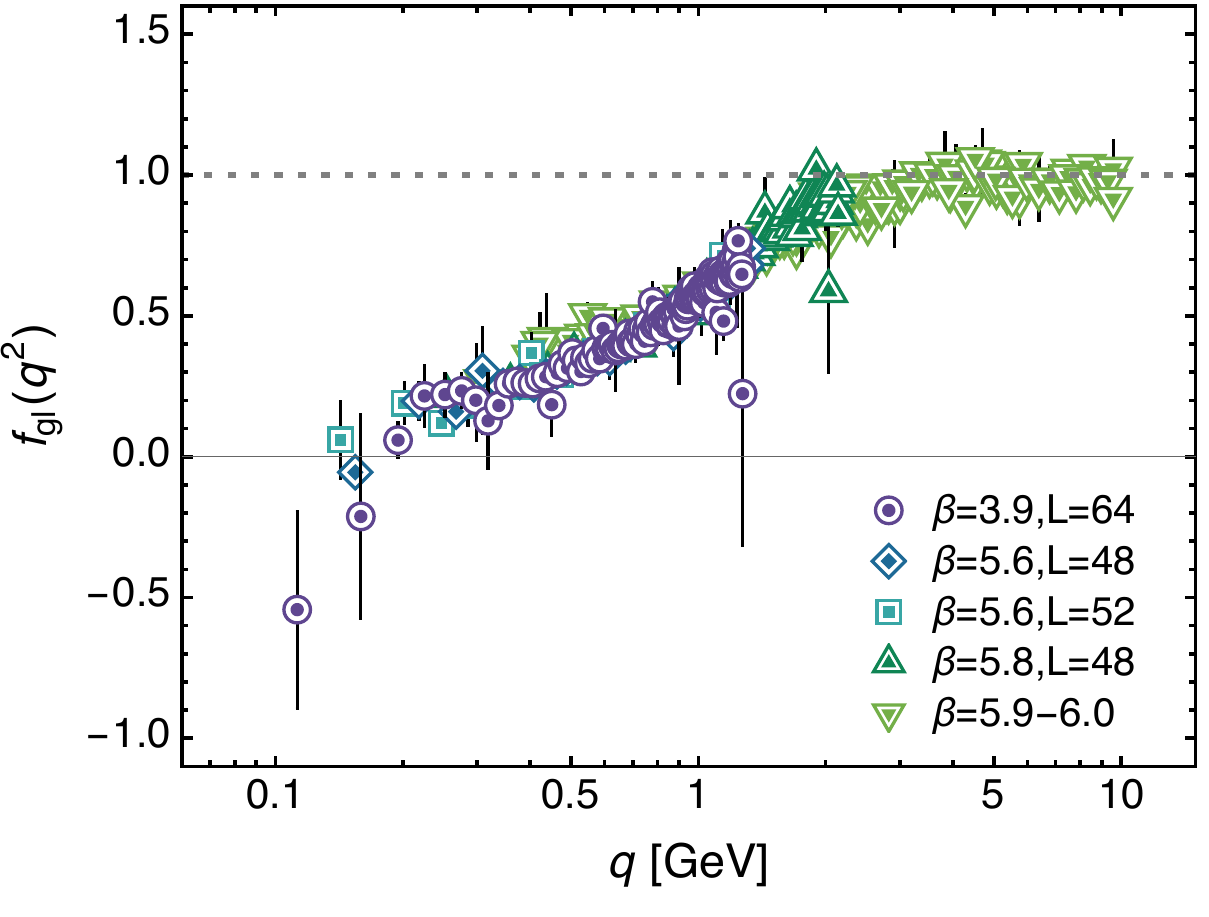}
	\caption{\label{fig:ghost}. ({\it Left panel})
	 $SU(3)$ lattice results for the ghost dressing function $F$~\cite{Boucaud:2018xup}. To guide the eyes we plot in gray the solution of the corresponding ghost DSE within the truncation scheme defined in~\cite{Aguilar:2021okw}. ({\it Right panel}) $SU(3)$ lattice results for the form factor $f_\mathrm{gl}$ multiplying the 3-gluon vertex tree-level tensor structure in the symmetric configuration. Notice the three signatures of an IR massless free ghost: ({\it i}) the suppression at intermediate momenta with respect to the tree-level value $f_\mathrm{gl}^{(0)}=1$; ({\it ii}) the zero-crossing; and ({\it iii}) the logarithmic divergence in the IR.}
\end{figure}

As far as the ghost propagator in Landau gauge is concerned, lattice results in both three and four dimensions show that it is the dressing function $F(q^{2})$, see~\1eq{ghdressF}, that approaches a finite constant in the deep IR, see the left panel of Fig.~\ref{fig:ghost}. This IR finiteness has been confirmed with increasing lattice volumes and improved gauge fixing algorithms, indicating that it is not an artifact of finite size effects or poor sampling of Gribov copies. Lattice studies of the ghost propagator in $R_{\xi}$ gauges are technically demanding and currently limited to small values of $\xi$, with the only available result indicating that the ghost dressing function remains IR finite and shows only a mild dependence on the gauge parameter~\cite{Cucchieri:2018doy}. The fact that the ghost behaves as a free massless particle in the deep IR, $D(q^2)\sim F(0)/q^2$, has profound and unexpected consequences: it implies that the gluon propagator must display a maximum in the deep IR (Fig.~\ref{fig:gluon}, left panel); and that some of the 14 form factors~\cite{Ball:1980ax} of the 3-gluon vertex, such as the one multiplying the tree-level structure, $f_\mathrm{gl}(q, k_1)\Gamma^{(0)}_{\mu\nu\rho}(q,k_1)$ in characteristic kinematic configurations ({\it e.g.}, the symmetric configuration, $k_1^2=k_2^2=q^2$) are suppressed with respect to its tree-level value ($f_\mathrm{gl}^{(0)}=1$), reverse its sign for relatively small momenta (an effect known as ``zero crossing''), and finally diverges logarithmically at the origin~\cite{Alkofer:2008dt,Tissier:2011ey,Pelaez:2013cpa,Aguilar:2013vaa,Blum:2014gna,Eichmann:2014xya,Williams:2015cvx,Cyrol:2016tym} (Fig.~\ref{fig:ghost}, right panel).     

\begin{figure}
\centering
	\includegraphics[scale=0.33]{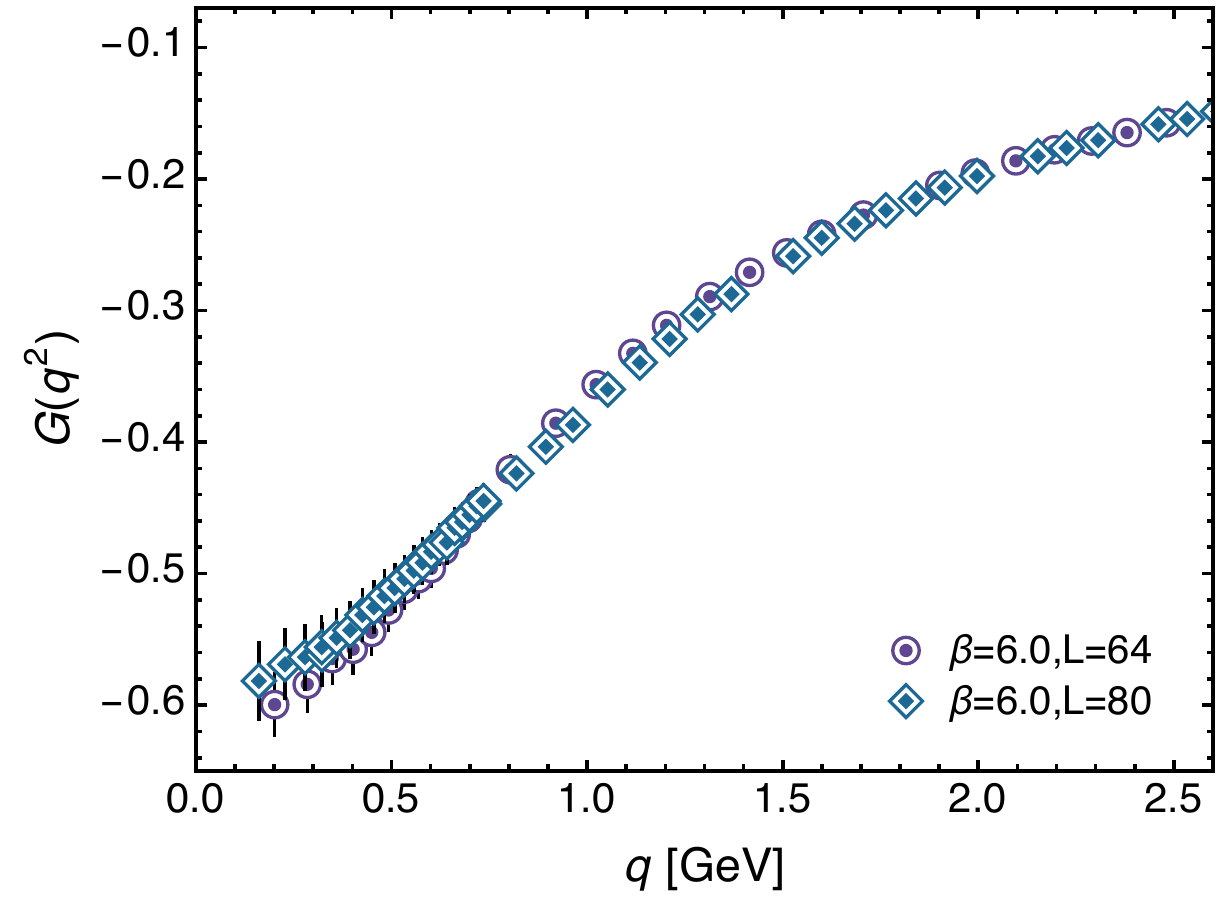}
	\includegraphics[scale=0.33]{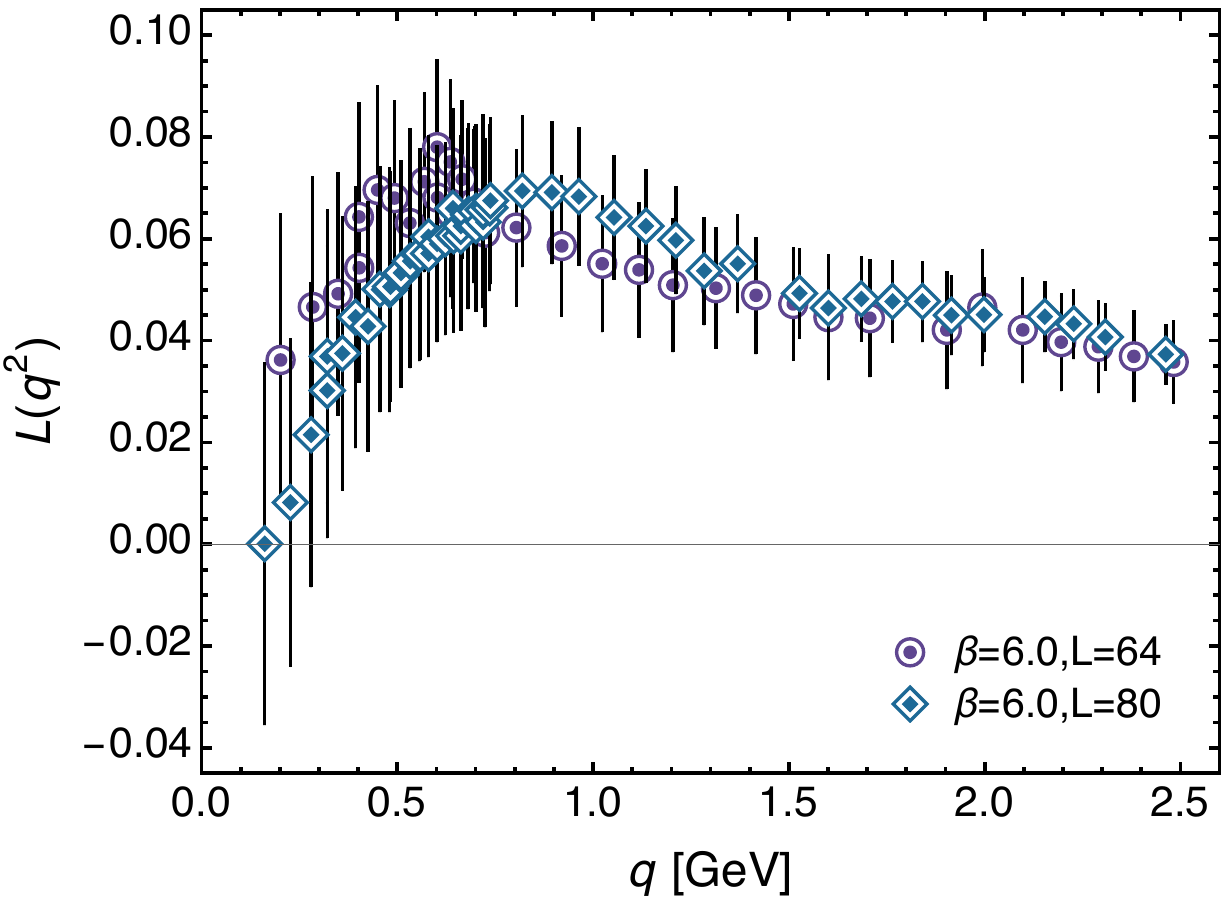}
	\caption{\label{fig:GandL}. ({\it Left panel})
	 $SU(3)$ lattice results for the functions $G(q^2)$ (left) and  $L(q^2)$ (right) in the Landau gauge~\cite{Aguilar:2024bwp}. According to~\1eq{F1GL}, the fact that $L(0)=0$ and $G(0)\neq-1$ signals a finite ghost dressing function $F(0)<\infty$.}
\end{figure}

Finally, lattice determinations of the Kugo--Ojima function $u(q^{2})=G(q^2)$ in the Landau gauge shows how the function becomes negative and approaches a finite value in the IR, but does not reach the value $G(0) = -1$ required by the Kugo--Ojima criterion, see the right panel of Fig.~\ref{fig:GandL}. The deviation from $G(0) = -1$, together with the fact that $L(0)=0$, see Fig.~\ref{fig:GandL} right panel, is consistent with the IR finiteness of the ghost dressing function, as implied by~\1eq{IRG}, and the saturation of the gluon propagator. It also tells us  a final piece of information, {\it i.e.,} the RGI combination $\mbox{}\hspace{0.13cm}\widehat{\hspace{-0.13cm}\widehat{d}}\mbox{\ }(q^2)$  of~\1eq{dhathat} is defined for all $q^2$ from the UV to the IR including the crucial point $q^2=0$.

\section{\label{five}Dynamical gluon mass generation}
\subsection{Masslessness of the gluon propagator}

The combined lattice evidence therefore supports a coherent IR picture characterized by a finite gluon propagator, a finite ghost dressing function, and a Kugo--Ojima function that falls short of the confinement condition $G(0) = -1 $. This picture is robust under changes of dimension, lattice spacing, and inclusion of dynamical quarks, and extends smoothly from Landau gauge to small values of the gauge parameter in linear covariant gauges. 

Taken at face value, these results are telling us that IR interactions generate dynamically a mass for the gluon. But how can this practically happen? We know that gauge symmetry forbids the addition of an explicit tree-level mass term $mA^{a\mu}A_\mu^a$ in the Yang-Mills action~\1eq{S_YM}; but does this still hold at the level of the DSE satisfied by the gluon 2-point function? 

This question can most-easily be answered by analyzing the gluon's mixed $Q\widehat{A}$ self-energy. According to the DSE~\1eq{PhiPhiDSE} and the BF gauge Feynman rules there are 3 possible contributions: the gluon and ghost 1-loop dressed classes  obtained when considering the first line in~\1eq{PhiPhiDSE} and replacing internal particles with gluons and ghosts respectively; the 2-loop dressed gluon class obtained when replacing all internal lines with gluons in the second line in~\1eq{PhiPhiDSE}.

\begin{detailedcalc}
	Obviously there is no 2-loop dressed ghost contribution in this case, as there is no tree-level coupling~$QQc\bar{c}$. 
\end{detailedcalc}    

Let us concentrate on the one loop dressed diagrams (recall that owing to the block-wise realization of the WI we can do this gauge invariantly). Then, for the gluon diagrams one has:

\noindent\makebox[\textwidth][c]{%
\includegraphics[scale=0.6]{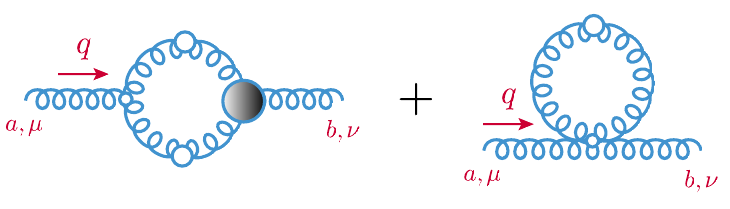}%
}
\begin{align}
	\widehat{\Pi}^{1\ell;\mathrm{gl}}_{\mu\nu}(q;\mu)&=\frac12g^2C_A\int_k\!\Gamma^{(0)}_{\mu\alpha\beta}(q,k)\Delta^{\alpha\rho}(k)\Delta^{\beta\sigma}(k+q)\widehat{\Gamma}_{\nu\sigma\rho}(q,-k-q)\nonumber \\
	&\mathrel{\phantom{=}}-ig^2C_A\int_k\left[g_{\mu\nu}\Delta^\rho_\rho(k)-\Delta_{\mu\nu}(k)\right], 
\end{align}
whereas ghost diagrams provide the contribution: 

\noindent\makebox[\textwidth][c]{%
\includegraphics[scale=0.6]{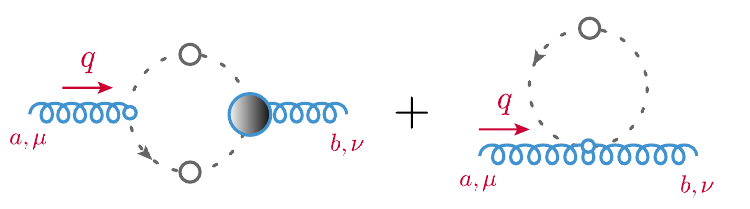}%
}
\begin{align}
	\widehat{\Pi}^{1\ell;\mathrm{gh}}_{\mu\nu}(q;\mu)&=-g^2C_A\int_k\!\Gamma^{(0)}_{\mu}(q,k-q)D(k-q)D(k)\widehat\Gamma^{(0)}_\nu(-q,k)\nonumber \\
	&\mathrel{\phantom{=}}-ig^2C_Ag_{\mu\nu}\int_k\!D(k).
\end{align}

A dynamical generation of a gluon mass would be signaled by obtaining a nonzero value for the self-energy as $q^2\to0$, a fact that can be checked by setting $q=0$ directly in the equations above. Then, since only the $g_{\mu\nu}$ term can appear, each contribution to the $Q\widehat{A}$ self-energy reads
\begin{align}
	\widehat{\Pi}^{i;j}_{\mu\nu}(0) &= \widehat{\Pi}^{i;j}(0) g_{\mu\nu};&
	d\widehat{\Pi}^{i;j}(0) = g^{\mu\nu}\widehat{\Pi}^{i;j}_{\mu\nu}(0).
\label{PiQA(0)}
\end{align}
 
\begin{detailedcalc}
	While this fact may appear obvious, it needs to be further discussed. Block-wise transversality of $\widehat{\Pi}^{i,j}_{\mu\nu}(q)$ implies that the coefficients of the $g_{\mu\nu}$ and $q_\mu q_\nu/q^2$ terms are equal in magnitude and opposite in sign. Therefore, $\widehat{\Pi}^{i,j}_{\mu\nu}(0)$ may in principle be obtained from the IR behavior of either tensor structure as $q^2 \to 0$. In practice, however, the two procedures are not equivalent. While the extraction of the $q_\mu q_\nu/q^2$ component is straightforward, the treatment of the $g_{\mu\nu}$ term is considerably more involved.
	This inequivalence is illustrated by the integral~\cite{Binosi:2012sj}
	\begin{align}
		I_{\mu\nu}(q) = \int_k k_\mu k_\nu f(k,q),
	\label{Imunu}
	\end{align}
	where $f(k,q)$ remains finite as $q \to 0$.
	By Lorentz covariance,
	\begin{align}
		I_{\mu\nu}(q) = g_{\mu\nu} A(q^2) + \frac{q_\mu q_\nu}{q^2} B(q^2),
	\end{align}
	with
	\begin{align}
	A(q^2) &= \frac{1}{d-1} \int_k \left[ k^2 - \frac{(k{\cdot}q)^2}{q^2} \right] f(k,q);&
	B(q^2) &= -\frac{1}{d-1} \int_k \left[ k^2 - d \frac{(k{\cdot}q)^2}{q^2} \right] f(k,q).
	\label{justI}
	\end{align}
	Using $(k{\cdot}q)^2 = q^2 k^2 \cos^2\theta$, and
	\begin{align}
		\int_k \cos^2\theta \, f(k^2) = \frac{1}{d} \int_k f(k^2),
	\label{costheta-rel}
	\end{align}
	one finds, in the limit $q \to 0$,
	\begin{align}
		A(0) &= \frac{1}{d} \int_k k^2 f(k^2);&
		B(0) = 0.
	\label{I0}
	\end{align}
	Thus while $B(0)$ vanishes identically, depending on the nature of $f(k^2)$, $A(0)$ may be divergent.

	A simpler way to reach the same conclusion would be indeed to set $q=0$ directly in Eq.~\eqref{Imunu},
	\begin{align}
		I_{\mu\nu}(0) = \int_k k_\mu k_\nu f(k^2),
	\label{Imunu0}
	\end{align}	
	which is necessarily proportional to $g_{\mu\nu}$.
	The absence of any $q_\mu q_\nu/q^2$ structure immediately implies $B(0)=0$, whereas $A(0)$ follows from the Lorentz trace, reproducing \1eq{I0} and vindicating~\1eq{PiQA(0)} above. 
\end{detailedcalc}

The key technical ingredient required for the calculation of~\1eq{PiQA(0)} is the Taylor expansion of a generic {\it regular} function $f(q,k_1,k_2)$ around $q=0$ (with $k_2=-k_1$), namely
\begin{align}
f(q,k_1,-k_1)=f(0,k_1,-k_1)+q^\mu\left.\frac{\partial}{\partial q^\mu}
f(q,k_1,k_2)
\right|_{q=0}
+
\mathcal{O}(q^2),
\label{Taylf}
\end{align}
where any Lorentz or color structure carried by $f$ has been suppressed for simplicity. After differentiating $f(q,k_1,-k_1-q)$ with respect to $q^\mu$ and setting $q=0$, the term in brackets on the right-hand side becomes a function of $k_1$ only. 

Let us then start by rewriting the WIs~\1eq{hatWIs} factoring out the color factors:
\begin{subequations}
\begin{align}
	q^\mu \widehat{\Gamma}_{\mu\nu\rho}(q,k_1) &= i\Delta_{\nu\rho}^{-1}(k_1) - i\Delta_{\nu\rho}^{-1}(k_2), \\
	q^\mu \widehat{\Gamma}_\mu(q,k_2) &= iD^{-1}(k_1^2) - iD^{-1}(k_2^2).
\end{align}
\label{BQids}
\end{subequations}
If we now {\emph{assume}} that the vertices of the theory are regular in $q=0$, then we can Taylor expand both sides of~\1eq{BQids}, and, through equating the coefficients of the linear terms in $q$, obtain the results 
\begin{subequations}
\begin{align}
	\widehat{\Gamma}_{\mu\nu\rho}(0,k_1) &= -i \frac{\partial }{\partial k_1^\mu}\Delta^{-1}_{\nu\rho}(k_1), \\
	\widehat{\Gamma}_\mu(0,k_2) &= i\frac{\partial }{\partial k_2^\mu}D^{-1}(k_2^2).
\end{align}	
	\label{WIsin0}
\end{subequations}

\begin{detailedcalc}
	To obtain the first of~\1eq{WIsin0}, let us expand the left- and right-hand side of the first identity in~\1eq{BQids} to obtain
\begin{subequations}
 \begin{align}
 	q^\mu \widehat{\Gamma}_{\mu\nu\rho}(q,k_1)&\xrightarrow[q\to0]{}q^\mu \widehat{\Gamma}_{\mu\nu\rho}(0,k_1),\\
 	i\Delta_{\nu\rho}^{-1}(k_1) - i\Delta_{\nu\rho}^{-1}(k_2)&\xrightarrow[q\to0]{}-iq^\mu\left.\frac{\partial}{\partial q^\mu}\Delta^{-1}(k_1+q)\right\vert_{q=0}+{\cal O}(q^2)\equiv iq^\mu\left.\frac{\partial}{\partial q^\mu}\Delta^{-1}(k_2+q)\right\vert_{q=0}+{\cal O}(q^2),
 \end{align} 	
\end{subequations}
so that we obtain
\begin{align}
\widehat{\Gamma}_{\mu\nu\rho}(0,-k_2) &= i \frac{\partial }{\partial k_2^\mu}\Delta^{-1}_{\nu\rho}(k_2);& \widehat{\Gamma}_{\mu\nu\rho}(0,k_1) &= -i \frac{\partial }{\partial k_1^\mu}\Delta^{-1}_{\nu\rho}(k_1),
\label{BQ2-id}	
\end{align}
and, similarly,
\begin{align}
	\widehat{\Gamma}_\mu(0,k_2) &= i\frac{\partial }{\partial k_2^\mu}D^{-1}(k_2^2);&
	\widehat{\Gamma}_\mu(0,-k_1) &= -i\frac{\partial }{\partial k_1^\mu}D^{-1}(k_1^2).
\label{Bc2-id}
\end{align}
 
\end{detailedcalc}

We can now start evaluating the contribution at $q=0$ of the various class of diagrams contributing to the 2-point function. Let's start with the one-loop dressed gluon diagrams; we have
\begin{align}
d\widehat{\Pi}^{1\ell;\mathrm{gl}}(0) &=
\frac{1}{2} g^2 C_A
\int_k
\Gamma^{(0)}_{\mu\alpha\beta}(0,k)
\Delta^{\alpha\rho}(k)\Delta^{\beta\sigma}(k)
\widehat{\Gamma}^\mu_{\sigma\rho}(0,-k)\nonumber \\
&\mathrel{\phantom{=}}- i g^2 C_A (d-1) \int_k \Delta^\alpha_{\ \alpha}(k).
\label{1l;gl@0}
\end{align}
Next, using the WI~\eqref{BQ2-id} 
we obtain
\begin{align}
\Delta^{\alpha\rho}(k)\Delta^{\beta\sigma}(k)
\widehat{\Gamma}^\mu_{\sigma\rho}(0,-k)
=- i \frac{\partial}{\partial k_\mu} \Delta^{\alpha\beta}(k),
\label{2propsvertex}
\end{align}
and integrating by parts,~\1eq{1l;gl@0} becomes
\begin{align}
d\widehat{\Pi}^{1\ell;\mathrm{gl}}(0) &=
-\frac{i}{2} g^2 C_A
\left[
\int_k \frac{\partial}{\partial k^\mu}
\big(
\Gamma^{(0)}_{\mu\alpha\beta}(0,k)
\Delta^{\alpha\beta}(k)
\big)
-
\int_k
\Delta^{\alpha\beta}(k)
\frac{\partial}{\partial k^\mu}
\Gamma^{(0)}_{\mu\alpha\beta}(0,k)
\right]\nonumber \\
&\mathrel{\phantom{=}}- i g^2 C_A (d-1) \int_k \Delta^\alpha_{\ \alpha}(k).
\label{1l;gl@0_1}
\end{align}
Then, since
\begin{align}
\frac{\partial}{\partial k^\mu}
\Gamma^{(0)}_{\mu\alpha\beta}(0,k)
=
2 (d-1) g_{\alpha\beta},
\end{align}
the second term in the first line of~\1eq{1l;gl@0_1} cancels exactly the tadpole contribution; and, as a result, one obtains
\begin{align}
d\widehat{\Pi}^{1\ell;\mathrm{gl}}(0)
&=
- g^2 C_A (d-1)
\int_k \frac{\partial}{\partial k_\mu}
{\cal F}^{1\ell;\mathrm{gl}}_\mu(k);&
{\cal F}^{1\ell;\mathrm{gl}}_\mu(k) &= k_\mu \Delta(k^2).
\label{F1QCD}
\end{align}  

The analysis of the 1-loop dressed ghost contributions proceeds in a similar fashion. At $q=0$ one has 
\begin{align}
d\widehat{\Pi}^{1\ell;\mathrm{gh}}(0) &=
g^2 C_A \int_k k_\mu D^2(k^2)\widehat{\Gamma}^\mu(0,k)
- i g^2 C_A d \int_k D(k^2).
\label{1l;gh@0}
\end{align}
Using the identity~\eqref{Bc2-id},
\begin{align}
D^2(k^2)\widehat{\Gamma}^\mu(0,k)
=
- i \frac{\partial D(k^2)}{\partial k^\mu},
\label{derD}
\end{align}
and integrating by parts, \1eq{1l;gh@0} becomes
\begin{align}
d\widehat{\Pi}^{1\ell;\mathrm{gh}}(0) &=- i g^2 C_A
\left[
\int_k \frac{\partial}{\partial k^\mu}
\big( k_\mu D(k^2) \big)
-
d \int_k D(k^2)
\right]- i g^2 C_A d \int_k D(k^2).
\label{1l;gh@0_1}
\end{align}
As before, the tadpole contribution cancels out, leaving
\begin{align}
d\widehat{\Pi}^{1\ell;\mathrm{gh}}(0) &=
- i g^2 C_A
\int_k \frac{\partial}{\partial k_\mu}
{\cal F}^{1\ell;\mathrm{gh}}_\mu(k),
\qquad
{\cal F}^{1\ell;\mathrm{gh}}_\mu(k) = k_\mu D(k^2).
\label{F2QCD}
\end{align}

Putting all together we get the final result~\cite{Aguilar:2016ock}
\begin{align}
	d\widehat\Pi(0)&=-g^2C_A\int_k\frac\partial{\partial k_\mu}{\cal F}_\mu(k),
\end{align} 
with 
\begin{align}
	{\cal F}_\mu(k)&=(d-1){\cal F}^{1\ell;\mathrm{gl}}_\mu(k)+i{\cal F}_\mu^{1\ell;\mathrm{gh}}=k_\mu\left[(d-1)\Delta(k^2)+iD(k^2)
	\right].
\end{align}

\begin{detailedcalc}
	The calculation of the 2-loop dressed gluon diagrams is doable but cumbersome. The contribution of this class of diagrams to the gluon self-energy reads 
\begin{align}
	\delta^{ab}\widehat{\Pi}^{2\ell;\mathrm{gl}}_{\mu\nu}(q;\mu)&=i\frac{g^4}6\Gamma^{(0)amnr}_{\mu\alpha\beta\gamma}\hspace{-0.15cm}\int_{k,\ell}\hspace{-0.2cm}\Delta^{\alpha\rho}(k+\ell)\Delta^{\beta\sigma}(\ell)\Delta^{\gamma\tau}(k+q)\widehat{\Gamma}_{\nu\tau\sigma\rho}^{brnm}(-q,k+q,\ell)\nonumber \\
	&\mathrel{\phantom{=}}+i\frac{g^4}2f^{bre}f^{emn}\Gamma^{(0)amnr}_{\mu\alpha\beta\gamma}\hspace{-0.11cm}\int_k\!Y^{\alpha\beta}_{\delta}(k)\Delta^{\delta\lambda}(k)\Delta^{\gamma\tau}(k+q)\widetilde{\Gamma}_{\nu\tau\lambda}(-q,k+q),
	\label{2ldrgl}
\end{align}
where 
\begin{align}
	Y^{\alpha\beta}_{\delta}(k)&=\hspace{-0.11cm}\int_\ell\!\Delta^{\alpha\rho}(k+\ell)\Delta^{\beta\sigma}(\ell)\Gamma_{\sigma\rho\delta}(\ell,-k-\ell,k).
	\label{defY}
\end{align}
To evaluate~\1eq{2ldrgl} at $q=0$, we start by writing the last WI in~\1eq{hatWIs} as
\begin{align}
	q^\mu \widehat{\Gamma}^{abcd}_{\mu\nu\rho\sigma}(q,k_1,k_2) &= f^{ade}f^{ecb} \Gamma_{\nu\rho\sigma}(k_1,k_2) + f^{abe}f^{edc}\Gamma_{\rho\sigma\nu}(k_2,k_3) \nonumber \\
	&+ f^{ace}f^{ebd} \Gamma_{\sigma\nu\rho}(k_3,k_1).
	\label{BQids-2l}
\end{align}

Next, to obtain the expansion of this identity at $q=0$ we consider $k_1$ and $k_2$ to be the independent momenta; then the first term in~\1eq{BQids-2l} can be directly evaluated at $q=0$, whereas for the remaining two terms we get
\begin{subequations}
	\begin{align}
		\Gamma_{\rho\sigma\nu}(k_2,k_3)&=\Gamma_{\rho\sigma\nu}(k_2,-k_1-k_2)+q^\mu\left.\frac\partial{\partial q^\mu}\Gamma_{\rho\sigma\nu}(k_2,-q-k_1-k_2)\right\vert_{q=0}+{\cal O}(q^2),\\
		\Gamma_{\sigma\nu\rho}(k_3,k_1)&=\Gamma_{\sigma\nu\rho}(-k_1-k_2,k_1)+q^\mu\left.\frac\partial{\partial q^\mu}\Gamma_{\sigma\nu\rho}(-q-k_1-k_2,k_1)\right\vert_{q=0}+{\cal O}(q^2).
	\end{align}
\end{subequations}
\end{detailedcalc}

\begin{detailedcalc}
As the three gluon vertex is Bose symmetric the zeroth order terms are all equal; therefore their total contribution vanishes owing to the triggering of the Jacobi identity; for the liner terms in $q$ one has instead 
\begin{subequations}
	\begin{align}
		\left.\frac\partial{\partial q^\mu}\Gamma_{\rho\sigma\nu}(k_2,-q-k_1-k_2)\right\vert_{q=0}&=\frac\partial{\partial k_1^\mu}\Gamma_{\rho\sigma\nu}(k_2,-k_1-k_2)=\frac\partial{\partial k_1^\mu}\Gamma_{\nu\rho\sigma}(k_1,k_2),\\
		\left.\frac\partial{\partial q^\mu}\Gamma_{\sigma\nu\rho}(-q-k_1-k_2,k_1)\right\vert_{q=0}&=\frac\partial{\partial k_2^\mu}\Gamma_{\sigma\nu\rho}(-k_1-k_2,k_1)=\frac\partial{\partial k_2^\mu}\Gamma_{\nu\rho\sigma}(k_1,k_2),
	\end{align}
	\label{BQ3-id}
\end{subequations} 
which gives the final result
\begin{align}
	\widetilde{\Gamma}^{abcd}_{\mu\nu\rho\sigma}(0,k_1,k_2) &= \left(f^{abe}f^{edc}\frac{\partial}{\partial k_1^\mu} + f^{ace}f^{ebd}\frac{\partial}{\partial k_2^\mu}\right)\Gamma_{\nu\rho\sigma}(k_1,k_2).
	\label{4gWI}
\end{align}	
\end{detailedcalc}
\begin{detailedcalc}
Making use of this result, at vanishing external momentum the 2-loop dressed gluon contribution reads
\begin{align}
d\delta^{ab}\widehat{\Pi}^{2\ell;\mathrm{gl}}(0) &=
-\frac{1}{6} g^4 \Gamma^{(0)amnr}_{\mu\alpha\beta\gamma}
\int_k \int_\ell
\Delta^{\alpha\rho}(k+\ell)
\Delta^{\beta\sigma}(\ell)
\Delta^{\gamma\tau}(k)
\widehat{\Gamma}^{brnm}_{\mu\tau\sigma\rho}(0,-k,-\ell)\nonumber \\
&\mathrel{\phantom{=}}
- i \delta^{ab}{\cal N}_{\mu\alpha\beta\gamma}
\int_k
Y^{\alpha\beta}_\delta(k)
\Delta^{\gamma\tau}(k)
\Delta^{\delta\lambda}(k)
\widehat{\Gamma}^\mu_{\tau\lambda}(0,-k,k),
\label{2l;gl@0}
\end{align}
with
\begin{align}
{\cal N}_{\mu\alpha\beta\gamma}
=
\frac{3}{4} g^4 C_A^2
\left( g_{\mu\alpha} g_{\beta\gamma}
- g_{\mu\beta} g_{\alpha\gamma} \right).
\label{Npre}
\end{align}
The Bose symmetry of $\Gamma_{\sigma\rho\delta}$ implies that $Y^{\alpha\beta}_\delta(k)$ assumes the form~\cite{Binosi:2012sj}
\begin{align}
Y^{\alpha\beta}_\delta(k)
&=
\left( k^\alpha g^\beta_\delta - k^\beta g^\alpha_\delta \right) Y(k^2);&
Y(k^2)&=\frac1{d-1}\frac{1}{k^2}\,k_\alpha g_\beta^\delta Y_{\delta}^{\alpha\beta}(k),
\label{Yint2}
\end{align}
so that, using the identity~\eqref{2propsvertex}, the second term in the expression above becomes
\begin{align}
- i {\cal N}_{\mu\alpha\beta\gamma}
\int_k
Y^{\alpha\beta}_\delta(k)
\Delta^{\gamma\tau}(k)
\Delta^{\delta\lambda}(k)
\widehat{\Gamma}^\mu_{\tau\lambda}(0,-k,k)&=- {\cal N}_{\mu\alpha\beta\gamma}
\int_k
Y^{\alpha\beta}_\delta(k)
\frac{\partial \Delta^{\gamma\delta}(k)}{\partial k^\mu}.
\label{a6zero}
\end{align}

Although this contribution will cancel completely against part of the first term in~\1eq{2l;gl@0}, it is instructive to note that, in the Landau gauge, it reads
\begin{align}
\frac{3}{2} i (d-1) g^4 C_A^2
\int_k
\left[ \Delta(k^2) + 2 k^2 \Delta'(k^2) \right] Y(k^2),
\end{align}
which vanishes perturbatively but yields a quadratically divergent contribution nonperturbatively.

Turning to the first term in~\1eq{2l;gl@0}, employing the WI~\1eq{4gWI}, and making use of the identities
\begin{subequations}
\begin{align}
 \label{ffG}
f^{brx}f^{xmn}\Gamma_{\mu\alpha\beta\gamma}^{(0)amnr} &= \frac{3}{2}C_A^2\delta^{ab}(g_{\mu\alpha}g_{\beta\gamma} - g_{\mu\beta}g_{\alpha\gamma});\\
f^{bnx}f^{xrm}\Gamma_{\mu\alpha\beta\gamma}^{(0)amnr} &= \frac{3}{2}C_A^2\delta^{ab}(g_{\mu\gamma}g_{\alpha\beta} - g_{\mu\alpha}g_{\beta\gamma}),
\end{align}	
\end{subequations}
we obtain
\begin{align}
&-\frac{1}{6} g^4 \Gamma^{(0)amnr}_{\mu\alpha\beta\gamma}
\int_k \int_\ell
\Delta^{\alpha\rho}(k+\ell)
\Delta^{\beta\sigma}(\ell)
\Delta^{\gamma\tau}(k)
\widehat{\Gamma}^{brnm}_{\mu\tau\sigma\rho}(0,-k,-\ell)\nonumber \\
&=- {\cal N}^\mu_{\alpha\beta\gamma}\delta^{ab}
\left[
\frac{2}{3}
\int_k
\frac{\partial}{\partial k^\mu}
\big( \Delta^{\gamma\delta}(k) Y^{\alpha\beta}_\delta(k) \big)
-
\int_k
Y^{\alpha\beta}_\delta(k)
\frac{\partial \Delta^{\gamma\delta}(k)}{\partial k^\mu}
\right].
\label{a5Yint}
\end{align}
(This is where the calculation gets cumbersome, as to get to this result requires, among other things, integrating by parts, carrying out appropriate shifts in the integration momenta, and repeated use of the Jacobi identity).
\end{detailedcalc}
\begin{detailedcalc}
Evidently, the second term cancels exactly the contribution of~\1eq{a6zero},
leaving us with the final result
\begin{align}
d\widehat{\Pi}^{2\ell;\mathrm{gl}}(0)
&=
i (d-1) g^4 C_A^2
\int_k
\frac{\partial}{\partial k^\mu}
{\cal F}^{2\ell;\mathrm{gl}}_\mu(k),
\qquad
{\cal F}^{2\ell;\mathrm{gl}}_\mu(k) = k_\mu \Delta(k^2) Y(k^2),
\label{F3QCD}
\end{align}
which is exactly of the same type found for the  1-loop dressed contributions.
	 
\end{detailedcalc}

Now, observe that since ${\cal F}_\mu(k)$ is an odd function of $k$, it follows immediately that
\begin{equation}
\int_k {\cal F}_\mu(k) = 0.
\label{intFvect}
\end{equation}
Moreover, within dimensional regularization (or any scheme preserving translational invariance), the integration variable may be shifted by an arbitrary momentum $q$ without affecting the result in~\1eq{intFvect}. Expanding ${\cal F}_\mu(k+q)$ around $q=0$, one has
\begin{align}
{\cal F}_\mu(k+q)
&=
{\cal F}_\mu(k)
+
q^\nu
\left.
\frac{\partial}{\partial q^\nu}
{\cal F}_\mu(k+q)
\right|_{q=0}
+
\mathcal{O}(q^2)
\nonumber\\
&=
{\cal F}_\mu(k)
+
q^\nu
\frac{\partial {\cal F}_\mu(k)}{\partial k^\nu}
+
\mathcal{O}(q^2).
\label{TaylFvectq0}
\end{align}
Integrating both sides and using Eq.~\eqref{intFvect}, the vanishing of each order in $q$ implies
\begin{align}
q^\nu
\int_k
\frac{\partial {\cal F}_\mu(k)}{\partial k^\nu}
=
0.
\label{projectq}
\end{align}
Since the integral carries two free Lorentz indices and no external scale, it can only be proportional to $g_{\mu\nu}$; and because $q^\nu$ is arbitrary, this condition is satisfied if and only if
\begin{align}
\int_k
\frac{\partial}{\partial k^\mu}
{\cal F}_\mu(k)
=
0,
\label{seagull}
\end{align}
which represents the so-called seagull identity~\cite{Aguilar:2009ke,Aguilar:2016vin}.

As $\widehat\Delta^{-1}(q^2) = q^2 + i \widehat{\Pi}(q^2)$, this identity leads to the conclusion that  
\begin{align}
	\widehat{\Delta}^{-1}(0) =0.
	\label{hatD0}
\end{align}
In order to extract from~\1eq{hatD0} the behavior of the 
conventional (quantum) propagator the additional step of employing the \noeq{intBQI}, which, at $q=0$ yields
\begin{align}
	\Delta^{-1}(0) = \frac{\widehat{\Delta}^{-1}(0)}{1+G(0)}.
\label{D1G}
\end{align} 
If we now additionally assume that $1+G(0)$ is finite for every $\xi$ (see discussion below), then we reach the final conclusion that, in the absence of poles,  
\begin{align} 
\Delta^{-1}(0) = 0.
\end{align}

\begin{detailedcalc}
	It is important to emphasize that the proof presented is valid for \emph{any} value of the gauge-fixing parameter $\xi$ within the class of linear covariant ($R_\xi$) gauges: indeed, at no stage was it necessary to specialize to a particular value of $\xi$.
	
	Closely related to this observation, recall that in an $R_\xi$ gauge one has the ghost sector constraint~\1eq{ghost-sector-constraint}. In the Landau gauge, $\xi=0$, one knows that $F(0)\neq 0$ while $L(0)=0$~\cite{Aguilar:2009pp}; consequently, the constraint reduces to $F^{-1}(0) = 1 + G(0)$ which guarantees that the quantity $1+G(0)$ is finite. For $\xi \neq 0$, the situation is more involved. Analytical studies~\cite{Aguilar:2015nqa,Huber:2015ria} indicate that in this case the ghost dressing function vanishes in the IR (something that however is not currently supported by the preliminary lattice simulations of~\cite{Cucchieri:2018doy}), suggesting a divergence in the left-hand side of~\1eq{ghost-sector-constraint}. At the same time, these studies show that the gluon propagator continues to saturate in the IR, a result that, as we have discussed, has been corroborated by lattice simulations~\cite{Bicudo:2015rma}. If this IR suppression of the ghost dressing function is confirmed, these findings point to highly nontrivial dynamics in the ghost sector, ensuring that the functions $K$ (and possibly $L$) conspire to cancel the divergence, leaving a finite value for $1+G(0)$ even when $\xi \neq 0$.
	\end{detailedcalc}
	
	\begin{detailedcalc}
	Finally, we stress that no particular assumption has been made regarding the explicit form of the gluon or ghost propagators $\Delta$ and $D$ entering the one- and two-loop dressed diagrams. Even if one were to \emph{assume} a massive form for the gluon propagator, for example, a Cornwall-type propagator~\cite{Cornwall:1981zr}, $\Delta(q^2)=1/[q^2+m^2(q^2)]$, or a Stingl form~\cite{Stingl:1985hx}, $\Delta(q^2)=s_3(1+s_1 q^2)/[(q^2+m^2)^2+s_2^2]$ (with suitable parameters $s_i$), the seagull identity,~\1eq{seagull}, would still enforce the cancellation of the individual contributions $\widehat{\Pi}^{i;j}(0)$, leading to $\Delta^{-1}(0)=0$ and thus contradicting the original assumption. The main lesson is that, when treated consistently, the theory resists the generation of a mass through a highly nontrivial cancellation mechanism. This mechanism relies crucially on the assumed regularity of the vertices as $q\to0$ required to obtain the expressions~\1eq{WIsin0}; relaxing this assumption is precisely what ultimately allows for the emergence of an IR-finite gluon propagator, as will be discussed next.
\end{detailedcalc}

\subsection{Dynamical gluon mass generation}

In order to obtain an IR-finite gluon propagator in a self-consistent manner, one must therefore allow for the appearance of massless pole terms in the relevant vertices. In particular, such poles must occur in the kinematic channel associated with the momentum $q$ flowing into the gluon DSE, since this is the momentum carried by the vertices inserted in the corresponding diagrams. According to the conventions adopted so far, this requirement implies that (some of) the vertices $\widehat{\Gamma}_{\mu\nu\rho}$, $\widehat{\Gamma}_\mu$, and $\widehat{\Gamma}^{abcd}_{\mu\nu\rho\sigma}$ must contain pole contributions of the form $q_\mu/q^2$.

\begin{detailedcalc}

Let us see why this is the case. According to Goldstone's theorem~\cite{Nambu:1960tm,Goldstone:1961eq,Goldstone:1962es}, the spontaneous breaking of a continuous global symmetry is accompanied by the appearance of massless excitations, known as Goldstone particles. Although Goldstone's theorem is most commonly realized through a scalar field acquiring a nonvanishing vacuum expectation value, Nambu and Jona-Lasinio~\cite{Nambu:1961tp,Nambu:1961fr} introduced the notion of a \emph{dynamical} Goldstone boson, demonstrating that the Goldstone mechanism may operate even in the absence of fundamental scalar fields in the Lagrangian. In this case, when chiral symmetry is broken, the associated Goldstone boson (the pion) is not an elementary excitation but rather a quark--antiquark bound state~\cite{Maris:1997hd}.

Goldstone's theorem, however, does not apply when the symmetry that is spontaneously broken is local rather than global, and the theory contains a massless gauge boson mediating the interaction (equivalently, when gauge-invariant long-range forces are present). In this situation the Goldstone phenomenon is replaced by the Higgs mechanism: no massless excitations appear in the physical spectrum, because the would-be Goldstone bosons combine with the gauge field to provide the longitudinal degree of freedom of a massive spin-one particle~\cite{Higgs:1964pj,Higgs:1964ia,Englert:1964et,Guralnik:1964eu,Higgs:1966ev}. In addition, a massive scalar particle, the Higgs boson, emerges and plays a crucial role in ensuring the renormalizability of the theory.

Both mechanisms may coexist: if a local and a global symmetry are simultaneously broken, massive gauge bosons may appear together with massless Goldstone modes. The converse situation may also occur: the breaking of a global symmetry may generate Goldstone bosons which, in the absence of elementary scalar fields, are absorbed by gauge bosons, thereby rendering them massive.

Before turning to the specific case of non-Abelian gauge theories, it is useful to recall a different mechanism capable of generating masses for gauge bosons without introducing explicit mass terms in the Lagrangian.

Independently of these considerations, and prior to the formulation of the Higgs mechanism, Schwinger argued that gauge invariance of a vector field does not necessarily imply a vanishing mass for the associated particle, provided that the vector current coupling is sufficiently strong~\cite{Schwinger:1962tn}. Schwinger's key observation was that if the vacuum polarization tensor $\Pi_{\mu\nu}(q)$ develops a pole at zero momentum transfer, then the vector boson becomes massive, even though a mass term is forbidden at the level of the fundamental Lagrangian by gauge symmetry~\cite{Schwinger:1962tp}. Indeed, since
\begin{align}
\Delta_{\mu\nu}(q)=-iP_{\mu\nu}(q)\frac1{q^2\left[1+i\Pi(q^2)\right]},
\label{JS}
\end{align}
it is evident that if $\Pi(q^2)$ possesses a pole at $q^2=0$ with positive residue $m^2$, the vector boson acquires a mass $m$, even though it is massless in the absence of interactions ($g=0$, $\Pi=0$).

There is, in fact, no fundamental principle forbidding $\Pi(q^2)$ from developing such a pole. Since bound states are expected to occur in many physical systems and to generate poles in $\Pi_{\mu\nu}$ at timelike momenta, one may envisage that sufficiently strong binding could drive the mass of a bound state to zero, thereby inducing a mass for the vector boson without violating gauge invariance.

\end{detailedcalc}
\begin{detailedcalc}

A particularly clear realization of this mechanism occurs in two-dimensional massless QED, the so-called Schwinger model. In this theory the special properties of the Dirac algebra in $d=2$ allow an exact solution. The vacuum polarization tensor $\Pi_{\mu\nu}$ indeed develops a pole at $q^2=0$, and the photon acquires a mass $m^2=e^2/\pi$ (in two dimensions the coupling $e$ has dimensions of mass). The appearance of this pole is related to the axial anomaly: the anomalous divergence of the axial current prevents the appearance of a massless Goldstone boson and instead generates a mass for the gauge field. In this way Goldstone's theorem is evaded, consistently with Coleman's theorem~\cite{Coleman:1973ci}, which excludes Goldstone bosons in $d=2$.

From a broader perspective, the Higgs mechanism may be regarded as a particular realization of the Schwinger mechanism. In the Higgs case the pole required by the Schwinger mechanism is generated through the vacuum expectation value $v$ of a scalar field, which produces tadpole contributions to $\Pi(q^2)$. The residue of the pole is then saturated by $v^2$, yielding a gauge-boson mass $m^2=2 g^2 v^2$, while the would-be Goldstone bosons decouple from the physical spectrum.

We now return to the case of non-Abelian gauge theories, where the Schwinger mechanism may operate dynamically without the presence of elementary scalar fields.

In realistic field theories the crucial challenge in implementing the Schwinger mechanism is to demonstrate the dynamical generation of the required pole. While the Higgs mechanism ensures this at the classical or semiclassical level, its realization in the absence of fundamental scalar fields is subtler but also more conceptually appealing, as it avoids postulating elementary scalars not observed in Nature.

Motivated by this idea, considerable effort in the 1970s was devoted to combining the dynamical Goldstone mechanism, {\it i.e.}, the Nambu--Jona-Lasinio mechanism with composite Goldstone bosons, with gauge theories in order to generate gauge-invariant masses for vector bosons. In this two-step scenario, the spontaneous breaking of a global (typically chiral) symmetry produces composite Goldstone bosons, which supply the pole in $\Pi_{\mu\nu}$, while gauge invariance ensures that these massless excitations decouple from physical scattering amplitudes. The explicit realization of this decoupling proceeds through a subtle mechanism, illustrated in detail in a number of models~\cite{Jackiw:1973tr,Cornwall:1973ts}.

Although much of this activity was originally motivated by alternatives to the Higgs mechanism in the electroweak sector and in grand unified model building, these ideas have had a lasting impact on our understanding of strong interactions. When applied to pure Yang--Mills theories, such as quarkless QCD, the underlying idea is the following~\cite{Eichten:1974et}. In a strongly coupled non-Abelian gauge theory, longitudinally coupled, massless bound-state excitations may be dynamically generated. Demonstrating the existence of such bound states, particularly massless ones, is a highly nontrivial dynamical problem. In practice this question is typically investigated using Bethe--Salpeter equations and related continuum or lattice approaches~\cite{Poggio:1974qs}. Strong coupling, arising from IR dynamics, is an essential ingredient in this scenario.

These excitations resemble Nambu--Goldstone bosons in that they are massless, composite, and longitudinally coupled; however, they differ crucially in origin, as they do not arise from the spontaneous breaking of any global symmetry. Their primary role is to activate the Schwinger mechanism by providing the necessary pole in the gluon self-energy, thereby generating a gauge-invariant dynamical mass for the gluons. A further essential step is to show that these Goldstone-like excitations are removed from the $S$-matrix, either through cancellations among massless poles or as a consequence of current conservation.

At the level of Green functions and the corresponding DSEs, such composite excitations manifest themselves as poles in off-shell Green functions associated with fields absent from the classical action. These poles arise dynamically as bound-state solutions of the DSEs and encode the nonperturbative physics underlying the Schwinger mechanism.
	
\end{detailedcalc}

Let us then analyze how the inclusion of such pole terms evades the seagull identity, thereby opening the possibility for a nonvanishing inverse propagator at the origin, $\Delta^{-1}(0)\neq 0$~\cite{Jackiw:1973tr,Eichten:1974et,Poggio:1974qs,Smit:1974je}.

To this end, we assume that the relevant nonperturbative vertices can be decomposed into a no-pole part (np) and a pole part (p); additionally, and according to the discussion above, we require that such pole terms be longitudinally coupled, so that the physical requirement that these contributions act as ``dynamical Nambu--Goldstone bosons'' and decouple from physical observables is satisfied~\cite{Jackiw:1973tr,Jackiw:1973ha,Cornwall:1973ts,Eichten:1974et,Poggio:1974qs,Aguilar:2011xe,Ibanez:2012zk}. Accordingly, we write
\begin{subequations}
\begin{align}
\widehat{\Gamma}_{\mu\nu\rho}(q,k_1)&=\g_{\mu\nu\rho}(q,k_1)
+\gp_{\mu\nu\rho}(q,k_1),\\
\widehat{\Gamma}_{\mu}(q,k_2)&=
\g_{\mu}(q,k_2)+\gp_{\mu}(q,k_2),
\end{align}
\label{altwr}	
\end{subequations}
with the pole parts that can be written in full generality as
\begin{subequations}
\begin{align}
\gp_{\mu\nu\rho}(q,k_1)
&=
\frac{q_\mu}{q^2}\,\widehat C_{\nu\rho}(q,k_1),\\
\gp_{\mu}(q,k_2)
&=
\frac{q_\mu}{q^2}\,\widehat C(q,k_2).
\end{align}
\label{UIB}
\end{subequations}

Next, in order to preserve BRST symmetry, we demand that the STIs maintain their exact form in the presence of these poles. Therefore,~\1eq{BQids} will be replaced by
\begin{subequations}
\begin{align}
q^\mu \g_{\mu\nu\rho}(q,k_1) + \widehat C_{\nu\rho}(q,k_1)
&=
i\Delta^{-1}_{\nu\rho}(k_1) - i\Delta^{-1}_{\nu\rho}(k_2),\\
q^\mu \g_{\mu}(q,k_2) + \widehat C(q,k_2)
&=
iD^{-1}(k_1^2) - iD^{-1}(k_2^2).
\end{align}
\label{BQinew}	
\end{subequations}
\begin{detailedcalc}
	Note that if $\widehat{\Gamma}_{\mu\nu\rho}(q,k_1)$ contains poles in $q^2$, then, by virtue of the corresponding BQI~\cite{Binosi:2008qk}, the conventional three-gluon vertex $\Gamma_{\mu\nu\rho}$ contains analogous pole terms. 
\end{detailedcalc}

The above decomposition clarifies the role of the pole terms in the self-energy: the no-pole parts $\g$ contribute to the $g_{\mu\nu}$ component of $\widehat{\Pi}_{\mu\nu}$ and therefore participate in the seagull cancellations; in contrast, the pole parts $\gp\sim q_\mu/q^2$ feed exclusively the $q_\mu q_\nu/q^2$ component, which is not affected by the seagull identity. Since the STIs retain their form, block-wise transversality is preserved: after the seagull cancellation, the surviving contribution to the $g_{\mu\nu}$ part is exactly equal (and opposite in sign) to that multiplying $q_\mu q_\nu/q^2$.

To make these statements explicit, we repeat the analysis of the previous subsection in the limit $q\to 0$. We first derive the analogs of Eqs.~\eqref{BQ2-id}, \eqref{Bc2-id}, and~\eqref{BQ3-id} by expanding both sides of~\1eq{BQinew} around $q=0$. The zeroth-order terms vanish,
\begin{align}
\widehat C_{\alpha\beta}(0,k_1) &= 0;&
\widehat C(0,k_2) &= 0,
\label{C0s}
\end{align}
while the first-order terms yield
\begin{subequations}
\begin{align}
\g_{\mu\nu\rho}(0,k_1)
&=
-i\frac{\partial}{\partial k_1^\mu}\Delta^{-1}_{\nu\rho}(k_1)
-\left.\frac{\partial}{\partial q^\mu}\widehat C_{\nu\rho}(q,k_1)\right|_{q=0},\\
\g_{\mu}(0,k_2)
&=i\frac{\partial}{\partial k_2^\mu}D^{-1}(k_2^2)
-\left.\frac{\partial}{\partial q^\mu}\widehat C(q,k_2)
\right|_{q=0}.
\end{align}
\label{dis-1}
\end{subequations}

\begin{detailedcalc}
	Similarly one finds

\begin{subequations}
\begin{align}
\g_{\mu\nu\rho}(0,-k_2)
&=
i\frac{\partial}{\partial k_2^\mu}\Delta^{-1}_{\nu\rho}(p)
-
\left.
\frac{\partial}{\partial q^\mu}\widehat C_{\nu\rho}(q,-k_2-q)
\right|_{q=0},\\
\g_{\mu}(0,-k_1)
&=-i\frac{\partial}{\partial k_1^\mu}D^{-1}(k_1^2)
-\left.\frac{\partial}{\partial q^\mu}\widehat C(q,-k_1-q)\right|_{q=0}.
\end{align}
\label{dis-2}
\end{subequations}
\end{detailedcalc}

The crucial difference with respect to the pole-free case is now clear: in addition to the terms already present in the identities \3eqs{BQ2-id}{Bc2-id}{BQ3-id}, the new WIs contain derivatives of the functions $\widehat C$ (sometimes referred to as the ``displacement'' terms~\cite{Ferreira:2025anh}). The former trigger again the seagull identity and vanish as before, whereas the latter escape the total annihilation. In particular, one finds
\begin{subequations}
\begin{align}
d\widehat{\Pi}^{1\ell;\mathrm{gl}}(0)
&=
-\frac{1}{2} g^2 C_A
\int_k
\Gamma^{(0)}_{\mu\alpha\beta}(0,k)
\Delta^{\alpha\rho}(k)\Delta^{\beta\sigma}(k)
\left.
\frac{\partial}{\partial q^\mu}
\widehat C_{\sigma\rho}(q,-k-q)
\right|_{q=0},
\label{0a1a2mass}
\\
d\widehat{\Pi}^{1\ell;\mathrm{gh}}(0)
&=
- g^2 C_A
\int_k
\Gamma^{(0)}_{\mu}(0,k)
D^2(k^2)
\left.
\frac{\partial}{\partial q^\mu}
\widehat C(-q,k)
\right|_{q=0}.
\label{0a3a4mass}
\end{align}
\end{subequations}

\begin{detailedcalc}
	It is instructive to verify that the same result for $\widehat{\Pi}(0)$ is obtained by focusing directly on the longitudinal component of the self-energy, which is not subject to seagull cancellations. For definiteness, let us consider the one-loop dressed gluon diagrams. The relevant contribution is obtained by replacing the full $\widehat AQ^2$ vertex by its pole part; one finds
\begin{align}
\widehat{\Pi}^{1\ell;\mathrm{gl}}(q^2)
=
\frac{1}{2} g^2 C_A \frac{q^\mu}{q^2}
\int_k
\Gamma^{(0)}_{\mu\alpha\beta}(q,k)
\Delta^{\alpha\rho}(k)\Delta^{\beta\sigma}(k+q)
\widehat C_{\sigma\rho}(q,-k-q)
+\cdots,
\end{align}
where the ellipses denote terms vanishing as $q\to 0$. Expanding $\widehat C$ around $q=0$ and using~\1eq{C0s}, one obtains
\begin{align}
\widehat{\Pi}^{1\ell;\mathrm{gl}}(q^2)
=\frac{1}{2} g^2 C_A \frac{q^\mu q^\nu}{q^2} L_{\mu\nu},
\label{X}
\end{align}
with
\begin{equation}
L_{\mu\nu}
=
\int_k
\Gamma^{(0)}_{\mu\alpha\beta}(0,k)
\Delta^{\alpha\rho}(k)\Delta^{\beta\sigma}(k)
\left.
\frac{\partial}{\partial q^\nu}
\widehat C_{\sigma\rho}(q,-k-q)
\right|_{q=0}.
\end{equation}
Since $L_{\mu\nu}$ carries no dependence on $q$, it must be proportional to $g_{\mu\nu}$, and one recovers immediately ~\1eq{0a1a2mass}. The same reasoning applies to the remaining contributions, thereby establishing explicitly the transversality of the gluon self-energy even in the presence of massless poles.

\end{detailedcalc}

To continue our analysis further, let us 
consider then a generic scalar function $f(q,k_1,k_2)$ that is antisymmetric under $k_1\leftrightarrow k_2$ and expand it around $q=0$ (so that $k_2=-k_1-q$). Since antisymmetry enforces $f(0,k_1,-k_1)=0$, the leading term in the expansion is linear in $q$,
\begin{align}
f(q,k_1,k_2) = 2 (q{\cdot}k_1) f'(0,k_1,-k_1) + \mathcal{O}(q^2),
\label{Taylor}
\end{align}
where the prime denotes differentiation with respect to $(q+k_1)^2$, followed by the limit $q\to 0$,
\begin{align}
f'(0,k_1,-k_1) \equiv \lim_{q\to 0} \frac{\partial}{\partial (q+k_1)^2} f(q,k_1,-k_1-q).
\label{Der}
\end{align}
By Lorentz invariance, $f'(0,k_1,-k_1)$ depends only on $k_1^2$, {\it i.e.}, $f'(0,k_1,-k_1)=f'(k_1^2)$; so that we get
\begin{subequations}
\label{Der2}
\begin{align}
\left.
\frac{\partial}{\partial q^\mu}
\widehat C_{\sigma\rho}(q,k_1)
\right|_{q=0}&=2k_1^\mu g_{\rho\sigma}\widehat C'_{\mathrm{gl}}(k_1^2),\\
\left.
\frac{\partial}{\partial q^\mu}\widehat{C}(q,k_1)
\right|_{q=0}
&=
2 k_1^\mu \widehat{C}'_{\mathrm{gh}}(k_1^2),
\end{align}%
\end{subequations}
where we have also assumed that that among the five possible tensor structures of $\widehat{C}_{\rho\sigma}$~only the component proportional to $g_{\rho\sigma}$ develops a pole in $q^2$. Setting $\xi=0$, we then get
\begin{subequations}
	\begin{align}
	\widehat{\Pi}^{1\ell;\mathrm{gl}}(0)
&=-2g^2C_A\left(\frac{d-1}d\right)\int_k\!k^2\Delta^2(k^2)\widehat C'_\mathrm{gl}(k^2),\\
\widehat{\Pi}^{1\ell;\mathrm{gh}}(0)
&=2g^2C_A\frac1d\int_k\!k^2 D^2(k^2)\widehat C'_\mathrm{gh}(k^2).
	\end{align}
\end{subequations}
At this point we can use the BQI~\1eq{BQIprop} together with the relation~\1eq{IRG} to obtain
\begin{align}
	m^2&=-i\Pi(0)=-iF(0)\widehat\Pi(0)\nonumber \\
	&=2ig^2C_A\left(\frac{d-1}d\right)F(0)\left[\int_k\!k^2\Delta^2(k^2)\widehat C'_\mathrm{gl}(k^2)-\frac1{d-1}\int_k\!k^2 D^2(k^2)\widehat C'_\mathrm{gh}(k^2)\right].
\end{align}
The expression above can be written solely in terms of of  quantum Green functions, by invoking the additional BQIs~\cite{Ferreira:2023fva}
\begin{align}
	C_\mathrm{gl}(q^2)
&=F(0)\widehat C_\mathrm{gl}(q^2);&
C_\mathrm{gh}(q^2)
&=F(0)\widehat C_\mathrm{gh}(q^2),
\end{align}
so that, setting $d=4$, we are left with the result
\begin{align}
	m^2
	&=\frac{3}2g^2C_A\left[\int_k\!k^2\Delta^2(k^2) C'_\mathrm{gl}(k^2)-\frac1{3}\int_k\!k^2 D^2(k^2) C'_\mathrm{gh}(k^2)\right].
	\label{gluonmass}
\end{align} 
Then $m^2\neq0$ provided that at least one between $C'_\mathrm{gl}$ and $C'_\mathrm{gh}$ is different from zero, in which case a gluon mass has been generated dynamically.	That this is indeed the case for both amplitudes amplitudes has been established in a large number of publications (see~\cite{Ferreira:2025anh} and references therein); in addition, it is known that the ghost amplitude is sub-leading with roughly $C'_\mathrm{gl}\sim5C'_\mathrm{gh}$~\cite{Aguilar:2017dco}. 

\begin{detailedcalc}
	Notice that the procedure followed to arrive at  the result in~\1eq{gluonmass} elucidates how longitudinally coupled poles can evade the seagull identity to generate a non-vanishing value for the gluon propagator. Indeed, the transversality of the gluon self-energy allows for its unrestricted contraction with the projector $P_{\mu\nu}(q)$, which automatically removes the corresponding pole contributions from all vertices. This leads to an apparent paradox: if the poles no longer contribute explicitly, what generates the gluon mass? Its resolution: the fact that the WIs get displaced according to~\2eqs{dis-1}{dis-2}.
\end{detailedcalc}

\section{\label{six}The Process Independent QCD Effective Charge}

Let us go back now to the RGI combination introduced in~\1eq{dhathat}:
\begin{align}
	\widehat{\hspace{-0.13cm}\widehat{d}}\mbox{\ }(q^2)&=\alpha(\mu^2)\frac{\Delta(q^2;\mu^2)}{[1+G(q^2;\mu^2)]^2}.
\end{align}
Our detailed understanding of the IR sector of QCD---particularly of the gauge-sector dynamics and the emergent IR behavior of the gluon propagator and ghost dressing function---has established that both quantities saturate to a non-vanishing value in the deep IR. As a result, the combination above extends over the entire momentum region, {\it i.e.}, $\forall q^2\in\mathbb{R},\ 0\leq\hspace{0.13cm}\widehat{\hspace{-0.13cm}\widehat{d}}\mbox{\ }(q^2)<\infty$.

To be sure,
\begin{align}
\widehat{\hspace{-0.13cm}\widehat{d}}\mbox{\ }(0)&=
\frac{\alpha(\mu^2)}{\widehat{\widehat{m}}_{\mathrm{gl}}^2(\mu^2)}
=
\frac{\alpha_0}{m_0^2},
\qquad
\widehat{\widehat{m}}_{\mathrm{gl}}^2(\mu^2)
=
\frac{m_{\mathrm{gl}}^2(0;\mu^2)}{F^2(0;\mu^2)}.
\label{d0}
\end{align}

\begin{figure}
	\includegraphics[scale=0.345]{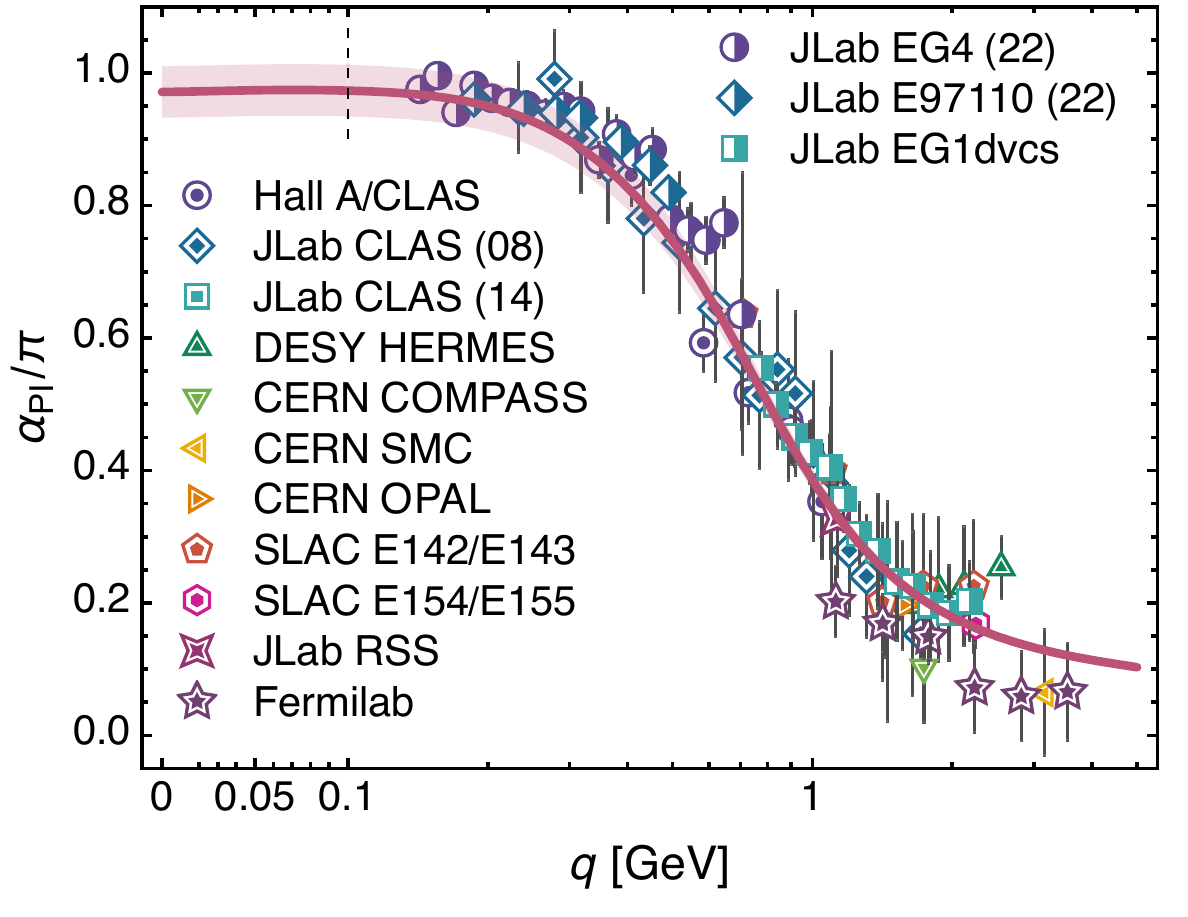}
	\includegraphics[scale=0.345]{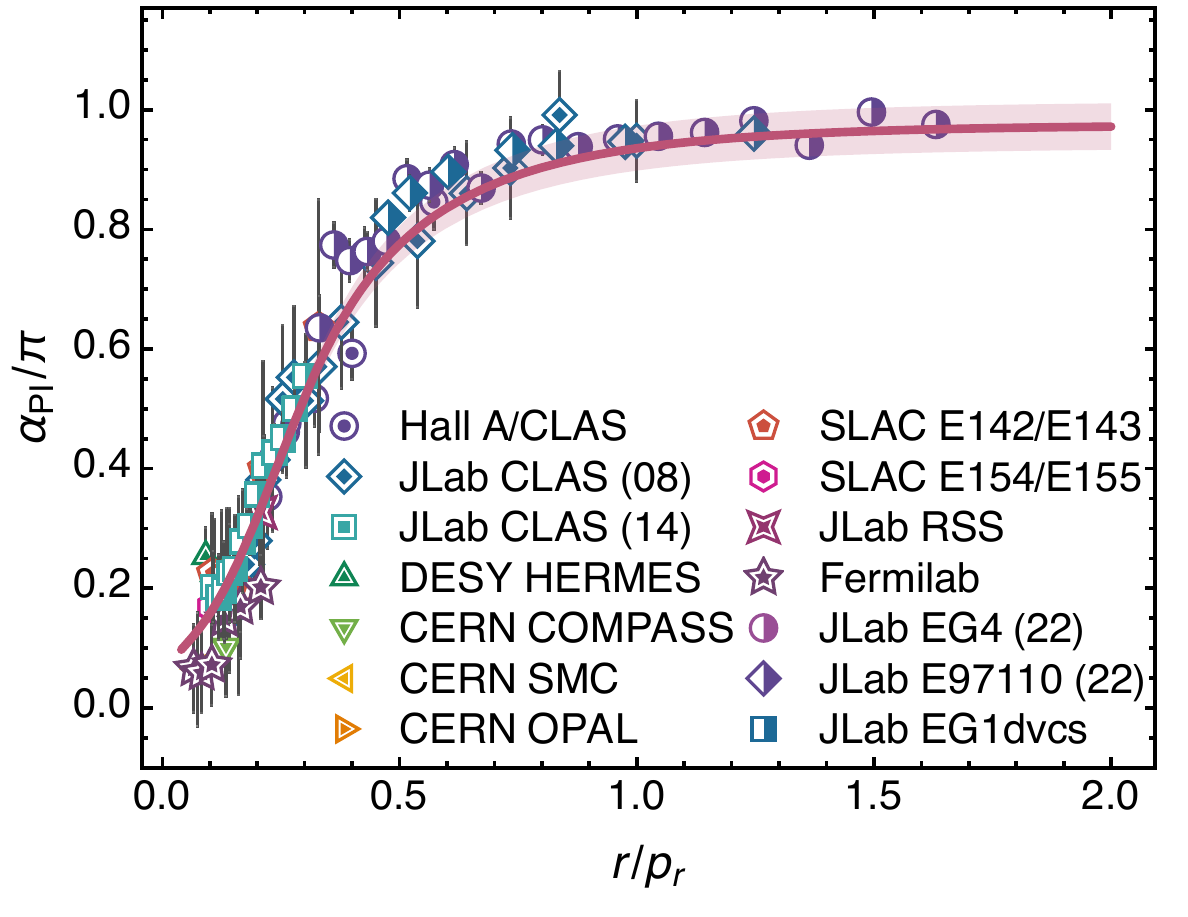}
	\caption{\label{fig:eff-chrg}The predicted QCD all-order effective charge as a function of $q$ (left) and $r/p_r$ (right, with $p_r$ the proton radius), obtained using~\1eq{QCD-effchrg} and the most precise unquenched lattice results for QCD's gauge sector~\cite{Cui:2019dwv}. Plotted are also world data on the process-dependent $\alpha_{g_1}$ defined via the Bjorken sum rule, see~\1eq{alpha-g1}.}
\end{figure}

Since $\hspace{0.13cm}\widehat{\hspace{-0.13cm}\widehat{d}}$ carries dimensions of inverse mass squared, a dimensionless effective charge is obtained by factoring out an RGI mass scale. To this end, consider the RGI quantity~\cite{Binosi:2016nme,Rodriguez-Quintero:2018wma,Cui:2019dwv}
\begin{align}
\label{calD}
\mathcal{D}(k^2) & = \frac{\Delta_{\rm F}(k^2;\zeta)}{ m_0^2\Delta_{\rm F}(0;\zeta)},
\end{align}
employing for $\Delta_{\rm F}$ a parametrization of continuum- and/or lattice-QCD calculations of the canonical gluon two-point function built such that  
\begin{eqnarray}
\label{eq:Dasym}
\frac 1 {\mathcal{D}(k^2)} = \left\{ 
\begin{array}{lr}
m_0^2+\mathrm{O}(k^2 \ln{k^2}) & k^2 \gg m_0^2 \\
k^2+ \mathrm{O}(1) & k^2 \ll m_0^2
\end{array} 
\right., 
\end{eqnarray}
so that the nonperturbative IR behaviour is preserved and the UV anomalous dimension remains in $\hspace{0.13cm}\widehat{\hspace{-0.13cm}\widehat{d}}(k^2)$. Then the dimensionless ratio 
\begin{equation}
\alpha_\mathrm{PI}(q^2) = \frac{\widehat{\hspace{-0.13cm}\widehat{d}}(q^2)}{{\cal D}(q^2)},
\end{equation}
defines a process independent and RGI running coupling valid for all $q^2$. Using \2eqs{F1GL}{IRG}, one finally finds~\cite{Aguilar:2009nf,Aguilar:2010gm,Binosi:2014aea,Binosi:2016nme,Rodriguez-Quintero:2018wma,Zafeiropoulos:2019flq,Cui:2019dwv}
\begin{align}
\alpha_\mathrm{PI}(q^2)
&=
\alpha_0
\frac{\Delta(q^2)}{{\cal D}(q^2)}
\left[
\frac{F(q^2;\mu^2)/F(0;\mu^2)}{1-L(q^2;\mu^2)F(q^2;\mu^2)}
\right]^2
&\overset{q^2\lesssim\mu^2}{=}
\alpha_0
\left[
\frac{F(q^2;\mu^2)/F(0;\mu^2)}{1-L(q^2;\mu^2)F(q^2;\mu^2)}
\right]^2.
\label{QCD-effchrg}
\end{align}
This effective charge has a number of notable properties: it is RGI, process independent, coincides with the perturbative QCD coupling in the UV, saturates to $\alpha_0$ in the IR, and is parameter free, being entirely expressed in terms of quantities that can be computed using continuum and/or lattice methods. Indeed, using lattice results for the gluon and ghost propagators obtained with $N_f=3$ dynamical quarks and a physical pion mass, and choosing a renormalization point $\mu=3.6$~GeV, one finds~\cite{Zafeiropoulos:2019flq,Cui:2019dwv}
\begin{align}
g^2(\mu^2) &= 4.44,
\qquad
m_{\mathrm{gl}}(0;\mu^2) = 0.445~\mathrm{GeV},
\qquad
m_0 = 0.428~\mathrm{GeV},
\end{align}
yielding
\begin{align}
\widehat{\hspace{-0.13cm}\widehat{d}}(0) &= 16.6(4)~\mathrm{GeV}^{-2},
\qquad
\alpha_0 = m_0^2\, \hspace{0.13cm} \widehat{\hspace{-0.13cm}\widehat{d}}(0) = 0.97(4)\pi.
\end{align}
The resulting effective charge is shown in Fig.~\ref{fig:eff-chrg}, together with the process-dependent coupling $\alpha_{g_1}$ defined via the Bjorken sum rule~\cite{Bjorken:1966jh,Bjorken:1969mm}
\begin{equation}
\int_0^1\!dx\,
\big[g_1^p(x,q^2)-g_1^n(x,q^2)\big]
=
\frac{g_A}{6}
\left[
1-\frac{\alpha_{g_1}(q^2)}{\pi}
\right].
\label{alpha-g1}
\end{equation}
Remarkably, the two effective charges exhibit near-perfect agreement over the entire momentum range. While the UV matching is guaranteed by asymptotic freedom, the agreement in the IR is highly nontrivial and reflects the correct incorporation of nonperturbative gluon-ghost dynamics~\cite{Binosi:2016nme,Rodriguez-Quintero:2018wma,Deur:2023dzc,Brodsky:2024zev}. This correspondence identifies the Bjorken sum rule as a practical probe of a QCD analog of the Gell-Mann--Low effective charge~\cite{GellMann:1954fq}.

For phenomenological applications it is useful to provide a smooth interpolation for $\alpha_\mathrm{PI}(q^2)$. This may be achieved through
\begin{align}
\alpha_\mathrm{PI}(q^2)
&=
\frac{1}
{\frac{\beta_m}{4\pi}\log\frac{{\cal Q}^2(q^2)}{\Lambda_{\mathrm{QCD}}^2}};&
{\cal Q}^2(q^2)
&=
\frac{a_0^2 + a_1 q^2 + q^4}{b_0 + q^2},
\label{nptcoupl}
\end{align}
with $\beta_m = \beta_0-2N_f/3$.
Fixing the parameters in the momentum-subtraction scheme and choosing $N_f=4$ with $\Lambda_{\mathrm{QCD}}=0.52$~GeV, a least square fit yields (in GeV$^2$)
\begin{align}
a_0 &= 0.5138;&
a_1 &= 0.4814;&
b_0 &= 0.5952,
\end{align}
or, equivalently~\cite{Binosi:2022djx},
\begin{align}
a_0 &= 1.9\,\Lambda_{\mathrm{QCD}}^2;&
a_1 &= 1.7805\,\Lambda_{\mathrm{QCD}}^2;&
b_0 &= 2.2010\,\Lambda_{\mathrm{QCD}}^2.
\end{align}

Evidently, nonperturbative dynamics replaces the perturbative argument $q^2/\Lambda_{\mathrm{QCD}}^2$ of the logarithm in~\1eq{thed} with ${\cal Q}^2(q^2)/\Lambda_{\mathrm{QCD}}^2$, eliminating the Landau pole and introducing the ``hadron scale''
\begin{equation}
\zeta_H
=
{\cal Q}(\Lambda_{\mathrm{QCD}}^2)
\approx
\sqrt2\,\Lambda_{\mathrm{QCD}},
\label{Hscale}
\end{equation}
which marks the transition between hard and soft dynamics. For $q^2\lesssim\zeta_H^2$ the effective charge bends toward its IR fixed point $\alpha_0$, the interaction becomes screened, and the theory is driven toward a conformal regime.

\begin{detailedcalc}
We can finally go back to the issue of Gribov copies in the $R_\xi$ gauges we have worked with.

A restriction of the functional integral to the first Gribov region $\Omega$ can be implemented by introducing a horizon term governed by a dynamically determined parameter $\gamma$, see~\cite{Vandersickel:2012tz} and references therein. This term suppresses gauge field configurations for which the Faddeev--Popov operator develops near zero modes, namely configurations that probe regions where $\det \mathcal{M}$ changes sign. The parameter $\gamma$ thus sets an IR scale, $m^4_\gamma\sim g^2\gamma$ associated with the proximity of typical configurations to the Gribov horizon in field space. 

Clearly the dynamical generation of a gluon mass provides an equivalent IR screening mechanism by suppressing long wavelength gauge fluctuations. The scale associated with the horizon condition can be compared with this dynamically generated gluon mass $m_0$, and one finds $m_\gamma\approx m_0$~\cite{Gao:2017uox}. In this regime, the IR dynamics do not require an additional, independent horizon scale beyond $m_0$: IR gluon modes are already efficiently screened, and the field configurations that would otherwise accumulate near the Gribov horizon are dynamically disfavored.

 In this sense, the dynamical gluon mass generation effectively mitigates, if not solves, the Gribov problem: by suppressing the IR fluctuations responsible for driving the system toward $\partial \Omega$, it prevents the functional integral from being dominated by near horizon configurations where gauge copies proliferate, vindicating our use of the $R_\xi$/BF gauges, the only gauges in which the calculations put forward in these notes can be effectively carried out, to obtain the most complete description of QCD's IR dynamics. 

\end{detailedcalc}

\section{Instead of Conclusions}

The reader that has made it this far, should now be able to engage in a meaningful discussion on the dynamical generation of a gluon mass, while avoiding the common misconceptions and pitfalls that often arise when addressing such a subtle and technically delicate subject.

The dynamical generation of a gluon mass is a cornerstone of the broader phenomenon known as emergent hadron mass (EHM)~\cite{Roberts:2020hiw}. EHM describes how the vast majority of the mass in the visible Universe was generated approximately $10^{-6}$ seconds after the Big Bang. Although the QCD Lagrangian is written in terms of massless gluon and (in the chiral limit) quark fields, strong interactions, as we have explicitly seen at least in the case of gluons, dress these elementary degrees of freedom into complex quasiparticles. These quasiparticles are characterized by dynamically generated, momentum-dependent mass functions that become large in the IR, reaching values of roughly one-half (for gluons) and one-third (for quarks) of the proton mass, $\sim 1\,\mathrm{GeV}$~\cite{Roberts:2021nhw,Binosi:2022djx,Ding:2022ows, Roberts:2022rxm,Raya:2024ejx}. 

Subsequently, these massive quasiparticles bind to form hadrons, whose properties and observable characteristics carry the imprint of emergent hadron mass. In this way, mass generation in QCD is not the consequence of explicit mass terms in the fundamental Lagrangian, but rather the outcome of nonperturbative dynamics operating at IR scales.

We have only begun to uncover the full implications of EHM. It is our hope that some of you will now feel prepared to contribute to this ongoing effort, especially in view of the intense experimental program being pursued worldwide in hadron physics with the operation, construction and planning of facilities like the Jefferson Lab running at 12 GeV, AMBER at CERN, and the electron ion colliders planned in the USA (EIC) and China (EicC). Such centres are in operation \cite{Ent:2015kec, Adams:2018pwt, Andrieux:2020, Roberts:Strong2020, Quintans:2022utc}, under construction \cite{Aguilar:2019teb, Arrington:2021biu}, or being planned \cite{Anderle:2021wcy, Wang:2022xad, Accardi:2023chb}.
 
\begin{acknowledgments}

	More than twenty years separate my first contribution to the subject of this paper from the most recent one. I am deeply indebted to the many people who have accompanied me throughout this journey, in particular J.~Papavassiliou and C.~D.~Roberts.

\end{acknowledgments}


\end{document}